\documentclass[aps,prb,twocolumn,superscriptaddress,nofootinbib,final]{revtex4-2}

\pdfoutput=1

\usepackage[colorlinks=true,allcolors=blue]{hyperref}
\usepackage{graphicx}
\usepackage{amsmath,amssymb,mathrsfs,array,longtable,delarray}
\usepackage{dcolumn}
\usepackage{bm}
\usepackage{textcomp} 

\begin{document}

\title{Enhanced effective masses, spin-orbit polarization, and dispersion relations \\ in 2D hole gases under strongly asymmetric confinement}

\author{N. A. Cockton}
\thanks{These authors contributed equally to this work}
\affiliation{Department of Physics and Astronomy, University of Waterloo, Waterloo N2L 3G1, Canada}

\author{F. Sfigakis}
\thanks{These authors contributed equally to this work}
\email{francois.sfigakis@uwaterloo.ca}
\affiliation{Institute for Quantum Computing, University of Waterloo, Waterloo N2L 3G1, Canada}
\affiliation{Department of Electrical and Computer Engineering, University of Waterloo, Waterloo N2L 3G1, Canada}
\affiliation{Department of Chemistry, University of Waterloo, Waterloo N2L 3G1, Canada}

\author{M. Korkusinski}
\thanks{These authors contributed equally to this work}
\affiliation{Emerging Technologies Division, National Research Council of Canada, Ottawa K1A 0R6, Canada}

\author{S. R. Harrigan}
\affiliation{Department of Physics and Astronomy, University of Waterloo, Waterloo N2L 3G1, Canada}
\affiliation{Institute for Quantum Computing, University of Waterloo, Waterloo N2L 3G1, Canada}
\affiliation{Waterloo Institute for Nanotechnology, University of Waterloo, Waterloo N2L 3G1, Canada}

\author{G. Nichols}
\affiliation{Department of Physics and Astronomy, University of Waterloo, Waterloo N2L 3G1, Canada}
\affiliation{Institute for Quantum Computing, University of Waterloo, Waterloo N2L 3G1, Canada}

\author{\\Z. D. Merino}
\affiliation{Department of Physics and Astronomy, University of Waterloo, Waterloo N2L 3G1, Canada}
\affiliation{Institute for Quantum Computing, University of Waterloo, Waterloo N2L 3G1, Canada}

\author{T. Zou}
\affiliation{Department of Physics and Astronomy, University of Waterloo, Waterloo N2L 3G1, Canada}

\author{A. C. Coschizza}
\affiliation{Department of Physics and Astronomy, University of Waterloo, Waterloo N2L 3G1, Canada}

\author{T. Joshi}
\affiliation{Institute for Quantum Computing, University of Waterloo, Waterloo N2L 3G1, Canada}
\affiliation{Department of Electrical and Computer Engineering, University of Waterloo, Waterloo N2L 3G1, Canada}

\author{A. Shetty}
\affiliation{Institute for Quantum Computing, University of Waterloo, Waterloo N2L 3G1, Canada}
\affiliation{Department of Chemistry, University of Waterloo, Waterloo N2L 3G1, Canada}

\author{M. C. Tam}
\affiliation{Department of Electrical and Computer Engineering, University of Waterloo, Waterloo N2L 3G1, Canada}
\affiliation{Waterloo Institute for Nanotechnology, University of Waterloo, Waterloo N2L 3G1, Canada}

\author{\\Z. R. Wasilewski}
\affiliation{Department of Physics and Astronomy, University of Waterloo, Waterloo N2L 3G1, Canada}
\affiliation{Institute for Quantum Computing, University of Waterloo, Waterloo N2L 3G1, Canada}
\affiliation{Department of Electrical and Computer Engineering, University of Waterloo, Waterloo N2L 3G1, Canada}
\affiliation{Waterloo Institute for Nanotechnology, University of Waterloo, Waterloo N2L 3G1, Canada}

\author{S. A. Studenikin}
\author{D. G. Austing}
\affiliation{Emerging Technologies Division, National Research Council of Canada, Ottawa K1A 0R6, Canada}

\author{J. Baugh}
\email{baugh@uwaterloo.ca}
\affiliation{Department of Physics and Astronomy, University of Waterloo, Waterloo N2L 3G1, Canada}
\affiliation{Institute for Quantum Computing, University of Waterloo, Waterloo N2L 3G1, Canada}
\affiliation{Department of Chemistry, University of Waterloo, Waterloo N2L 3G1, Canada}
\affiliation{Waterloo Institute for Nanotechnology, University of Waterloo, Waterloo N2L 3G1, Canada}

\author{J. B. Kycia}
\email{jkycia@uwaterloo.ca}
\affiliation{Department of Physics and Astronomy, University of Waterloo, Waterloo N2L 3G1, Canada}
\affiliation{Institute for Quantum Computing, University of Waterloo, Waterloo N2L 3G1, Canada}


\begin{abstract}
The dispersion of Rashba-split heavy-hole subbands in GaAs two-dimensional hole gases (2DHGs) is difficult to access experimentally because strong heavy-hole–light-hole mixing produces nonparabolicity and breaks the usual correspondence between carrier density and Fermi wave vector. Here we use low-field magnetotransport ($B<1$~T) to reconstruct the dispersions of the two spin-orbit-split heavy-hole branches (HH$-$, HH$+$) in undoped (100) GaAs/AlGaAs single heterojunction 2DHGs operated in an accumulation-mode field-effect geometry. The dopant-free devices sustain out-of-plane electric fields up to 26 kV/cm while maintaining mobilities up to 84 m$^2$/Vs and exhibiting a spin-orbit polarization as large as 36\%. Fourier analysis of Shubnikov–de Haas (SdH) oscillations resolves the individual HH$-$/HH$+$ subband densities; fitting the temperature dependence of the corresponding Fourier amplitudes yields both branch-resolved SdH effective masses over the same magnetic field window. SdH regimes in which reliable subband parameters can be extracted are delineated. Over 2DHG densities $(0.76–1.9) \times 10^{15}$ /m$^2$, the HH$-$ mass is nearly density independent ($\approx 0.34m_e$), implying a near-parabolic HH$-$ dispersion below the first LH$+$/HH$-$ anticrossing, whereas HH$+$ exhibits strong nonparabolicity with an effective mass that increases with density. Combining the extracted dispersions yields a transport-based determination of the spin-orbit splitting energy $\Delta_\textsc{hh}$ between HH$-$ and HH$+$ as a function of in-plane wave vector. Parameter-free Luttinger-model calculations reproduce the qualitative trends but underestimate both masses by a common factor $\approx$ 2, suggesting a many-body renormalization of the heavy-hole mass in this strongly asymmetric regime.
\end{abstract}

\maketitle

\section{Introduction}

Due to their strong spin-orbit interactions (SOI), two-dimensional hole gases (2DHGs) are interesting for fundamental studies \cite{shayeganbook,manfra2007impact,croxall2019orientation}, as well as applications in spintronics \cite{Zutic2004spintronics, Datta1990electronic, Chuang2015allelectric} and quantum information processing \cite{loss1998quantum, bulaev2005spin, bulaev2007electric}. In particular, GaAs holes are attractive candidates for spin-based qubits \cite{Burkard2023review, studenikin2021review}. Furthermore, applications such as spin-to-photon conversion \cite{Hsiao2020singlephoton} or photon-to-spin conversion \cite{Fujita2019angular, gaudreau2017entanglement} are possible owing to the direct bandgap of GaAs. Fundamentally, a deeper understanding of SOI in GaAs holes paves the way to a better understanding in other material systems with broad application potential such as InAs, InSb, and Ge.

Unlike the conduction band, the valence band in GaAs is profoundly affected by SOI. One important manifestation is that the effective mass $m^*$ varies with 2DHG density: the dispersion relation in $k$-space can no longer be approximated by a parabola and the 2D density of states is no longer a constant, rendering invalid two common assumptions of semiconductor transport theory. Another striking manifestation of SOI is that, in some cases of broken symmetry, holes from the same subband but with opposite spins experience different effective masses, and are not degenerate in energy, even at zero magnetic field: they are ``spin-orbit split.'' That regime is the focus of this article.

\begin{figure*}[t]
    \includegraphics[width=1.5\columnwidth]{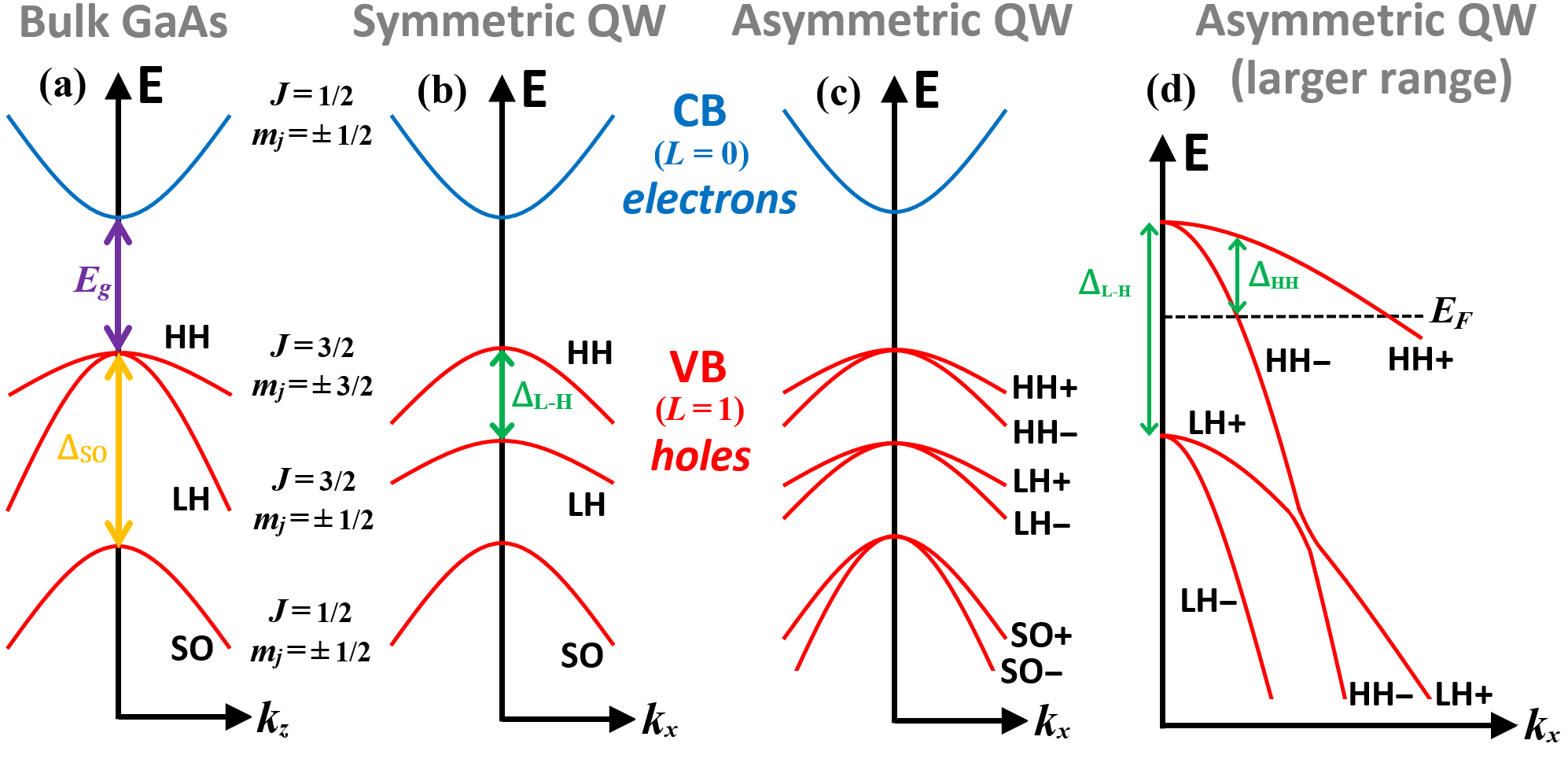}
    \caption{Schematic energy dispersion diagram of the conduction band (CB) and valence band (VB) at small wave vector $k_x$ values for a direct bandgap semiconductor: (a) bulk (3D), (b) symmetric quantum well (2D), and (c) asymmetric quantum well (QW). Values of the orbital angular momentum $L$, the total angular momentum $J$, and the projection $m_j$ of the latter for each subband are indicated. (d) Enlarged energy dispersion diagram for the LH and HH subbands versus in-plane $k_x$ values, illustrating the anticrossing of HH$-$/LH$+$ when taking into account LH/HH mixing. The HH$+$ and HH$-$ dispersions are shown near a Fermi energy $E_F$ typically observed in experiments. }
    \label{Fig:VB}
\end{figure*}

Past transport studies in GaAs spin-orbit split 2DHGs have mostly focused on the (311)A crystal orientation of the 2DHG plane due to the fact that, historically, higher mobilities were achieved compared to the (100) 2DHG plane orientation. The studies in (311)A 2DHGs, all performed in the weak spin-orbit splitting regime, often agreed with predictions from Luttinger theory \cite{lu1998tunable,Chiu2011}. However, transport studies with 2DHGs in GaAs (100) have given inconsistent results \cite{habib2004spin,nichele2014spin}, and disagreed with theoretical predictions. Furthermore, the cyclotron resonance effective mass in (100) GaAs single heterojunctions \cite{lu2008cyclotron} appear to be inconsistent with those measured by transport \cite{habib2004spin,nichele2014spin}. These fundamental gaps in our understanding of hole transport and band structure of the valence band hamper progress towards applications.

In this article, we used accumulation-type devices in GaAs (100) single heterojunctions with very high electric fields (up to 26 kV/cm) across the 2DHG. These fields enabled the observation of the second largest spin-orbit polarization ($\sim$\,36\% for $B<1$~T) in GaAs 2DHGs to our knowledge, whether in a single heterojunction or quantum well, (100) or (311) crystal orientations. Despite the strongly asymmetric confining potential, we also observed the highest 2DHG mobilities (up to $\sim$\,84~m$^2$/Vs at a density of 1.7$\times$10$^{15}$/m$^2$) reported in a GaAs single heterojunction to our knowledge. Thus, our experiments are performed in the regime of both high mobility and high spin-orbit splitting. Previous studies were performed in one of these two regimes, but not both.

We report the values of the two effective masses for spin-orbit split heavy holes from transport experiments over a wide range of 2DHG densities, using techniques pioneered by Refs.~\onlinecite{nichele2014spin}, without assuming a form of the band structure.
We observe the smaller effective mass to be independent of 2DHG density, while the larger mass varies approximately linearly over a range of 2DHG densities from 0.7$\times$10$^{15}$/m$^2$ to 1.9$\times$10$^{15}$/m$^2$. Comparing experiments with numerical predictions from the Luttinger model, we observe an enhancement of both effective masses by a common factor, which is itself nearly independent of 2DHG density. Knowledge of both masses allows the empirical derivation of analytical expressions for the dispersion relations for both branches of spin-orbit split heavy holes. The branch with the smaller effective mass is parabolic in $k$-space, while the one with the larger effective mass is non-parabolic. Finally, we reconcile results from transport and cyclotron resonance studies on hole effective masses in GaAs (100) single heterojunction 2DHGs.

The paper is organized as follows. Section~\ref{sec:background} provides a brief overview of relevant theory and prior experiments concerning the GaAs valence band. Section~\ref{sec:methods} details the experimental methods, including sample growth, device fabrication, measurement techniques, and general transport characterization. Section~\ref{sec:experiments} presents the Fourier transform analysis of Shubnikov–de Haas oscillations used to measure the spin–orbit splitting, along with the direct extraction of the two corresponding effective masses. Section~\ref{sec:discussion} discusses the broader implications of our results, comparisons with previous work, and the applicability of the Luttinger model. The heavy-hole band structure is also mapped in this section. The Appendix outlines the numerical implementation of our Luttinger model simulations.

\section{Background}
\label{sec:background}

\textbf{Theoretical framework}. Spin-orbit interactions can be grouped into two types. Dresselhaus SOI stem from bulk inversion asymmetry (BIA), mostly determined by the material's crystal structure. Rashba SOI stem from structural inversion asymmetry (SIA), from electric fields generated either internally (e.g., with intentional doping \cite{koga2002rashba, pollanen2015heterostructure}) or externally (e.g., with gates \cite{Nitta1997gatecontrol, croxall2013demonstration}) to the material. Due to SOI, valence bands cannot be approximated by parabolic dispersion relations, and the effective mass has a non-trivial dependence on hole momentum (or carrier density).

It is convenient to describe the band structure near the bandgap ($E_g$) in terms of eigenstates for total angular momentum $J$ and its projection $m_j$, defined with respect to a quantization axis aligned with the growth direction $\hat{z}$. Holes with $J=3/2$ occupy both the light hole (LH) subband ($m_j=\pm 1/2$) and the heavy hole (HH) subband ($m_j=\pm 3/2$). The split-off (SO) subband contains holes with $J=1/2$. Values for $J$ and $m_j$ are obtained from $\vec{J} = \vec{L} + \vec{S}$, where holes have orbital angular momentum $L=1$ and spin $S=1/2$ in all three subbands involved (HH, LH, SO). Although $J$ and $m_j$ are only approximately good quantum numbers, they are used here nonetheless for ease of nomenclature.

In bulk (3D) GaAs [see Figure \ref{Fig:VB}(a)], the LH and HH subbands are degenerate for wave vector $k=0$ nm$^{-1}$, but are non-degenerate for $k\neq0$ \cite{winkler2003spin}. The LH and HH subbands are separated from the SO subband by the spin-orbit gap $\Delta_{\textsc{so}}$. The reduced dimensionality of a 2DHG hosted in a symmetric quantum well (QW) lifts the $k=0$ nm$^{-1}$ degeneracy between LH and HH subbands with energy gap $\Delta_{\textsc{l-h}}$ [see Figure \ref{Fig:VB}(b)], where $k_z$ is in the growth direction and $k_{x}$ is in the in-plane direction relative to the quantum well.

For a 2DHG in an asymmetric quantum well [see Fig.~\ref{Fig:VB}(c)], Rashba SOI lift the $m_j=\pm 3/2$ degeneracy in the HH subband, which spin-orbit splits into two subbands labeled HH$+$ ($m_j=+3/2$) and HH$-$ ($m_j=-3/2$) \cite{winkler2008spin} with energy gap $\Delta_{\textsc{hh}}$ [see Fig.~\ref{Fig:VB}(d)]. At the same time, these two spin-orbit-split subbands develop different effective masses, with a larger (smaller) mass for HH$+$ (HH$-$), that vary with $k$. Similarly, the total angular momentum degeneracy is also lifted in the LH and SO subbands, with LH$-$/LH$+$ and SO$-$/SO$+$ appearing [see Fig.~\ref{Fig:VB}(c)].

\begin{figure*}[t]
    \includegraphics[width=1.8\columnwidth]{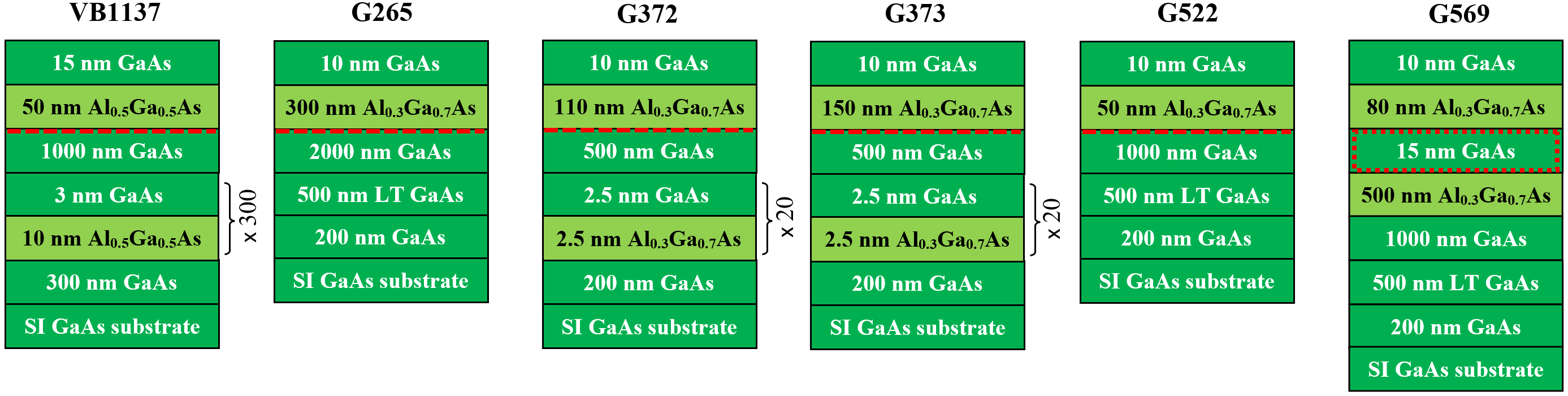}
    \caption{Layer structure of the six GaAs/AlGaAs heterostructures reported here (not to scale). None of the layers are intentionally doped. The red dashed line (box) indicates the location of the 2DHG in a single heterojunction (quantum well).}
    \label{Fig:MBE}
\end{figure*}

At high $k_{x}$ values corresponding to much higher 2DHG densities than those typically used in transport experiments, the HH$-$ and LH$+$ subbands anticross [see Figure \ref{Fig:VB}(d)], and their wavefunctions are hybridized. Although LH/HH mixing is maximized at the anticrossing and the LH subband is unoccupied, the hybridization is present to some extent for \textit{all} values of $k$. This mixing/hybridization between the HH and LH subbands is the principal reason why $J$ and $m_j$ are not good quantum numbers.

A clarifying remark about effective masses is necessary. In bulk (3D) GaAs, HH holes have a higher effective mass ($m^*_{\textsc{hh}}\approx 0.5 m_e$ \cite{Adachi1985GaAs,Pavesi1994photoluminescence}) than LH holes ($m^*_{\textsc{lh}} \approx 0.08 m_e$ \cite{Adachi1985GaAs,Pavesi1994photoluminescence}), where $m_e$ is the electron rest mass. However, in a 2DHG, the effective mass is not isotropic in space. In the (100) growth direction ($k_z$), the effective masses are $m^*_{\textsc{hh},\perp}\approx 0.35 m_e$ and $m^*_{\textsc{lh},\perp} \approx 0.09 m_e$ \cite{Winkler1995excitons,Vurgaftman2001}, whereas they are $m^*_{\textsc{hh},\parallel}\approx 0.11 m_e$ and $m^*_{\textsc{lh},\parallel} \approx 0.21 m_e$ \cite{Winkler1995excitons,HerbertLi2001} within the (100) plane ($k_x, k_y, k_{||}$). We note that the in-plane effective mass of heavy holes is \textit{smaller} than the in-plane effective mass of light holes, which is one of the factors driving the LH/HH anticrossing in Figure \ref{Fig:VB} for 2D systems.

\textbf{Experimental framework}. Conventional Zeeman spin-splitting in GaAs electrons or heavy holes with a single effective mass (i.e., a 2DHG in a symmetric quantum well) results in the simple doubling of the frequency in Shubnikov-de Haas (SdH) oscillations in magnetic field $B$. However, the Rashba spin-orbit splitting of the HH subband into HH$+$ and HH$-$ subbands in an asymmetric quantum well manifests itself by the appearance of a complex beating pattern in SdH oscillations at low $B$ \cite{stormer1983energy,eisenstein1984effect,Papadakis1999effect,habib2004spin,nichele2014spin,yuan2009landau,grbic2004,Rendell2022,grbic2008strong,habib2009spin},
driven by the difference in effective masses between HH$+$ and HH$-$. This complex beating pattern prevents the conventional method for measuring the effective mass, the temperature dependence of a single SdH oscillation at a fixed filling factor, from being used in an asymmetric quantum well.

The effective mass of both HH$+$ and HH$-$ can be extracted from Fourier transform (FT) analysis of the temperature dependence of the complex SdH oscillation beating pattern \cite{habib2004spin,nichele2014spin}, without the need for any assumptions on the parabolicity of the underlying band structure \cite{stormer1983energy,eisenstein1984effect}. Only two studies have been reported to use this technique in the literature \cite{habib2004spin,nichele2014spin}. While these two pioneering studies broadly gave many results consistent with the SOI description above, there were significant discrepancies between them. One study \cite{habib2004spin}, using a gated modulation-doped single heterojunction, demonstrated a very strong dependence of the lower (HH$-$) effective mass on magnetic field. It reported HH$+$/HH$-$ effective masses extrapolated to $B=0$~T, which agreed with Luttinger model predictions. The other study \cite{nichele2014spin}, using a gated asymmetrically-doped quantum well, reported both effective mass values over a wide range of 2DHG densities spanning 2 to $3 \times 10^{15}$/m$^2$, and did not observe any magnetic field dependence. The effective mass values did not agree with Luttinger model predictions.

We investigated the HH$-$/HH$+$ spin-orbit splitting and effective masses in dopant-free (100) GaAs/AlGaAs single heterojunctions to resolve the discrepancies outlined above, and to test theory with data extending beyond previously-achieved experimental parameters, namely larger mobilities and larger spin-orbit splitting. For example, we found no magnetic field dependence of either effective masses for $B<1$ T. Also, both HH$-$/HH$+$ effective masses were resolved over a wider range of 2DHG densities (from $p_{2d} = 0.7 \times 10^{15}$/m$^2$ to $2 \times 10^{15}$/m$^2$) and at much lower densities than previous reports.

Generally in semiconductor transport, theorists describe their work in terms of momentum and energy, whereas experimentalists describe theirs in terms of resistivity and carrier density. With a parabolic band structure, one can easily map results from experiments to theory and vice versa, through the use of the well-known relation $k_{\textsc{f}}^2 = 2\pi p_{2d}$ where $p_{2d}$ is the 2DHG density and $k_{\textsc{f}}$ is the Fermi wavevector. This relation conveniently does not require knowledge of the effective mass. With a non-parabolic band structure, this relation is no longer valid, and directly comparing experiments to theory becomes not straight forward. We developed an approach for mapping the dispersion relation of non-parabolic HH subbands from transport experiments. Our use of single heterojunctions was pivotal in mapping the HH$-$/HH$+$ dispersion relations (shown in Section~\ref{sec:discussion}). It led to the observation that the HH$-$ branch is parabolic in $k$\nobreakdash-space, allowing conventional analysis techniques to be applied. Since both HH$-$/HH$+$ dispersion relations share the same Fermi energy, the direct mapping of the non-parabolic HH$+$ band structure becomes possible. This mapping approach can be extended to other material systems with non-parabolic band structures.

\begin{table}[b]
    \begin{ruledtabular}
    \begin{tabular}{cccccc}
    Sample& Wafer & Type & Superlattice & Parallel & Gate \\
    ID & ID & ~ & in buffer & conduction & dielectric \\
    \hline \vspace{-3mm} \\
    ~A$^\dagger$ & VB1137 & SH & Yes & Yes & HfO$_2$\\ 
    ~B$^\dagger$ & VB1137 & SH & Yes & Yes & Al$_2$O$_3$\\ 
    ~C$^\dagger$ & VB1137 & SH & Yes & Yes & SiO$_2$\\ 
    D & VB1137 & SH & Yes & Yes & SiO$_2$\\ 
    ~E$^\dagger$ & G372 & SH & Yes & Yes & SiO$_2$\\ 
    F & G373 & SH & Yes & Yes & SiO$_2$\\ 
    ~H$^\dagger$ & G265 & SH & No & No & SiO$_2$\\ 
    J & G522 & SH & No & No & SiO$_2$\\ 
    K & G522 & SH & No & No & SiO$_2$\\ 
    M & G569 & QW & No & No & SiO$_2$\\ 
    \end{tabular}
    \end{ruledtabular}
    \caption{Index of all Hall bars reported here. Those marked with a dagger ($\dagger$) are ambipolar: they can host either a 2DEG or 2DHG, depending on the polarity of the top-gate voltage. Wafers are either single heterojunctions (SH) or square quantum wells (QW). Parallel conduction is diagnosed by the lifting of SdH oscillation minima in 2DHGs at low temperatures ($T<100$~mK) and high magnetic fields ($B>2$~T).}
    \label{tab:samples}
\end{table}

\textbf{Dopant-free field effect transistors (FET)}. The semiconductor-insulator-semiconductor field effect transistor (SISFET) \cite{kane1993, Saku1998high, Hirayama1998two, Kawaharazuka2001free, Hirayama2001backgated, Lilly2003resistivity, Valeille2008highly} or the
heterostructure-insulator-gate field effect transistor (HIGFET) \cite{Harrell1999fabrication, Willett2006simple, Wendy2010distinguishing, pan2011impact, ChenJCH2012fabrication, croxall2013demonstration, WangDQ2013influence, Sebastian2016gating, croxall2019orientation} are accumulation devices: they do not conduct unless a voltage is applied via a gate. They are a versatile platform from which many types of devices can be fabricated, such as quantum wires \cite{Klochan2006ballistic, sarkozy2009zero}, quantum dots \cite{see2010algaas, klochan2011observation, mak2013ultra, bogan2017consequences, bogan2018landau, bogan2019single, studenikin2019electrically, ducatel2021single,bogan2021single, padawer2022characterization}, single electron pumps \cite{Brandon2021nonadiabatic}, lateral p-n junctions \cite{dai2013lateral,chung2019quantized,dobney2023formation,tian2023stable}, and single photon sources \cite{hsiao2020single}. Relative to their modulation-doped counterparts, dopant-free devices have exceptional reproducibility and lower disorder \cite{sarkozy2009zero, see2012impact, mak2013ultra, srinivasan2020improving, shetty2022effects}, and enable functionalities not otherwise possible (e.g., ambipolar operation and lateral p-n junctions). This exceptional reproducibility and low disorder have enabled the results reported here. Last but not least, the ability to achieve
the highest reported mobility in a gated 2DHG hosted in a GaAs/AlGaAs single heterojunction while applying a very large electric field allowed us to gain detailed information about the valence band hole spectrum.

\begin{figure*}[t]
    \includegraphics[width=2.0\columnwidth]{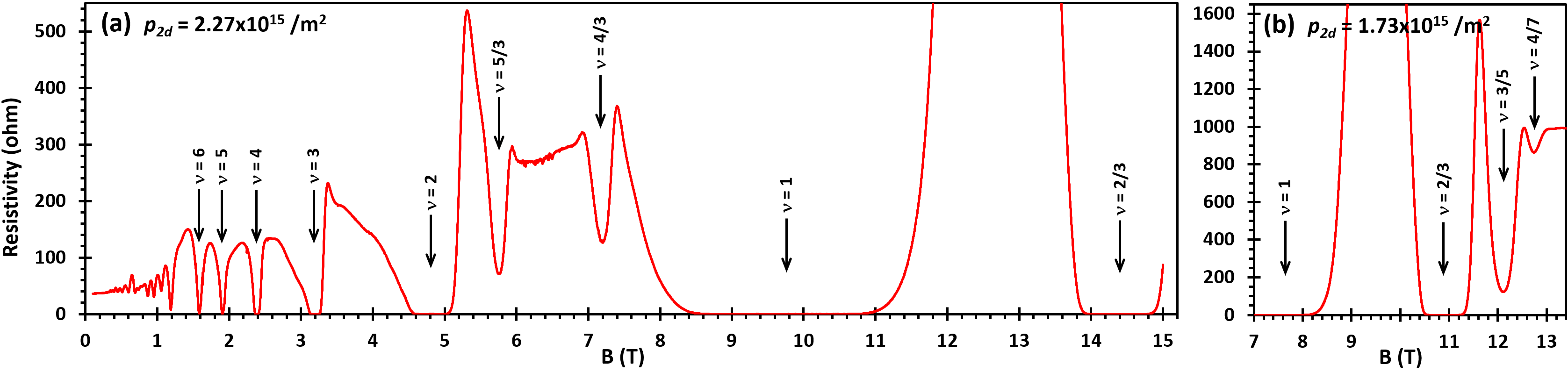}
    \caption{High-field SdH oscillations in sample K at $T=100$ mK for hole densities: (a) $p_{2d} = 2.27 \times 10^{15}$/m$^2$ and (b) $p_{2d} = 1.73 \times 10^{15}$/m$^2$. The latter density corresponds to the mobility peak, 84 m$^2$/Vs, for this sample. In both panels, many fractional quantum Hall states can be observed at filling factors $\nu=$ 5/3, 4/3, 2/3, 3/5, and 4/7, easily identifiable by their quantum Hall plateau (not shown). Even at the highest $p_{2d}$, the high-field SdH oscillation at $\nu=$ 2/3 reaches $\rho_{xx}=0$~\textohm.}
    \label{Fig:FQHE}
\end{figure*}

\section{Experimental methods}
\label{sec:methods}

\textbf{Molecular beam epitaxy (MBE).} Five GaAs/AlGaAs single heterojunctions of varying depths and one quantum well were grown by MBE.\footnote{Wafers G265, G372, G373, G522 and G569 were grown at the University of Waterloo. Wafer VB1137 was grown at Sandia National Laboratories.} There was no intentional doping anywhere in any of the heterostructures, and all were grown on semi-insulating (SI) GaAs (100) substrates. Some wafers (VB1137, G372, and G373) were grown with a smoothing superlattice in the buffer layer, while others (G265, G522, and G569) used low-temperature (LT) GaAs instead. Figure~\ref{Fig:MBE} shows the corresponding MBE layer structure and AlGaAs composition for each wafer.

\textbf{Sample fabrication.} Gated Hall bars were fabricated on all wafers (see Table~\ref{tab:samples} for the list of samples). The Hall bar channel was oriented in the high mobility crystal direction $[1\bar{1}0]$, and was typically 300~\textmu m long and 50~\textmu m wide. Details for the fabrication of recessed ohmic contacts for dopant-free two-dimensional electron gases (2DEGs) and 2DHGs are otherwise identical to and extensively described in Refs.\,\onlinecite{Wendy2010distinguishing, ChenJCH2012fabrication, Deepyanti2016}. Briefly, a large mesa was defined using a solution 1:8:120 (by volume) of H$_2$SO$_4$:H$_2$O$_2$:H$_2$O. Unlike in conventional modulation-doped 2DHGs, the mesa here does not define the shape of the 2DHG, but merely ensures no 2DHG can form underneath bondpads and interconnects. After the deposition of recessed AuBe (1\% Be by weight) \textit{p}\nobreakdash-type ohmic contacts, these were annealed at 520$^\circ$C for 3 minutes in forming gas (5\% of H$_2$ in N$_2$ by vol.). In ambipolar Hall bars, recessed NiAuGe (12\%/88\% of Ge/Au by weight) \textit{n}\nobreakdash-type ohmic contacts were deposited and annealed at 450$^\circ$C for 3 minutes in forming gas (5\% of H$_2$ in N$_2$ by vol.). Next, a 300 nm SiO$_2$ layer was deposited by plasma-enhanced chemical vapor deposition (PECVD). Above this gate dielectric, a patterned Ti/Au top-gate is deposited, overlaps the ohmic contacts, and entirely determines the shape of the 2DHG. Using the field effect from the insulated top-gate, a 2DHG is induced (accumulation mode) at the GaAs/AlGaAs interface located 60$-$310 nm below the GaAs/SiO$_2$ interface. Typical ohmic contact resistances were less than 800 \textohm ~at $B=0$~T, and 5$-$7 k\textohm ~at $B=5$ T.

\begin{figure}[b]
    \includegraphics[width=1.0\columnwidth]{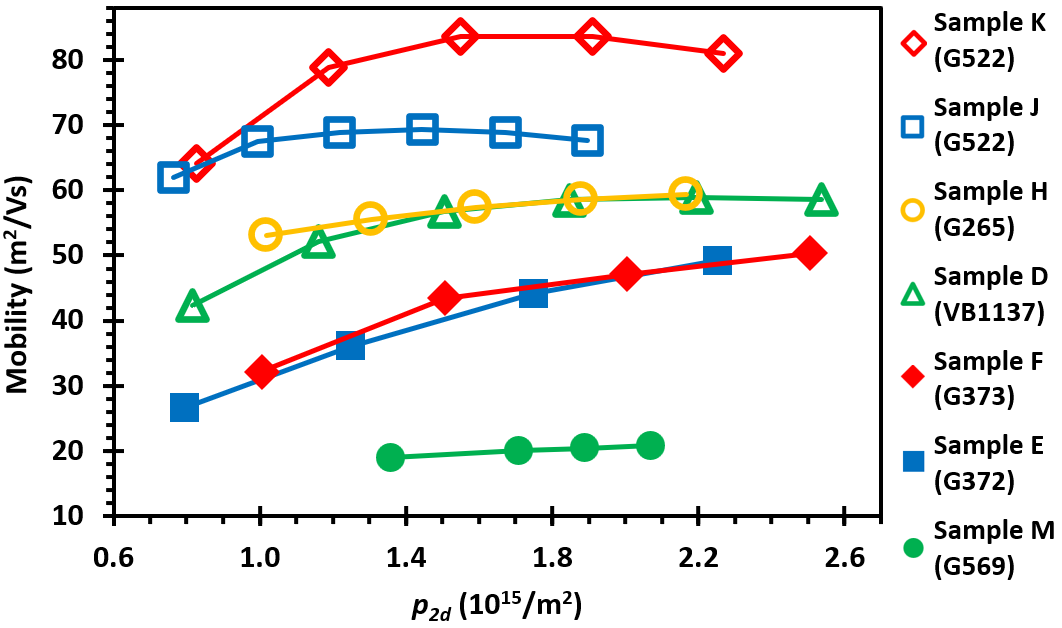}
    \caption{High mobilities at $T<0.1$ K for samples listed in Table~\ref{tab:samples}. Lines are guides to the eye. The mobility of all samples was measured at $B=50$~mT, except for samples K and M which were measured at $B \approx 10$~mT. }
    \label{Fig:mobility}
\end{figure}

\textbf{Transport experiments.} Samples were measured at low temperature in a $^3$He/$^4$He dilution refrigerator. Both cryogenic and room temperature low-pass filters were employed to achieve a hole (electron) temperature of $\sim\,$40~mK. Four-terminal measurements of the longitudinal resistivity $\rho_{xx}$ were probed using standard low frequency lock-in techniques. The temperature dependence data was collected using a small ac excitation current of 1~nA at 17.3~Hz through the Hall bar. The small current excitation prevented self-heating and was critical in achieving a hole temperature of 40~mK. At base temperature, self-heating was observed for currents greater than 3~nA at low magnetic fields, and for magnetic field sweep rates greater than 0.5 T/hour. At $T=77$~K, top-gate current leakage is less than 0.2~pA at a top-gate voltage of $V_g = -5$~V. The correct electron temperature is critical in accurately determining $m^*$. For example, the effective mass for the HH$-$ subband in sample H was measured to be $\sim$\,0.16$m_e$ before the addition of low-pass filters in the experimental setup, and $\sim$\,0.32$m_e$ afterwards.

\textbf{2DHG mobilities.} At high magnetic fields, Figure~\ref{Fig:FQHE} shows the minima of SdH oscillations reach $\rho_{xx}=0$~\textohm, thus excluding the possibility of parallel conduction from another channel. The presence of many fractional quantum Hall states is evidence of high-quality MBE growth. Figure~\ref{Fig:mobility} shows the mobilities of all heterostructures listed in Figure~\ref{Fig:MBE}. At lower densities, the mobility is limited by scattering from non-intentional background impurities, whereas it is limited by interface roughness scattering at higher densities \cite{shetty2022effects,Ando1982electronic}. The next 2D subband only populates at much higher $p_{2d}$, so intersubband scattering is not responsible for the peak in mobility for this range of densities.

The mobilities in Figure~\ref{Fig:mobility} are all lower than record values achieved in modulation-doped GaAs 2DHGs in symmetric quantum wells (up to $1.8\times 10^3$~m$^2$/Vs) \cite{Gupta2024,chung2022record,watson2012exploration,watson2011scattering,Gerl2006,Manfra2005}. However, the mobilities of four samples (K, J, H, and D) are among the highest reported in GaAs/AlGaAs single heterojunctions in this density range \cite{Proskuryakov2002,Manfra2005,Willett2007,Noh2003A}. Dopant-free devices suffer much less from Landau level broadening \cite{Sebastian2016gating} and can go to much lower carrier densities \cite{Noh2003A,Noh2003B,huang2006nonactivated} than modulation-doped devices. These two properties enabled the determination of effective masses for HH$-$/HH$+$ at low densities in Section~\ref{sec:experiments}.

\begin{figure}[t]
    \includegraphics[width=1.0\columnwidth]{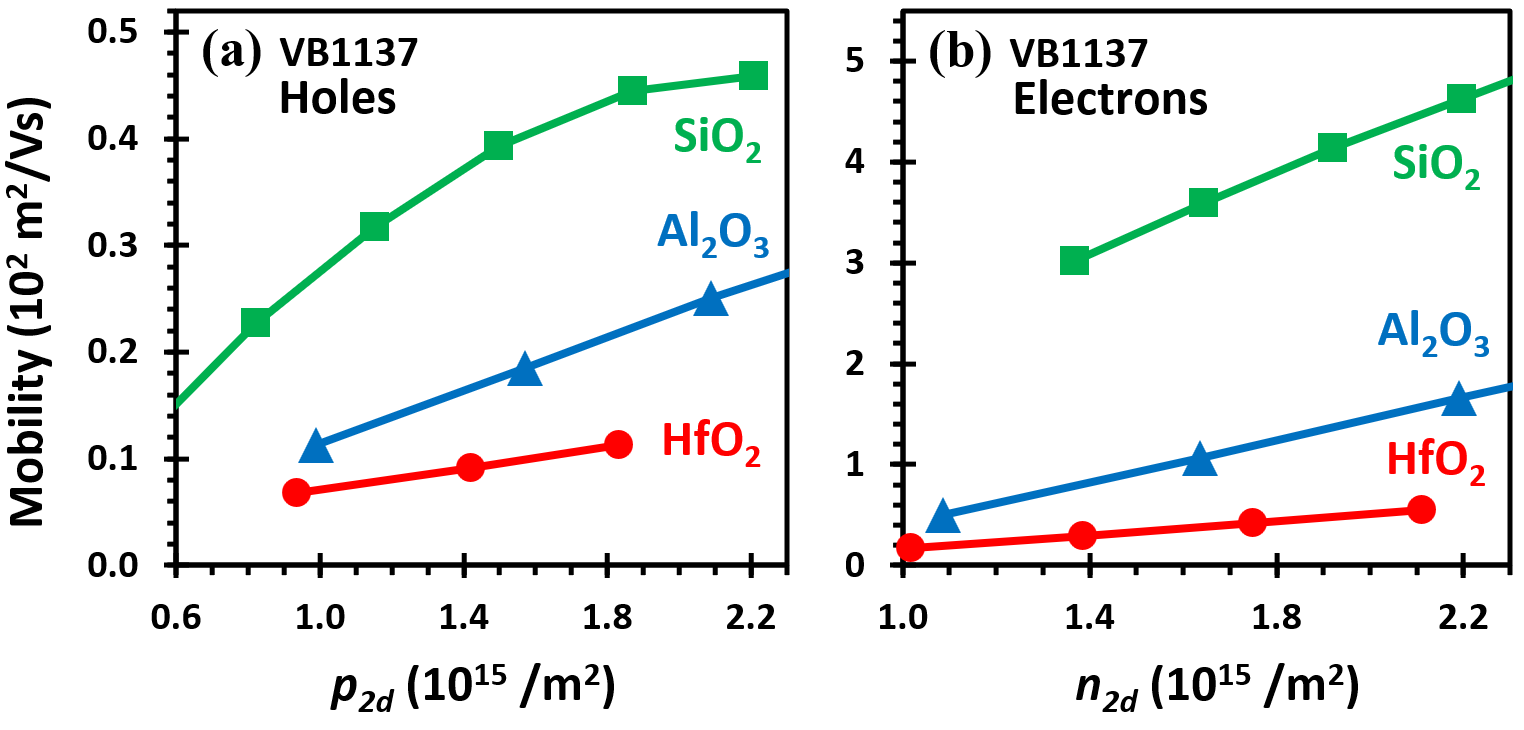}
    \caption{(a) Hole and (b) electron mobilities in ambipolar Hall bars (samples A, B, and C) at $T=1.6$~K, as a function of gate dielectric. Lines are guides to the eye.}
    \label{Fig:oxides}
\end{figure}

\textbf{Gate dielectric and mobilities.} MBE growth conditions are not the only variables determining the ultimate achievable mobilities in dopant-free devices. Sample fabrication processes can be detrimental. Figure~\ref{Fig:oxides} illustrates how the gate dielectric can have a drastic effect on mobility. A shallow wafer (VB1137) was used to emphasize the effects of the gate dielectric: the 2DEG/2DHG is only 65 nm away from the GaAs-oxide interface. Ambipolar Hall bars with both $p$-type and $n$-type ohmic contacts were fabricated with three types of gate dielectrics: 300 nm of SiO$_2$ by PECVD, 30 nm of Al$_2$O$_3$ by atomic layer deposition (ALD), or 30 nm of HfO$_2$ by ALD. The 2DEG mobility drops by an order of magnitude when using HfO$_2$ instead of SiO$_2$. The most likely culprit for the reduction in mobility is the presence of interface traps at the GaAs-oxide interface.\footnote{The quality of a deposited dielectric and of the semiconductor-dielectric interface depend strongly on the deposition technique and conditions, so we do not draw conclusions about the general suitability of SiO$_2$ versus Al$_2$O$_3$ or HfO$_2$ from our limited data.} Interface traps are suspected to affect some of the transport and optical properties of GaAs-based dopant-free devices \cite{Fujita2021distinguishing, shetty2022effects, tian2023stable}.

\begin{figure}[t]
    \includegraphics[width=0.80\columnwidth]{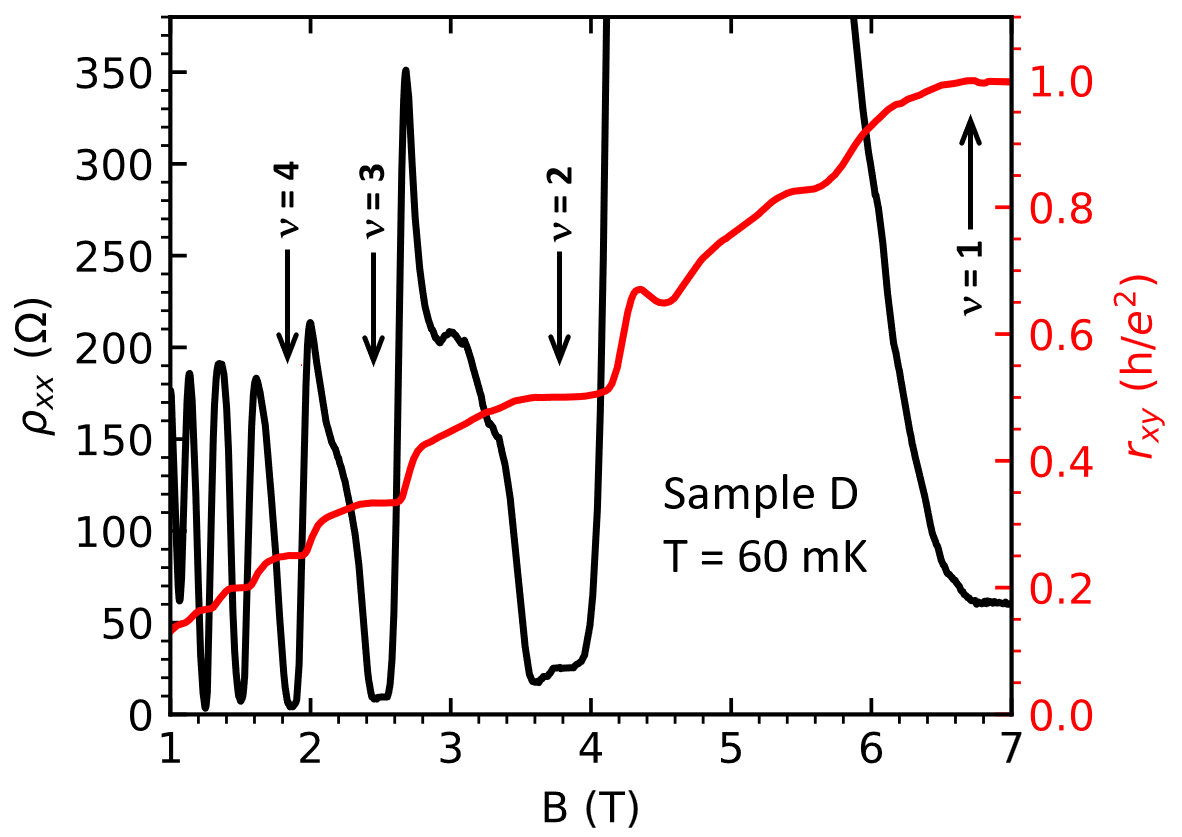}
    \caption{Parallel conduction in a 2DHG, manifested by the monotonically increasing lifting of SdH oscillations minima above $\rho_{xx}=0$ when $B>1.7$~T. Quantized Hall plateaus from $\nu=1$ to $\nu=7$ are visible.}
    \label{Fig:PC}
\end{figure}

\textbf{Parallel conduction.} Well-defined SdH oscillations at low temperatures were observed in all Hall bars. However, samples fabricated from wafers grown with a smoothing superlattice buffer (see VB1137, G372, G373 in Figure~\ref{Fig:MBE}) displayed evidence of parallel conduction. The most sensitive and universal sign for parallel conduction is the lifting of SdH oscillation minima above $\rho_{xx}=0$ at high magnetic fields (e.g., see Figure~\ref{Fig:PC}).

\begin{figure}[b]
    \includegraphics[width=1.0\columnwidth]{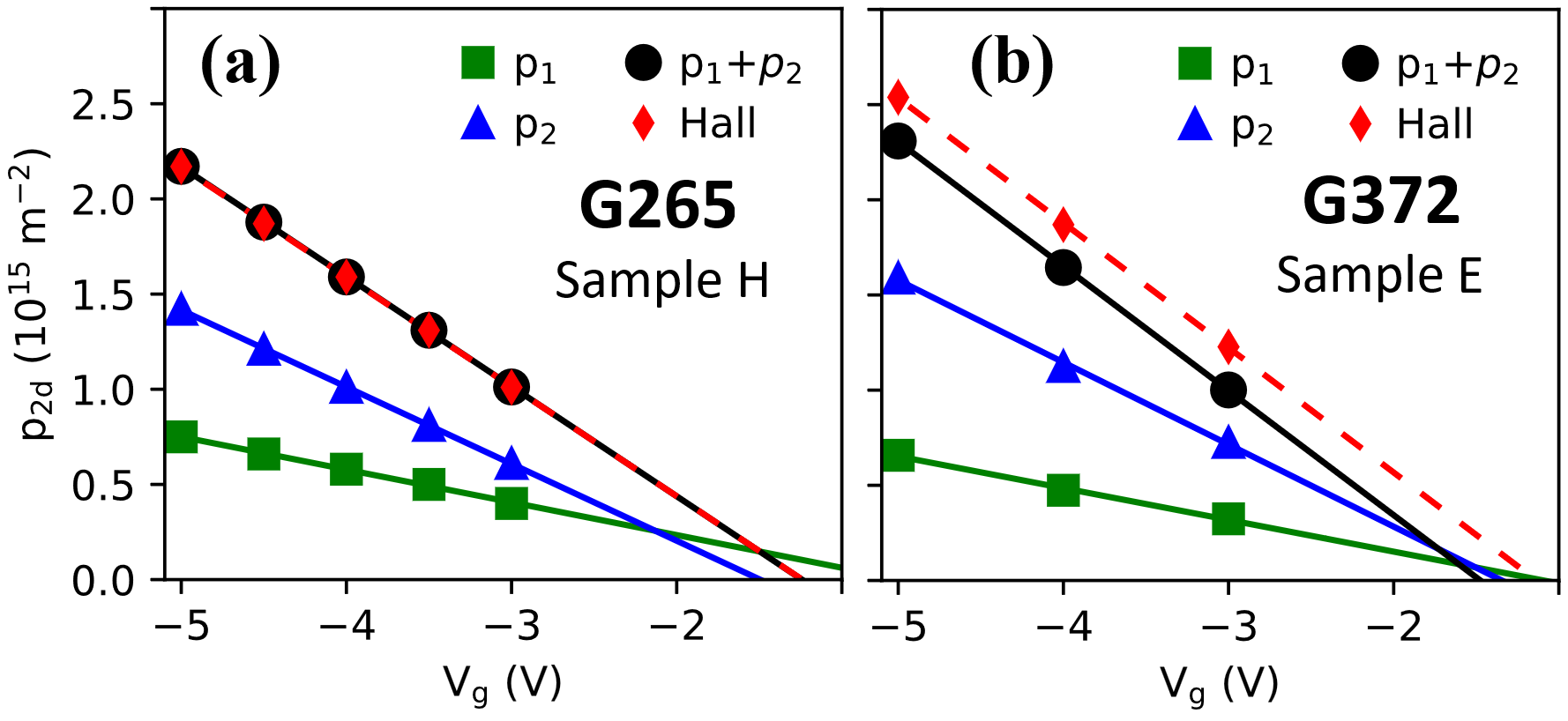}
    \caption{Comparisons between carrier densities extracted from the Hall effect and SdH oscillations for two wafer types: (a) without and (b) with a superlattice buffer. The quantities $p_1$ (squares), $p_2$ (triangles), and $p_1 + p_2$ (circles) were obtained from the Fourier transform (FT) of SdH oscillations measured at 40~mK. Diamond symbols represent the hole densities determined from the low-field Hall effect. Solid lines (red dashed lines) are least-squares linear fits to the FT (Hall) data. }
    \label{Fig:densities}
\end{figure}

Another, less sensitive, symptom of parallel conduction in a 2DHG is a mismatch between the total hole density extracted from the Hall effect and the 2DHG carrier density extracted from the Fourier transform (FT) analysis of SdH oscillations. In our single-heterojunction 2DHGs, the HH$-$/HH$+$ spin-splitting causes two frequencies to appear in the FT of the SdH oscillations, yielding two separate carrier densities (FT freqencies), $p_1$ and $p_2$. The sum of the two HH density components equals the total 2DHG carrier density, $p_1 + p_2$. Ideally, $p_1 + p_2$ should be equal to the Hall density, $p_1 + p_2 = p_{\text{Hall}}$. This is exactly the case for sample~H [see Fig.~\ref{Fig:densities}(a)] and sample~J [see Fig.~\ref{Fig:Subplot_all}(a)].

\begin{table*}[t]
    \begin{ruledtabular}
    \begin{tabular}{ccl}
    SdH regimes & Experimental conditions & Description \\
    \hline \vspace{-3mm} \\
    n/a~~ & $k_B T > (\frac{\hbar eB}{m_1}-h\Gamma) > (\frac{\hbar eB}{m_2}-h\Gamma)$ & SdH oscillations not observable \\
    Regime I~~ & $(\frac{\hbar eB}{m_1}-h\Gamma) > k_B T > (\frac{\hbar eB}{m_2}-h\Gamma)$ & Only $p_1$ peak visible in FT spectra; no other peaks \\
    Regime II~ & $(\frac{\hbar eB}{m_1}-h\Gamma) > g_{zz}^* \mu_B B > k_B T > (\frac{\hbar eB}{m_2}-h\Gamma)$ & Peaks $p_1$ and $2p_1$ visible in FT spectra; no $p_2$ peaks \\
    Regime III & $(\frac{\hbar eB}{m_1}-h\Gamma) > (\frac{\hbar eB}{m_2}-h\Gamma) > k_B T $ & Both $p_1$ and $p_2$ peaks visible in FT spectra \\
    Regime IV & SdH minima reaching $\rho_{xx}=0$ & Quantum Hall effect; single peak ($p_1+p_2$) at high $B$
    \end{tabular}
    \end{ruledtabular}
    \caption{Summary of the four possible SdH regimes, and the experimental conditions in which they are observable.}
    \label{tab:SdHregimes}
\end{table*}

On the other hand, Figure~\ref{Fig:densities}(b) shows that $p_1 + p_2 \neq p_{\text{Hall}}$ for G372, where the difference amounts to at least 10\% of the 2DHG density. The difference between the Hall and SdH densities is also at least 10\% for G373, but is at most 1\% for VB1137 [sample~D in Fig.~\ref{Fig:Subplot_all}(b)]. The most likely culprit for the observed parallel conduction is the low-density population of one or more of the 2.5$-$3~nm wide GaAs quantum well nearest the intended 2DHG, contained in the smoothing GaAs/AlGaAs superlattice of G372, G373, and VB1137. It is possible that a minisubband (a.k.a. superlattice subband) has formed and populated \cite{Bastard1981, Ando1982electronic}. We note that the carrier density of the parallel conducting channel does not appear to change with top-gate voltage once the intended 2DHG populates, as evidenced by the constant offset between $p_{\text{Hall}}$ and $p_1 + p_2$ in Fig.~\ref{Fig:densities}(b). This is classic behavior of hole bilayers with two 2DHGs located in two different quantum wells, with the lower 2DHG not responding to a surface gate once the upper 2DHG populates \cite{Hamilton1996fractional}. The much smaller difference between the Hall and SdH densities in wafer VB1137 relative to wafers G372 and G373 is consistent with its larger AlGaAs barrier, in both width and valence band offset (10~nm  of Al$_{0.5}$Ga$_{0.5}$As instead of 2.5~nm  of Al$_{0.3}$Ga$_{0.7}$As), between the intended 2DHG and the nearest narrow GaAs quantum well in the smoothing superlattice.

\textbf{Gate hysteresis.} At $V_{g}=0$~V, dopant-free devices do not conduct. For all Hall bars reported here with 300~nm of SiO$_2$, $V_{g}<-1$~V is necessary to induce a 2DHG. When $-$5~V~$<V_{g}<-$3~V, the hole density can be linearly and reproducibly tuned from 0.75~$\times$~10$^{15}$/m$^2$ to 2.5~$\times$~10$^{15}$/m$^2$ [e.g., see Fig.~\ref{Fig:densities}]. If the top-gate voltage is taken to more negative values, then the entire voltage-density linear relation $p_{2d}(V_{g})$ shifts towards more negative top-gate values for the remaining duration of a given cooldown, by the voltage difference exceeding $-$5~V. For example, if $V_g$ is taken to $-$7~V, then the maximum density achieved would remain 2.5~$\times$~10$^{15}$/m$^2$ and the lowest density would now occur at $V_g = -5$~V (instead of $-$3~V previously). By cycling the sample to room temperature for several hours with all contacts grounded, the original $p_{2d}(V_{g})$ can be recovered in a subsequent cooldown. The mechanism for the observed top-gate voltage shift is most likely due to the population of charge traps at the GaAs-SiO$_2$ interface. The charge traps populate when holes from the 2DHG tunnel through the AlGaAs barrier while $V_{g}<-5$~V. The same $V_{g}$ shift and mechanism for $n_{2d}(V_{g})$ occurs in dopant-free 2DEGs when $V_{g}>5.5$~V with 300 nm of SiO$_2$ (the higher voltage threshold is accounted for by the larger conduction band offset than the valence band offset). Factors influencing the value of the maximum top-gate voltage without incurring a shift in $n_{2d}(V_{g})$ include the AlGaAs barrier thickness and aluminum composition, the gate dielectric thickness, and the gate dielectric permittivity.

\begin{figure*}
    \begin{center}
    \includegraphics[width=2.0\columnwidth]{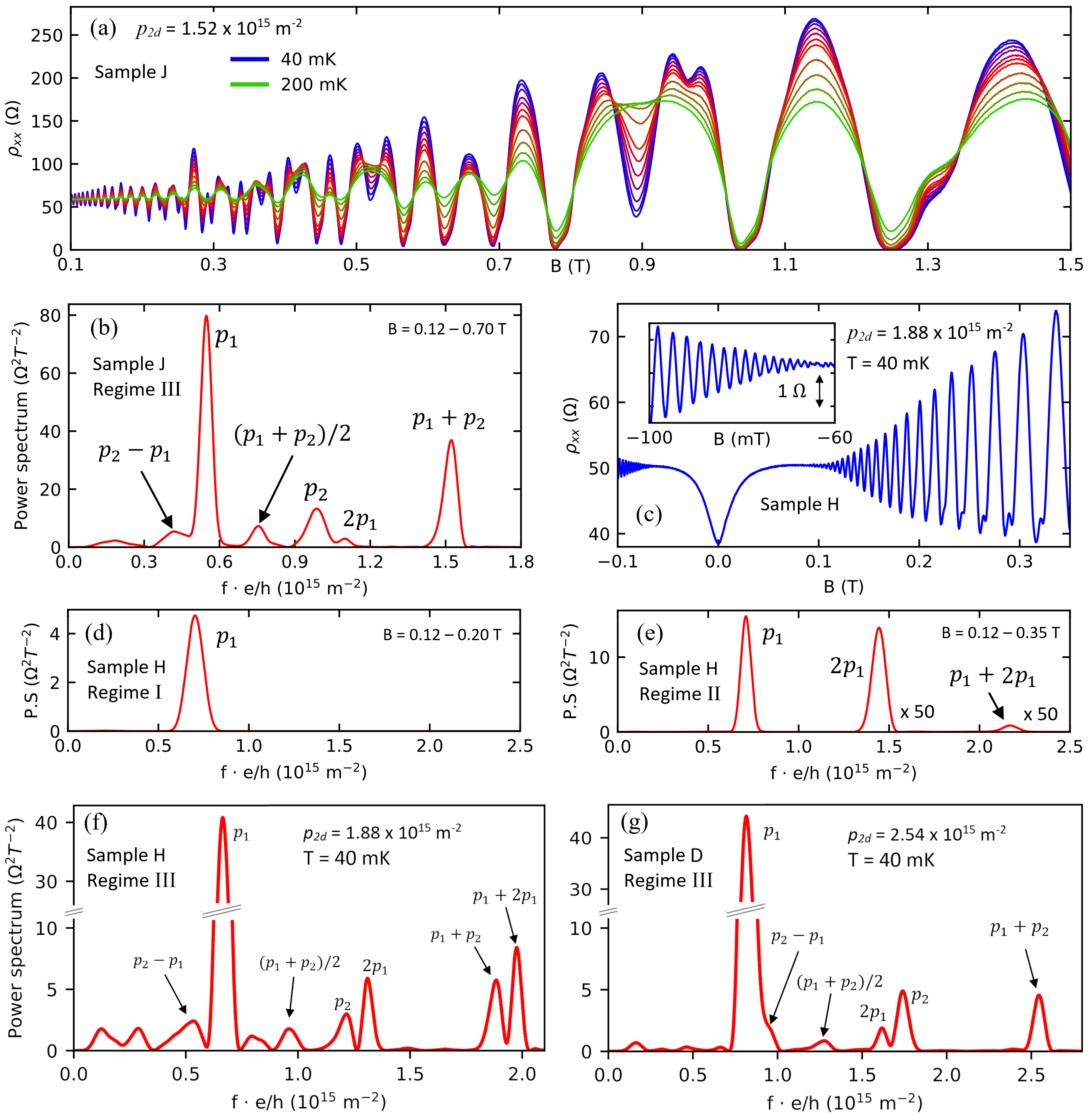}
    \end{center}
    \caption{(a) Shubnikov-de Haas (SdH) oscillations at various temperatures. (b) Corresponding Fourier transform of the SdH oscillations at base temperature of 40 mK from panel (a). (c) SdH oscillations at low magnetic fields, with the inset providing a closer view of the same dataset. Power spectrum (P.S.) of the Fourier transforms on the data from panel (c) to illustrate: (d) Regime I, and (e) Regime II (see main text). For $f \cdot e/h > 1 \times 10^{15}$/m$^2$, all FT peaks in panels (d) and (e) were scaled up by $\times$50. The field range over which the FT was performed is indicated in the upper right corner of each FT spectrum. Fourier spectra of SdH oscillations in Regime III for: (f) sample H and (g) sample D.}
    \label{Fig:MainExpt}
\end{figure*}

\section{Experimental results}
\label{sec:experiments}

Only samples D, H, and J were used for reporting quantitative results in this section. Samples E and F were not used, because of the large mismatch ($>$\,10\%) between the densities extracted from the Hall effect and SdH oscillations, and because of parallel conduction even at $B<2$~T. Figure~\ref{Fig:MainExpt}(a) shows the longitudinal resistivity $\rho_{xx}$ as a function of magnetic field and temperature in sample J, and is representative of all samples without parallel conduction. Collectively, SdH oscillations at 40~mK were measured at overlapping hole densities ranging from  $p_{2d} = 0.75 \times 10^{15}$/m$^2$ to $p_{2d} = 2.5 \times 10^{15}$/m$^2$ in samples D, H, and J.

\subsection{Spin-orbit polarization $\Delta p/p$}

To reveal the individual HH$-$ and HH$+$ populations from the resistivity oscillations, a Fourier transform of SdH oscillations with respect to $1/B$ is performed. The hole density $p$ can be obtained from $p = f \cdot e/h$, where $f$ is the frequency of a peak appearing in the Fourier transform, where $h$ is the Planck constant, and $e$ is the elementary charge. We use the formula $p = f \cdot e/h$ throughout the entire range of magnetic field values (0$-$15~T). The formula does not include a factor of 2 to take into account spin degeneracy \textit{before} the onset of spin(-orbit) splitting. Thus, in our framework, Zeeman spin splitting is heralded by the doubling of the SdH frequency $p$.

Carriers from each hole subband have a corresponding density $p=p_i$, effective mass $m_i^*$, and quantum scattering time $\tau_{qi}$ with $i=1,2$. We identify holes with the lower carrier density $p_1$ and lighter effective mass $m_1$ as HH$-$, also known as light-heavy-holes (HHl) in the literature \cite{habib2004spin,nichele2014spin}. We identify holes with the higher density $p_2$ and mass $m_2$ as HH$+$, also known as heavy-heavy-holes (HHh) in the literature.

Figure~\ref{Fig:MainExpt}(b) shows the corresponding Fourier transform of the data at base temperature from Fig.~\ref{Fig:MainExpt}(a) in sample~J. Several distinct peaks can be resolved. The peak labeled $(p_1$ + $p_2)$ matches the Hall density $p_{Hall}$, indicating no parallel conduction is present, and matches the sum of the individual peaks $p_1$ and $p_2$. The peak labeled 2$p_{1}$ arises from the low-field Zeeman splitting. The peak labeled $(p_1+p_2)/2$ corresponds to the average of the peaks $p_1$ and $p_2$. This anomalous peak has been reported before \cite{habib2009spin,nichele2014spin}, and has been interpreted as arising due to non-adiabatic effects \cite{keppeler2002anomalous}.

\begin{figure*}
    \includegraphics[width=2.0\columnwidth]{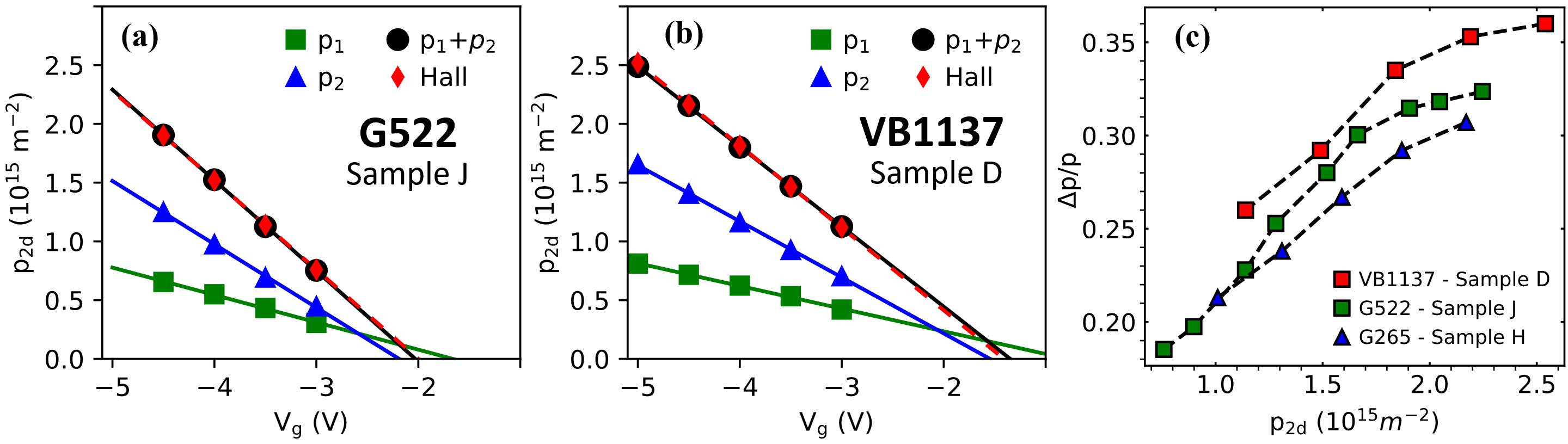}
    \caption{Carrier densities versus top-gate relation for: (a) sample J and (b) sample D. The quantities $p_1$ (squares), $p_2$ (triangles), and $p_1 + p_2$ (circles) were obtained from the Fourier transform of SdH oscillations measured at 40~mK. Diamonds symbols represent the hole densities determined from the low-field Hall effect. Solid lines (red dashed lines) are least-squares linear fits to the FT (Hall) data. (c) Rashba SOI strength in samples F, H, and J, expressed through $\Delta p/p$ using data from panels (a) and (b), as well as from Figure~\ref{Fig:densities}(a). The black dashed lines are guides to the eye.}
    \label{Fig:Subplot_all}
\end{figure*}

Four successive and distinct SdH regimes can be identified at different magnetic field ranges in Fig.~\ref{Fig:MainExpt}, in order of appearance with increasing $B$:\\
\indent\indent $\bullet$ \textbf{Regime I}, with a single SdH frequency ($p_1$);\\
\indent\indent $\bullet$ \textbf{Regime II}, with Zeeman splitting ($p_1; 2p_1$);\\
\indent\indent $\bullet$ \textbf{Regime III}, with spin-orbit splitting ($p_1; p_2$);\\
\indent\indent $\bullet$ \textbf{Regime IV}, with the quantum Hall effect.\\
Table~\ref{tab:SdHregimes} lists the experimental conditions and corresponding energy scales for each regime as a function of density, mobility, magnetic field, and temperature.

Regime I is best illustrated in sample~H when $0.07$~T~$<B<0.20$~T [see Fig.~\ref{Fig:MainExpt}(c)], where the first SdH oscillations become visible. Landau levels have formed, and can be resolved because their energy separation $\hbar \omega_c = \hbar eB/m^*$ is larger than $k_B T$, where $\omega_c$ is the cyclotron frequency and $k_B$ is the Boltzmann constant. At very low temperatures ($T < 50$~mK) however, mobility (i.e., disorder) dictates the onset of SdH oscillations. Disorder effectively reduces the Landau level energy separation. SdH oscillations are only visible once thermal energy is smaller than the reduced Landau level energy separation, $k_BT < (\hbar \omega_c - h \Gamma)$ where $h \Gamma$ represents energy broadening due to disorder. In all our samples, the onset of SdH oscillations occurs near $B\approx 70$~mT, a very strong indication of high mobility and low disorder. The resistance minimum at $B=0$~T mostly results from intersubband scattering between HH$-$ and HH$+$, not from weak anti-localization \cite{nichele2014spin}.

Significantly, only one SdH oscillation frequency $p_1$ is observed in Regime I; no other peaks are present in the Fourier spectrum from Fig.~\ref{Fig:MainExpt}(d). One possible explanation is that the energy splitting between HH$-$ and HH$+$ is too small to be resolved at this temperature. However, this same energy splitting is easily resolved at higher magnetic field, so this explanation is unlikely to be correct. Recalling that Landau energy level spacing have a direct dependence on effective mass, two sets of effective Landau levels might form, one set for each effective mass present in the 2DHG, $m_1$ and $m_2$. In this scenario, only one set of SdH oscillations is visible because $k_BT < (\frac{\hbar eB}{m_1}-h\Gamma)$ for HH$-$ whereas $k_BT > (\frac{\hbar eB}{m_2}-h\Gamma)$ for HH$+$, i.e. the Landau energy level separation for the higher effective mass $m_2$ is too small to be resolved at this temperature. Stated differently, HH$-$/HH$+$ spin-orbit splitting has already occurred at $B=0$~T, but SdH oscillations associated with HH$+$ are simply not resolved at this temperature and magnetic field. At higher magnetic fields (e.g., $B > 0.35$~T for sample H, and $B > 0.25$~T for sample J), the Landau energy level separation becomes larger, and SdH oscillations for both $m_1$ (HH$-$) and $m_2$ (HH$+$) are resolved.

Figure~\ref{Fig:MainExpt}(c) illustrates a generic instance with colder SdH oscillations for $B<0$ (earlier onset and larger amplitudes) than those for $B>0$. Generally, such asymmetry can occur for several reasons. These include sweeping the magnetic field too fast ($> 0.5$~T/hour; from magnetically-induced currents), crossing $B=0$ in either sweep direction (from spin re-alignment of magnetic impurities), or crossing $B=\pm 80$~mT while increasing $|B|$ (from quenching superconducting solder joints in a sample's leads). All experimental data used to measure $m_1$ and $m_2$ in this report avoided these conditions.

Regime II is best illustrated in sample~H when $0.20$~T~$<B<0.35$~T [see Fig.~\ref{Fig:MainExpt}(c)], where a second SdH oscillation frequency appears at $2p_1$ [see Fourier spectrum in Fig.~\ref{Fig:MainExpt}(e)], twice the value of the SdH frequency observed in Regime I. This second set of SdH oscillations is consistent with out-of-plane Zeeman spin-splitting $g_{zz}^* \mu_B B$, where $g_{zz}^*$ is the Land\'{e} effective g-factor in the growth direction $\hat{z}$ ($g_{zz}^* \approx 4-7$ for GaAs holes \cite{winkler2000highly, yuan2009landau, srinivasan2013using, komijani2013anisotropic, simion2014magnetic, miserev2017dimensional}) and $\mu_B$ is the Bohr magneton. The onset of Zeeman spin-splitting occurs at similar $B$ values observed in the literature \cite{marcellina2018electrical,Rendell2022gate}, when $g_{zz}^* \mu_B B > k_BT$. Regime~II occurs after the onset of the first SdH oscillations from Regime~I but before the onset of the complex SdH beating pattern from Regime~III (see Fig.~\ref{Fig:ZeemanSplit} in Appendix \ref{secA:spin-splitting}). Unlike the $2p_1$ peak, the $(p_1+2p_1)$ FT peak in Fig.~\ref{Fig:MainExpt}(e) is an artifact of the Fourier analysis, and does not correspond to any real SdH oscillations.\footnote{The artifact $p_1+2p_1$ FT peak results from the sum of the real $p_1$ and $2p_1$ FT peaks. It is not the 3$^\text{rd}$ harmonic $3p_1$ of $p_1$, otherwise the $3p_1$ peak would have been smaller than the $2p_1$ peak in Figure~\ref{Fig:MainExpt}(f).} Regime II occurs when $(\frac{\hbar eB}{m_1}-h\Gamma)> g_{zz}^* \mu_B B > k_BT > (\frac{\hbar eB}{m_2}-h\Gamma)$.

Regime III is best observed in sample J when $0.25$~T~$<B \lesssim 0.8$~T [see Fig.~\ref{Fig:MainExpt}(a)], where the SdH oscillations from HH$+$ become resolved [i.e., $k_BT < (\frac{\hbar eB}{m_2}-h\Gamma)$], the complex SdH beating pattern emerges, and the frequency $p_2$ appears in the FT spectrum for the first time [see Fig.~\ref{Fig:MainExpt}(b)]. Figures~\ref{Fig:MainExpt}(f) and \ref{Fig:MainExpt}(g) show the Fourier spectra for samples H and D in Regime III, respectively. This regime is the focal point of this paper, because it is the only SdH regime where both $m_1$ and $m_2$ can be measured simultaneously in the same experimental conditions. Other SdH regimes were used to correctly identify the FT peaks corresponding to $p_1$ and $p_2$.

In Regime III, the spin-orbit splitting $\Delta_{\textsc{hh}}$ between HH$-$ and HH$+$ is primarily driven by SOI due to the structural inversion asymmetry (SIA) of the confining potential, also known as Rashba SOI. The HH$-$/HH$+$ population imbalance normalized by the total carrier concentration is typically used as a proxy for Rashba SOI strength:
\begin{equation}
\frac{\Delta p}{p} = \frac{p_2 - p_1}{p_2 + p_1}.
\end{equation}
where $p_1$ and $p_2$ are experimentally measured through Fourier analysis of SdH oscillations,\footnote{Erroneous $\Delta p/p$ can be obtained at low values of $\Delta_\textsc{hh}$, when the $p_2$ FT peak is no longer individually resolvable or is shifted from the merging and larger $p_1$ FT peak. Parallel conduction ($p_\text{Hall} > p_1 + p_2$) invalidates the relation $\Delta p = p_2 - p_1 = p_\text{Hall} - 2p_1$. }  shown in Figs.~\ref{Fig:densities}(a), \ref{Fig:Subplot_all}(a), and \ref{Fig:Subplot_all}(b) for samples H, J, and D, respectively.

Figure \ref{Fig:Subplot_all}(c) shows $\Delta p/p$ monotonically increasing as a function of $p_{2d}$ for all samples and following the same parallel trend. In our samples' architecture, a more negative top-gate voltage produces a steeper confining potential for the 2DHG and, consequently, stronger Rashba SOI with larger $p_{2d}$ \cite{lu1998tunable, grbic2008strong, nichele2014spin}. The scaling of Rashba SOI strength between the three samples is consistent with their invidual potential profile. Sample~D (VB1137), with the largest $\Delta p/p$, has the steepest 2DHG potential, due to its close proximity to the top-gate and the largest AlGaAs barrier height [see Fig.~\ref{Fig:MBE}]. Sample~H (G265), with the weakest $\Delta p/p$, has the weakest 2DHG potential, from being the furthest away from to the top-gate and having a shallow AlGaAs barrier [see Fig.~\ref{Fig:MBE}]. Sample~J (G522), with a 2DHG close to the top-gate but a shallow AlGaAs barrier, ranks between the other two samples in $\Delta p/p$.

\begin{figure}[t]
    \includegraphics[width=1\columnwidth]{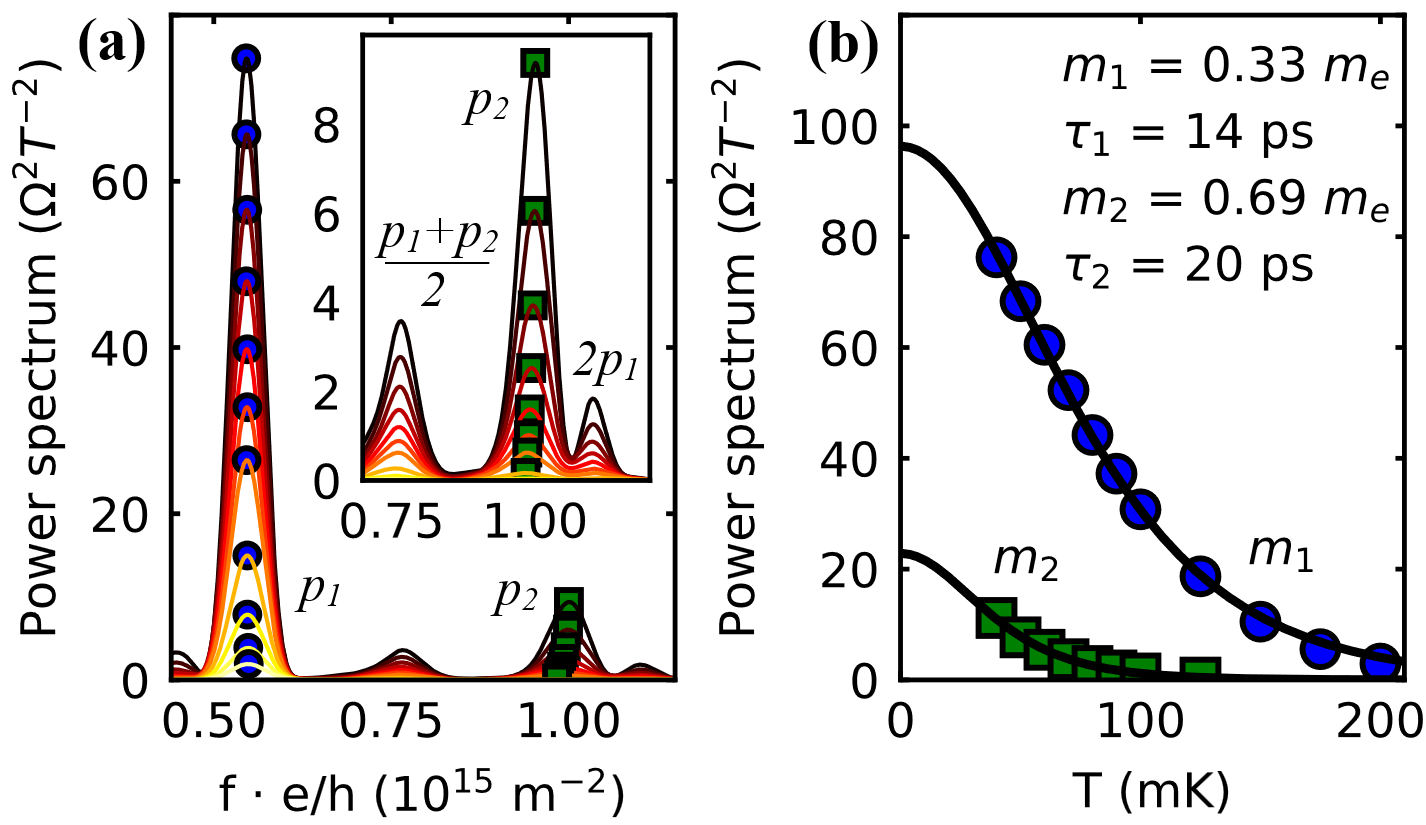}
    \caption{Sample J. (a) Temperature dependence (40$-$200 mK in eleven steps) of the Fourier peaks $p_1$ and $p_2$ at $p_{2d} = 1.52 \times 10^{15}$/m$^2$. (inset) Close view of the small FT peaks shown in the main panel. (b) Peak heights from (a) fitted to Eq.~(\ref{Eq:ando}), using $m^*$ and $\tau$ as fit parameters. }
    \label{Fig:FFT_T_FIT}
\end{figure}

Finally, Regime IV in sample J occurs when $B \gtrsim 0.8$~T, when SdH oscillation minima reach $\rho_{xx}=0$ [see Fig.~\ref{Fig:MainExpt}(a)]. This regime is not best suited for Fourier transform analysis for the purpose of determining the HH$-$/HH$+$ effective masses. Nevertheless, the spacing of integer and fractional filling factors plotted against $1/B$ remains constant and consistent with both the low-field Hall carrier concentrations $p_{\text{Hall}}$ and the $p_1+p_2$ FT peak from SdH oscillations. At high magnetic fields ($B>2$~T), the FT spectrum is dominated by the single peak $(p_1 + p_2)$ \cite{grbic2008strong}. High-field cyclotron resonance mass studies in quantum wells \cite{Rachor2009,Cole1997-PRB} have shown that effective masses increase dramatically with large magnetic fields. For the purposes of determining effective masses from transport experiments in this paper, Regime IV is altogether avoided.

\subsection{HH$-$/HH$+$ effective masses}

For a system with only a single electric subband occupied, the effective mass is typically extracted by analyzing the resistivity amplitude decay $\Delta\rho_{xx}/\rho_{xx}$ of SdH oscillations as a function of temperature \cite{coleridge1996effective}, which can be fit to \cite{nanostructures2010quantum}:
\begin{equation}
\frac{\Delta\rho_{xx}}{\rho_{xx}} = 2\,\exp\,\left(-\frac{\pi}{\omega_{c}\tau_{q}}\right)
\frac{2\pi^{2}k_{B}T/\hbar\omega_{c}}{\sinh(2\pi^{2}k_{B}T/\hbar\omega_{c})}.
\label{Eq:ando}
\end{equation}
However, due to the complex beating of the SdH oscillations at low magnetic field in asymmetric quantum wells, $\Delta \rho_{xx}$ cannot be simply normalized by a thermal dampening factor. Instead, we employed a technique pioneered in Refs. \cite{nichele2014spin,habib2004spin}. It consists of applying a Fourier transform to the SdH oscillations in $1/B$, and fitting the FT peak heights of HH$-$/HH$+$ as a function of temperature, using Eq.~(\ref{Eq:ando}). The SdH oscillations for a given temperature and B field range can thus be individually re-constructed for $m_1^*$ and $m_2^*$. The background resistivity and densities $p_1, p_2$ are directly obtained from experiments, while $m^{*}$ and $\tau_{q}$ are fitting parameters. The Fourier analysis presented here employs standard techniques such as windowing and zero-padding. These manipulations are valid as long as the same procedures are applied to the calculated resistivities as well. This method is robust in that it does not require any assumption regarding the form of the HH band structure.

Figure \ref{Fig:FFT_T_FIT}(a) shows the temperature dependence of the $p_1$ and $p_2$ Fourier peaks for 0.12 T~$ \leqslant B \leqslant 0.70$~T in sample J, from the dataset shown in Fig.~\ref{Fig:Subplot_all}(a). Figure \ref{Fig:FFT_T_FIT}(b) shows the corresponding peak height as a function of temperature, fitted to Eq.~(\ref{Eq:ando}). We find $m_1 = (0.33 \pm 0.003)m_{e}$ and $\tau_1 = (14 \pm 0.1)$ ps for HH$-$, and $m_2 = (0.69 \pm 0.01)m_{e}$ and $\tau_2 = (20 \pm 0.9)$ ps for HH$+$.

Due to thermal broadening when $T\gtrsim 0.1$~K, the FT peak $p_2$ disappears, while the $p_1$ only disappears for $T \gtrsim 0.2$~K. This is entirely consistent with $m_2$ being almost twice as large as $m_1$, which has a direct impact on the visibility of SdH oscillations through the cyclotron energy $\hbar \omega_c = \hbar eB/m^*$. For $T \gtrsim 0.2$~K, the SdH beating pattern is thermally smeared out, and only one SdH frequency, $p_1+p_2$, is observable in the Fourier spectrum. In itself, this does not imply $k_BT > \Delta_{\textsc{hh}}$, but rather that $\Delta_{\textsc{hh}}$ cannot be resolved through SdH oscillations.

\begin{figure}[t]
    \includegraphics[width=0.9\columnwidth]{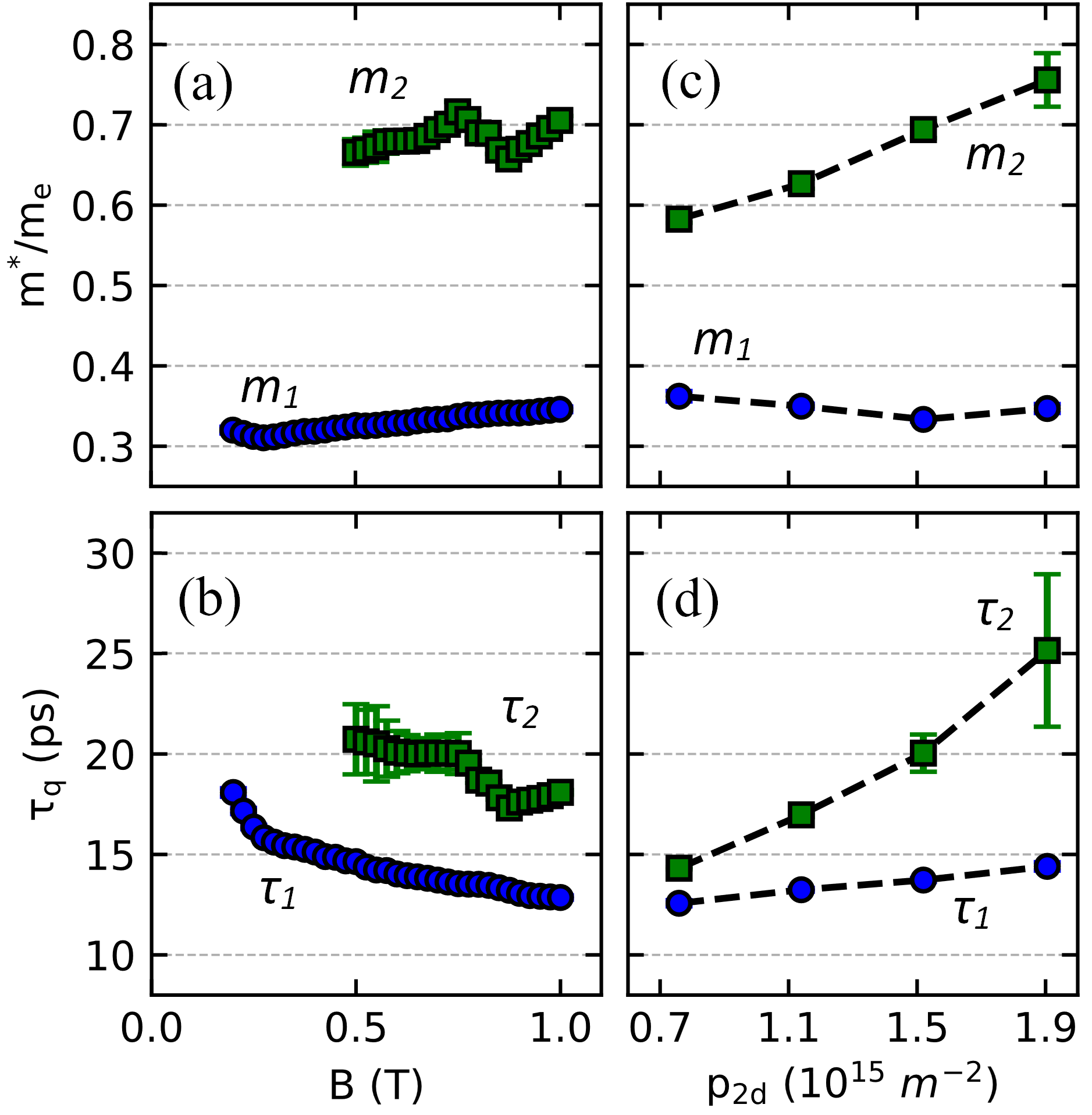}
    \caption{Sample J. (a) Effective masses and (b) quantum scattering times analyzed over different magnetic field ranges (see main text), keeping $p_{2d} = 1.52 \times 10^{15}$/m$^2$ constant. (c) Effective masses and (d) quantum scattering times for different $p_{2d}$, using SdH oscillations from $B=0.12$~T to 0.70~T. }
    \label{Fig:mass_vs_2d_B}
\end{figure}

To ensure that the effective hole temperature was not saturating and skewing results, the analysis in Figure~\ref{Fig:FFT_T_FIT} was also performed with the four lowest temperature points of $m_{1}$ removed: the results of the fit were unchanged (not shown), indicating that the achieved hole temperature is close to that of the thermometer mounted on the mixing chamber of the dilution refrigerator. The error bars reported reflect the statistical errors of the fits.

The same temperature dependence FT analysis as in Fig. \ref{Fig:FFT_T_FIT} was also performed over different ranges of magnetic fields, and is shown in Figs.~\ref{Fig:mass_vs_2d_B}(a) and \ref{Fig:mass_vs_2d_B}(b). To produce these plots, the ``start'' of a $B$ sweep of SdH oscillations is always $B=0.12$~T and the $x$-axis represents the ``end'' of the $B$ sweep over which the SdH oscillations are Fourier-transformed. The temperature dependence FT analysis is not strictly valid once SdH minima reach zero resistance in the quantum Hall regime. At $p_{2d} = 1.52 \times 10^{15}$/m$^2$, the first magnetic field where the SdH mimima reach zero resistance over a range of temperatures occurs at $B=0.78$~T, which limited our FT analysis to $B < 0.75$~T. The sudden and abrupt deviation of this trend occurring near $B \approx 0.75$~T for $m_{2}$ is likely due to artifacts of the Fourier analysis.

Over the range where the FT analysis is valid (i.e., 0.1~T~$< B <$~0.75~T), Figure \ref{Fig:mass_vs_2d_B}(a) shows that both $m_1$ and $m_2$ are only very slightly increasing with magnetic field, in broad agreement with the modulation-doped 2DHGs from Refs.~\cite{nichele2014spin} and \cite{eisenstein1984effect}  (but not with Ref.~\cite{habib2004spin}) and with Luttinger theory \cite{simion2014magnetic}. The latter predicts that both effective masses $m_1$ and $m_2$ only have a very weak dependence on magnetic field when $B<1$~T. For $B < 0.75$~T, Figure~\ref{Fig:mass_vs_2d_B}(b) shows both quantum lifetimes $\tau_1$ and $\tau_2$ decreasing almost linearly with magnetic field, indicating an increase in scattering due to cyclotron orbits becoming more localized.

Finally, the same temperature dependence FT analysis was performed in sample J at four 2DHG carrier concentrations, shown in Figs.~\ref{Fig:mass_vs_2d_B}(c) and \ref{Fig:mass_vs_2d_B}(d). The $m_{1}$ mass (HH$-$) shows no dependence on carrier concentration in the measured range, whereas the $m_{2}$ mass (HH$+$) shows a strong dependence on hole density. Both observations are in broad agreement with the gated modulation-doped 2DHG from Ref.~\cite{nichele2014spin}. The increase in quantum lifetimes with hole density is due to the increased Thomas-Fermi screening of charged background impurities.

\section{Discussion}
\label{sec:discussion}

This section discusses how the strength of spin-orbit polarization $\Delta p/p$ in our samples relates to literature, how our heavy hole effective masses $m_1$ and $m_2$ relate to literature,
the general nature of the HH$-$ and HH$+$ subbands, how $m_1$ and $m_2$ are used to map the dispersion relations of HH$+$/HH$-$ subbands, how the spin-orbit splitting energy is obtained, and the role of many-body interactions. Throughout this section, we discuss how our results fit within the Luttinger model.

\textbf{Luttinger model numerical calculations}. To guide the discussion, we modeled spin-orbit interactions in 2DHGs based on Luttinger theory \cite{Luttinger56} with well-known formulations \cite{Novik05, liu2018strong, Marcellina17, Szumniak12, Szumniak13, winkler2003spin, winkler2000rashba}. Nevertheless, since the numerical simulations are fairly involved to implement and contain many assumptions, we describe them in detail in Appendices \ref{secA:Luttinger}$-$\ref{secA:dispersion} for transparency and completeness. In order to fully capture the heavy-light hole subband mixing, a key driving feature of SOI, we fully retain the heavy hole and light hole subbands in all the calculations. It is not a perturbative approach, and the model contains no free parameters. As a basic check, our Luttinger model correctly predicts the absence of spin-orbit splitting of the HH subband in a symmetric quantum well [see Fig.~\ref{perturb_square} in the Appendix], and the spin-orbit splitting of the HH subband into HH$+$ and HH$-$ in an asymmetric quantum well [see Fig.~\ref{perturb_triang} in the Appendix], as described earlier in Section \ref{sec:background} [see Fig.~\ref{Fig:VB}]. Before proceeding further, two central results from the analysis of the calculations must be mentioned. The first result is that the apparent ``zero-magnetic field spin splitting'' is actually spin-orbit splitting due to structural inversion asymmetry (SIA). Bulk inversion asymmetry (BIA) alone cannot cause spin-orbit splitting. The second result is that the full-version bulk Hamiltonian [Eq.~(\ref{bulk_hamil}) in the Appendix] $-$ i.e, without sweeping simplifications $-$ already contains all the terms necessary to model SOI: there is no need to explicitly add extra Rashba and Dresselhaus terms to the Hamiltonian to do so.

\begin{figure}[t]
    \includegraphics[width=1.0\columnwidth]{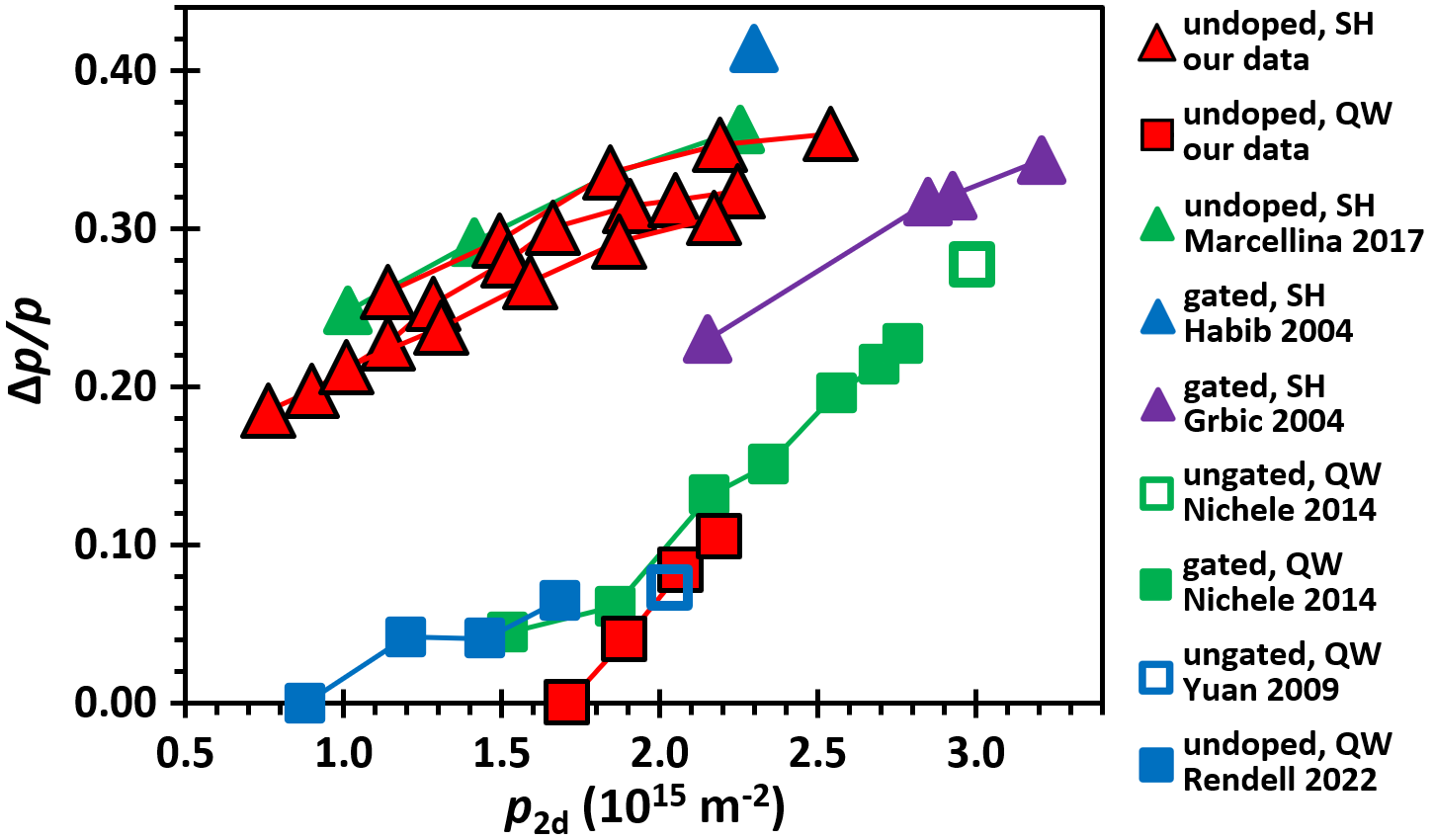}
    \caption{Comparison between $\Delta p/p$ in our SH heterostructures (red triangles; samples D, H, and J) and our QW heterostructures (red squares; sample M) with experimental values from literature \cite{habib2004spin, grbic2004, Marcellina17, nichele2014spin, yuan2009landau, Rendell2022}, covering (100) GaAs-based 2DHGs in QW heterostructures (squares) and SH heterostructures (triangles). Solid lines denote a group of points from a single gated sample, and are a guide to the eye. }
    \label{Fig:DeltaP-literature}
\end{figure}

\textbf{Strong spin-orbit polarization.} The term $\Delta p/p$ can be used as a proxy for $\Delta_{\textsc{hh}}$, the energy gap between HH$-$ and HH$+$ [see Fig.~\ref{Fig:VB}(d)]. Figure \ref{Fig:DeltaP-literature} shows the spin-orbit polarization $\Delta p/p$ from samples~D, H, J, and M as well as those from the literature in samples in (100) GaAs 2DHGs with high mobilities at hole densities similar to our samples \cite{habib2004spin, grbic2004, Marcellina17, nichele2014spin, yuan2009landau, Rendell2022}, using an experimental methodology based on SdH oscillations in single heterojunctions (triangles) or quantum wells (squares). All quantum well heterostructures shown in Figure \ref{Fig:DeltaP-literature} (ours and others') host their 2DHG in a 15 nm wide GaAs layer in an asymmetric confining potential.

Figure \ref{Fig:DeltaP-literature} highlights the extraordinary reproducibility of HIGFETs in dopant-free SH heterostructures (samples~D, H, J, and the sample from Ref.~\onlinecite{Marcellina17}): their $\Delta p/p$ are clustered together, despite having different MBE layer stacks (with 2DHG depths ranging from 60~nm to 310~nm), different Al$_x$Ga$_{1-x}$As composition (with $x = 0.3-0.5$), and different provenances (three MBE research groups). The data is remarkably consistent and reproducible, emphasizing the systemic advantages of dopant-free over modulation-doped gated devices.

The values of $\Delta p/p$ of the dopant-free HIGFETS are among the largest (up to 36\%) as far as we know for $p_{2d} < 2.3 \times 10^{15}$/m$^2$ and $B<1$~T in GaAs (100), whether in SH or QW heterostructure (of any well width). The reasons for this are twofold. First and foremost, single heterojunctions (SH) achieve much higher $\Delta p/p$ (and $\Delta_{\textsc{hh}}$) than quantum well (QW) heterostructures, due to their intrinsically larger structural inversion asymmetry (SIA). In other words, $\Delta_{\textsc{l-h}}$ (see Fig.~\ref{Fig:VB}) is much larger in SH heterostructures than in QW heterostructures. Measurements on a control dopant-free quantum well heterostructure (sample M; red squares in Fig.~\ref{Fig:DeltaP-literature}), yielding comparable $\Delta p/p$ to modulation-doped quantum wells, confirm this interpretation. Second, higher electric fields can be achieved in dopant-free devices, without risking parallel conduction in the doping layer, sacrificing mobility (by reducing the spacer layer), or rendering the device ungatable (by overdoping). The high electric fields achieved in Refs.~\onlinecite{habib2004spin, grbic2004} were at the expense of mobility, 16~m$^2$/Vs and 7.7~m$^2$/Vs respectively. In contrast, the peak mobilities of samples D, H, J, and K span 60$-$80~m$^2$/Vs. Minimizing disorder ($h\Gamma$) is important, as it smears out the SdH interference pattern.

\begin{figure}[t]
    \includegraphics[width=1.0\columnwidth]{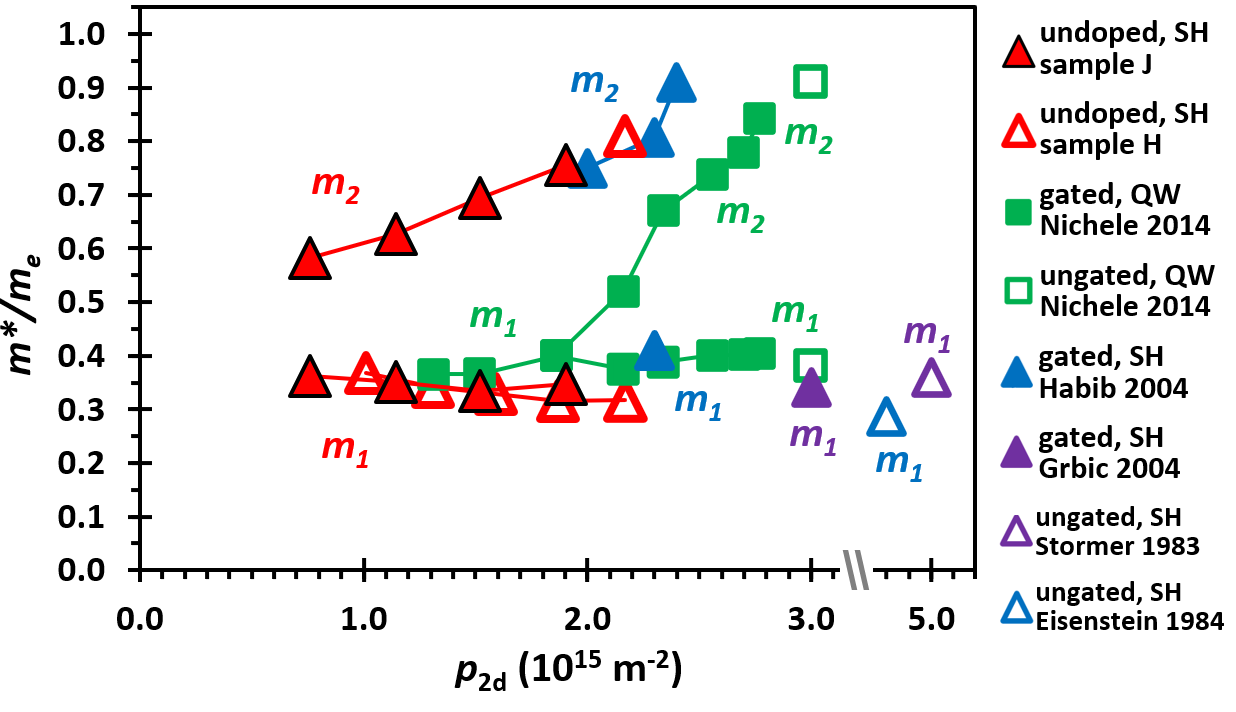}
    \caption{Comparison of $m_1, m_2$ between our data (red triangles) and literature, with gated/ungated QW samples (squares) \cite{nichele2014spin} and gated/ungated SH samples (non-red triangles) \cite{habib2004spin,grbic2004,stormer1983energy,eisenstein1984effect}. Solid lines denote a group of points from a single gated sample, and are a guide to the eye. }
    \label{Fig:SdH-literature-mass-experiments}
\end{figure}

An interplay between spin-orbit strength and disorder determines the threshold 2DHG density $p_t$ above which spin-orbit splitting can be resolved. In samples D, H, and J, it corresponds to $p_t \approx 0.5 \times 10^{15}$/m$^2$ when the HH$-$/HH$+$ densities are projected to be equal $p_1=p_2$ [see Figs.~\ref{Fig:densities}(a), \ref{Fig:Subplot_all}(a), and \ref{Fig:Subplot_all}(b)] and thus $\Delta p=0$ when $\Delta_{\textsc{hh}}-h\Gamma \approx 0$. Comparing two undoped QW heterostructures with a 15 nm QW width, the threshold density in sample M ($p_t \approx 1.7 \times 10^{15}$/m$^2$) is much higher than in the QW sample from Ref.\,\onlinecite{Rendell2022} ($p_t \approx 0.9 \times 10^{15}$/m$^2$) because the mobility of sample M is much smaller (20~m$^2$/Vs versus 76~m$^2$/Vs, respectively). However, mobility alone does not determine $p_t$. For example, the $p_t$ of the undoped QW heterostructure from Ref.\,\onlinecite{Rendell2022} is much higher than those of the SH heterostructures of samples D, H, and J because the spin-orbit strength is larger in SH heterostructures, for similar mobilities.

\begin{figure}[t]
    \includegraphics[width=1.0\columnwidth]{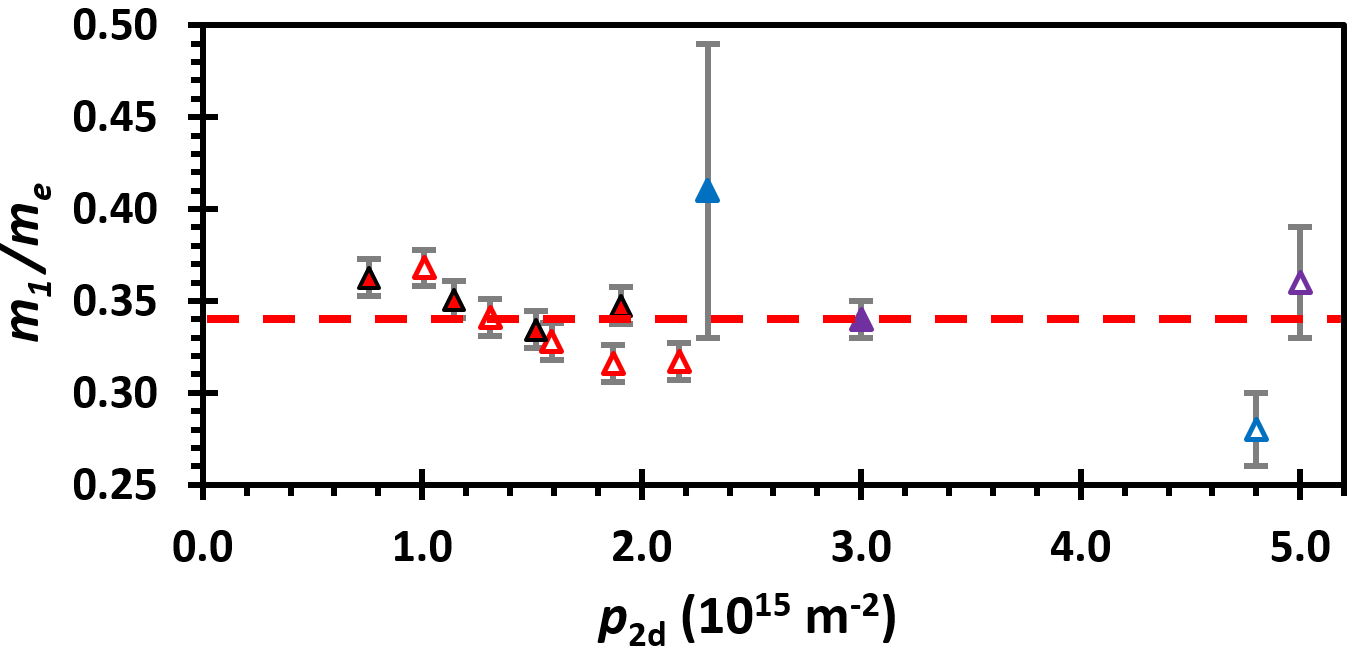}
    \caption{Enlarged view of the $m_1$ from SH heterostructures shown in Fig.~\ref{Fig:SdH-literature-mass-experiments}, with their uncertainty, and using the same symbols. Their average is $\overline{m_1} = (0.34\pm0.02)m_e$, indicated by the dashed line. }
    \label{Fig:m1-ErrorBars}
\end{figure}

\textbf{Near-parabolic HH$-$ subband and $m_1$.} Figure~\ref{Fig:SdH-literature-mass-experiments} compares our data to the effective masses $m_1$ of heavy holes in (100) GaAs 2DHGs hosted in single heterojunctions at low magnetic fields (before the onset of the quantum Hall effect) reported in the literature \cite{habib2004spin,grbic2004,stormer1983energy,eisenstein1984effect}. The $m_1$ from samples~J and H are near-independent of the 2DHG density over the range investigated. The median of these is $\widetilde{m_1} = (0.35\pm0.01)m_e$ for sample~J and $\widetilde{m_1} = (0.34\pm0.03)m_e$ for sample~H. These two measurements agree very well with each other. They also agree very well with the two published magnetotransport studies performed in Regime III: $m_1 = (0.36\pm0.03)m_e$ \cite{stormer1983energy}, and $m_1 = (0.34\pm0.01)m_e$ \cite{grbic2004}. Our measurements are consistent with one of the two published magnetotransport studies performed in Regime~I: $\widetilde{m_1} = (0.28\pm0.02)m_e$ \cite{eisenstein1984effect},\footnote{Ref.~\onlinecite{eisenstein1984effect} reported a range of  $m_1$ values for many magnetic fields at a fixed 2DHG density in SdH Regime I.  The reference value we stated here ($\widetilde{m_1}/m_e = 0.28\pm0.02$) is the median of all their $m_1$, and the error bars indicate the spread of the original values. We note a high electron temperature in a dilution refrigerator can cause measurements to underestimate the true value of $m_1$ (see paragraph ``Transport experiments'' in Section~\ref{sec:methods}). The difference between electron and lattice temperature in a 2DHG was an unknown concept in 1984.\label{foot:m1Eisenstein1984} } and $\widetilde{m_1} = (0.41\pm0.08)m_e$ \cite{habib2004spin}.\footnote{Ref.~\onlinecite{habib2004spin} reported a very unusually strong dependence of $m_1$ on magnetic field at a fixed 2DHG density in SdH Regime I. The reference value we stated here ($\widetilde{m_1}/m_e = 0.41\pm0.08$) is the median of all their $m_1$, and the error bars indicate the spread of the reported values. If we take into account the uncertainties stated in Ref.~\onlinecite{habib2004spin} for each data point, then the uncertainty becomes larger, $\widetilde{m_1} = (0.41\pm0.12)m_e$.\label{foot:m1Habib2004} } Figure~\ref{Fig:m1-ErrorBars} shows an enlarged view of all the $m_1$ values in SH heterostructures, with their uncertainty. Their median and average are equal, $\widetilde{m_1} = \overline{m_1} = (0.34\pm0.02)m_e$, with the error bars representing the standard deviation.\footnote{This standard deviation excludes the data points described in footnotes \ref{foot:m1Eisenstein1984} and \ref{foot:m1Habib2004}. Including/excluding them does not significantly change the average/median.}

The comparisons above illustrate two trends: (i) all published studies performed in the SdH Regime~III agree with our results, and (ii) all published studies that disagree with our results were performed in Regimes~I, II, or IV. These two sweeping statements are not limited to transport measurements of $m_1$ but also extend to transport measurements of $m_2$, as well as cyclotron resonance mass studies of $m_1$ and $m_2$ \cite{companionPRL2025}.

Taking into account all available data, Figures~\ref{Fig:SdH-literature-mass-experiments} and \ref{Fig:m1-ErrorBars} reveal that $m_1$ is near-independent of the 2DHG density over the very wide range $(0.7-5.0) \times$ 10$^{15}$/m$^2$ in Regime III. Only one type of band structure yields a density-independent effective mass: a parabolic dispersion in $k$\nobreakdash-space. This leads to a key claim of this work: the HH$-$ subband dispersion is generally near-parabolic $E \approx \hbar^2 k_1^2/2m_1$ in single heterojunctions with single subband occupation, where $m_{\textsc{hh$-$}}^{*} = \overline{m_1} \approx 0.34m_e$. This in turn enables an accurate conversion of the experimentally measured HH$-$ density $p_1$ (via the FT of SdH oscillations) into the Fermi energy of the 2D system:
\begin{equation}
E_{\textsc{f}}= 2\pi\hbar^2p_1/\overline{m_1}
\label{Eq:E_F(p1)}
\end{equation}
A near-parabolic dispersion for HH$-$ is predicted by the Luttinger model at Fermi energies below the anticrossing point between the HH and LH subbands (e.g., see Fig.~\ref{fermi_surfaces} in the Appendix), a prediction which we experimentally confirm in single heterojunctions.

Figure~\ref{Fig:SdH-literature-mass-experiments} also reveals a near-independence of $m_1$ on the 2DHG density in the QW heterostructure of Ref.~\onlinecite{nichele2014spin} in Regime~III, with an average value $\overline{m_1} = (0.39\pm0.02)m_e$ slightly higher but still very close to ours. The lack of density dependence strongly suggests the HH$-$ dispersion is parabolic in this single-side-doped quantum well. Thus we can further generalize our earlier claim, and declare that the spin-orbit-split HH$-$ subband of \textit{any} GaAs 2DHG in a strongly asymmetric confinement will have a near-parabolic dispersion below the LH/HH anticrossing.

\textbf{Non-parabolic HH$+$ subband and $m_2$.} Only two publications \cite{habib2004spin,nichele2014spin} directly measured the HH$+$ effective mass $m_2$ in SdH transport experiments, without explicitly assuming a parabolic dispersion for both the HH$-$ and HH$+$ subbands. Figure~\ref{Fig:SdH-literature-mass-experiments} compares their $m_2$ values to ours. All datasets were obtained in Regime~III, and all show $m_2$ changing as a function of 2DHG density, confirming the HH$+$ subband is non-parabolic.

The $m_2$ from sample~H follows the extrapolated trend set by the $m_2$ from sample~J at lower 2DHG densities. The $m_2$ values from the SH heterostructure in Ref.~\onlinecite{habib2004spin} agree with our data: $m_2 = (0.75\pm0.10)m_e$ at $p_{2d}=2.0 \times$ 10$^{15}$/m$^2$ from Ref.~\onlinecite{habib2004spin} agrees with our $m_2 = (0.76\pm0.03)m_e$ at $p_{2d}=1.9 \times$ 10$^{15}$/m$^2$ from sample~J, and  $m_2 = (0.81\pm0.10)m_e$ at $p_{2d}=2.3 \times$ 10$^{15}$/m$^2$ from Ref.~\onlinecite{habib2004spin} agrees with our $m_2 = (0.81\pm0.06)m_e$ at $p_{2d}=2.2 \times$ 10$^{15}$/m$^2$ from sample~H. All data appears to suggest $m_2$ tends towards $m_1$ as the 2DHG density decreases to $p_{2d}$\,=\,0.

Unlike samples from SH heterostructures, the QW sample from Ref.~\onlinecite{nichele2014spin} shows $m_1 \approx m_2$ at a finite 2DHG density, $p_{2d}=1.9 \times$ 10$^{15}$/m$^2$, because the energy gap between the HH$-$ and HH$+$ subbands has become quite small. This could be due to its $\Delta_\textsc{hh}$ being smaller than in SH heterostructures (its $\Delta p/p$ is smaller at the same 2DHG density in Fig.\ref{Fig:DeltaP-literature}), its $h\Gamma$ being larger (sample~J has the same mobility at half the 2DHG density), or a combination of both factors.

To independently confirm the non-parabolicity of the HH$+$ subband, one can plot $p_2$ as a function of $E_{\textsc{f}}$ or, equivalently, plot $p_2$ as a function of $p_1$, as shown in Fig.~\ref{Fig:dispersion-p(E)} in the Appendix. Any deviation from a linear relationship in either $p_2(E_{\textsc{f}})$ or $p_2(p_1)$ is direct evidence of non-parabolicity, and the coefficients of the non-linear terms quantify the non-parabolicity. Thus, even without directly measuring $m_1$ or  $m_2$, Figure~\ref{Fig:dispersion-p(E)} in the Appendix independently confirms the HH$+$ subband is non-parabolic in all our samples (D, H, and J).

Generally, in single heterojunctions or strongly asymmetric quantum wells, the Fermi energy of the HH$-$ and HH$+$ subbands are equal for all 2DHG densities ($E_{\textsc{f}} = E_{\textsc{f,hh+}} = E_{\textsc{f,hh-}}$; see Fig.~\ref{Fig:VB}). Therefore if $p_1$ is known, then the Fermi energy of the HH$+$ subband is also known.

\begin{figure}[t]
    \includegraphics[width=1.0\columnwidth]{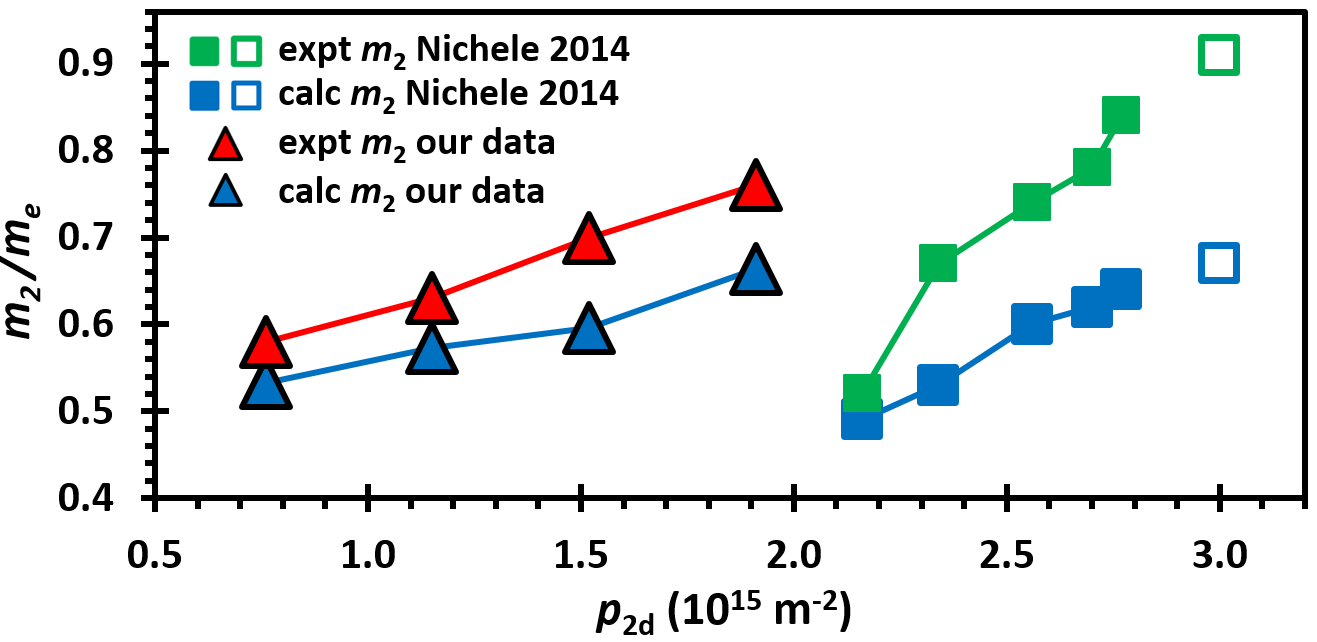}
    \caption{Difference between $m_2^\text{calc} = m_1^\text{expt} p_2/p_1$ (blue symbols) and $m_2^\text{expt}$ from sample~J (red triangles) and the QW samples from Ref.~\onlinecite{nichele2014spin} (green squares). Empty symbols denote a single-density measurement, whereas filled symbols linked by a solid line denote many densities from the same sample. Lines are a guide to the eye. }
    \label{Fig:Parabolic}
\end{figure}

\textbf{LH/HH hybridization and band parabolicity.} The Luttinger model (see Fig.~\ref{subbands_well_noD} in the Appendix) predicts the LH and HH subbands are more strongly hybridized when $\Delta_{\textsc{l-h}}$ at $k$\,=\,0 is smaller (see Fig.~\ref{Fig:VB}). Stronger hybridization causes more pronounced non-parabolicity in the band structure, which should be more pronounced in quantum wells than in single heterojunctions. Figure~\ref{Fig:Parabolic} illustrates this by quantifying the difference between the $m_2^\text{expt}$ values experimentally obtained from the temperature dependence of the FT of the SdH oscillations and the $m_2^\text{calc}$ values calculated with $m_2^\text{calc} = m_1^\text{expt} p_2/p_1$, assuming a parabolic dispersion for both HH$-$ and HH$+$ subbands. The difference, which is a direct measure of the HH$+$ subband's non-parabolicity, is most pronounced in the quantum well (up to $\sim$36\%) from Ref.~\onlinecite{nichele2014spin}, whereas it is much smaller in our undoped single heterojunction (9$-$15\%). The modulation-doped single heterojunction from Ref.~\onlinecite{stormer1983energy} has a much larger $\Delta_{\textsc{l-h}}$ at $k$\,=\,0 than in our undoped single heterojunction, and consequently the difference between its cyclotron masses $m_2^\text{expt}$ and $m_2^\text{calc}$ is much smaller ($<$\,1\%).\footnote{The $\Delta p/p$ ($\approx 0.20$) from Ref.~\onlinecite{stormer1983energy} is smaller than ours ($\Delta p/p \approx 0.36$) because its mobility is much smaller than ours, thus reducing the effective energy gap between its HH$-$ and HH$+$ subbands, $\Delta E=\Delta_\textsc{hh}-h\Gamma$.} Furthermore, for a given sample, calculations from our Luttinger model (see Fig.~\ref{fermi_surfaces} in the Appendix) predict the Fermi surface of the HH$+$ subband becomes more distorted and non-parabolic as the Fermi energy increases. This is confirmed by the difference between $m_2^\text{expt}$ and $m_2^\text{calc}$ increasing with 2DHG density in both SH and QW samples.

The Luttinger model predicts LH/HH hybridization and non-parabolicity are generally much weaker in single heterojunctions than in asymmetric quantum wells, that they are stronger in heterostructures with smaller $\Delta_{\textsc{l-h}}$ at $k$\,=\,0, and that they increase with 2DHG density for a given sample. Figure~\ref{Fig:Parabolic} shows these predictions are experimentally demonstrated.

\begin{figure}[t]
    \includegraphics[width=1.0\columnwidth]{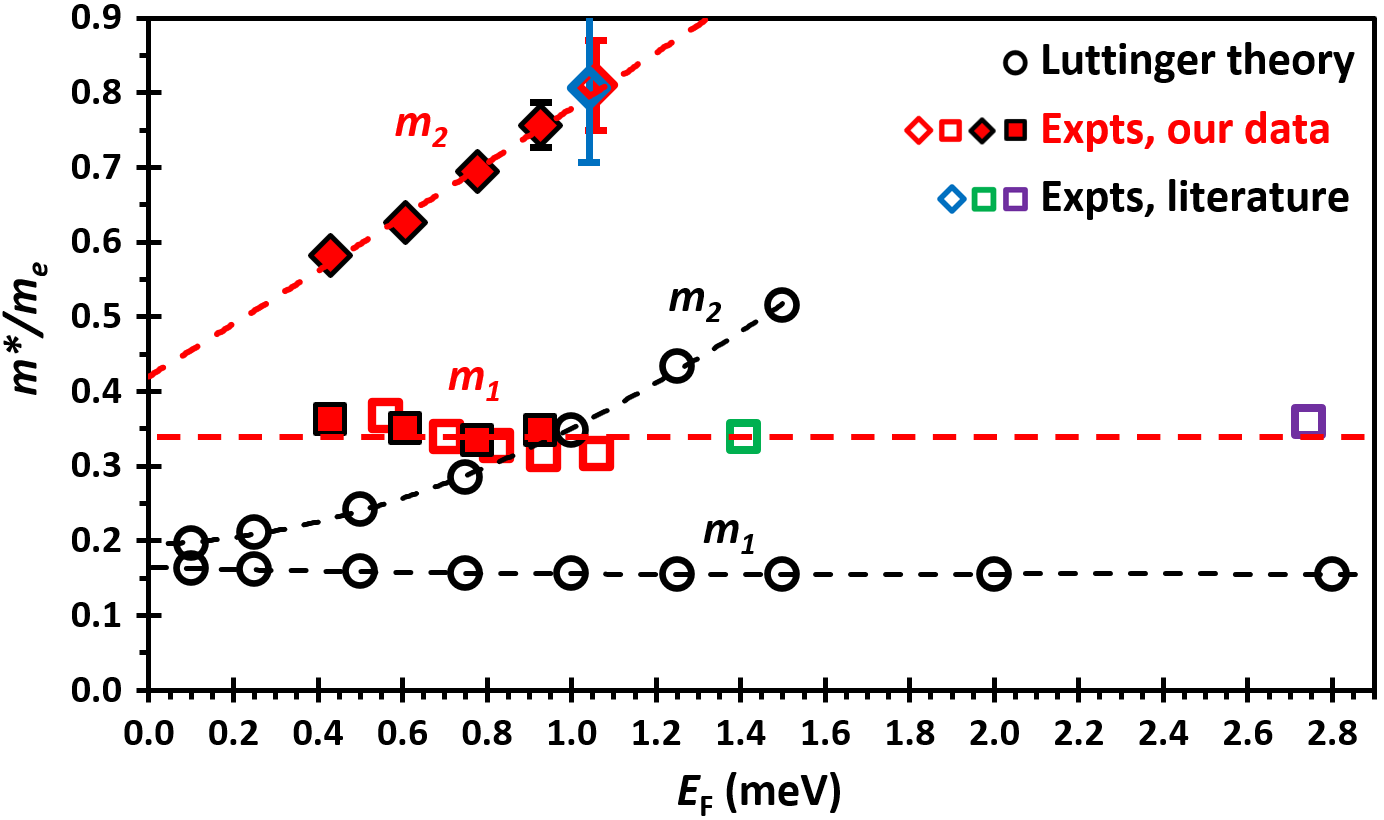}
    \caption{Comparison of experimental $m_1$ (squares) and $m_2$ (diamonds) obtained from magnetotransport between sample~J (filled red symbols), sample~H (empty red symbols), literature in the same experimental conditions \cite{stormer1983energy, habib2004spin, grbic2004} (non-red colored symbols), and Luttinger theory (black empty circles). All dashed lines are fits to data. Experimental error bars are only shown for $m_2$. }
    \label{Fig:Luttinger-vs-experiments}
\end{figure}

\textbf{Parameterizing $m_2$ from magnetotransport.} To make direct quantitative comparisons between experiments and theory, all effective masses $m_1,m_2$ in Fig.~\ref{Fig:Luttinger-vs-experiments} have been plotted against the Fermi energy, rather than the total 2DHG density or wavevectors. For consistency, all $E_{\textsc{f}}$ were calculated with Eq.~(\ref{Eq:E_F(p1)}), $\overline{m_1}=0.34m_e$, and $p_1$ from experiments. Figure~\ref{Fig:Luttinger-vs-experiments} includes magnetotransport results obtained in the SdH Regime~III from sample~J, sample~H, and literature \cite{stormer1983energy, habib2004spin, grbic2004}. We approximate the HH$+$ effective mass $m_2$ with the quadratic function:
\begin{equation}
m_2(E_{\textsc{f}}) = m_{2,0} + \kappa E_{\textsc{f}} + b E_{\textsc{f}}^2
\label{Eq:m2(E)}
\end{equation}
where $m_{2,0}$ is the zero-intercept, $\kappa$ is the linear coefficient, and $b$ is the bowing coefficient. Fitting only the $m_2$ data from sample~J in Fig.~\ref{Fig:Luttinger-vs-experiments} to a quadratic function gave the fit parameters $m_{2,0} = (0.420\pm0.010)m_e$, $\kappa = (0.350\pm0.007)m_e$/meV, and $b = (0.009\pm0.001)m_e$/meV$^2$. The extrapolated trend from the fit comfortably lies within the uncertainties of the $m_2$ values from sample~H and Ref.~\onlinecite{habib2004spin}.

\textbf{HH-/HH$+$ dispersions from magnetotransport.} We already established the dispersion relation for the HH$-$ subband is parabolic:
\begin{equation}
E_1 \approx \frac{\hbar^2 k_1^2}{2 m_1}\qquad\text{and}\qquad k_1\approx\sqrt{4\pi p_1}.
\label{eq:dispersion-E1}
\end{equation}
The equations above are valid for $0.4 < E_1 < 2.8$~meV (or $0.062 < k_1 < 0.158$~nm$^{-1}$) since $m_1$ is essentially constant over that range (see Fig.~\ref{Fig:Luttinger-vs-experiments}).

\begin{figure}[t]
    \includegraphics[width=1.0\columnwidth]{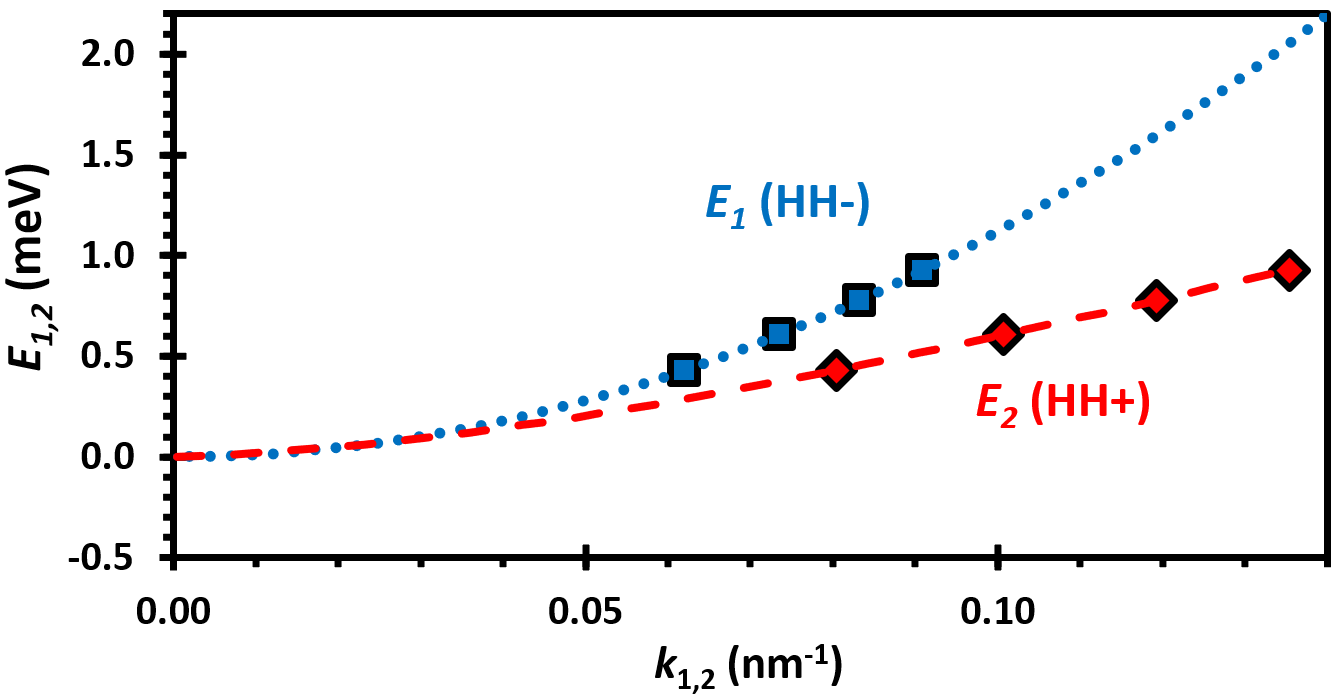}
    \caption{Dispersion relations for the HH$-$ subband (blue squares) and the HH$+$ subband (red diamonds) in sample J. The dotted line is Eq.~(\ref{eq:dispersion-E1}) with $m_1=0.34m_e$. The dashed line is Eq.~(\ref{eq:dispersion-E2expanded}) with $m_{2,0}=0.42m_e$, $\kappa = 0.350m_e$/meV, and $b=0.009m_e$/meV$^2$. Symbols (squares, diamonds) correspond to the $m_1, m_2$ data points in Fig.~\ref{Fig:Luttinger-vs-experiments}. }
    \label{Fig:dispersion-E(k)}
\end{figure}

Below the LH/HH anti-crossing point, we postulate that all non-parabolic effects of spin-orbit split HH$+$ can effectively be contained in the effective mass function $m_2(k_2)$, because the Luttinger Hamiltonian [see Eqs.~(\ref{bulk_hamil})$-$(\ref{bulk_hamil_4subband_asym})] essentially scales as $k^2$. Our postulate is essentially a generalized form of the original effective mass approximation, and is valid as long as $m_2(k_2) \neq 0$. Furthermore, in sample~J, the original parabolic approximation for the HH$+$ subband yields differences of only up to $\sim$\,13\% between the estimated and experimental $m_2$ values (see Fig.~\ref{Fig:Parabolic}). For these reasons, we approximate:
\begin{equation}
E_2\approx\frac{\hbar^2 k_2^2}{2 m_2}
\label{eq:dispersion-E2}
\end{equation}
which, using $m_2(E_2)$ from Eq.~(\ref{Eq:m2(E)}), can be re-written as:
\begin{equation}
k_2 = \sqrt{2[m_{2,0}E_2 + \kappa E_2^2 + b E_2^3]/\hbar^2},
\label{eq:dispersion-E2expanded}
\end{equation}
for $0.43 < E_2 < 0.93$~meV, corresponding to the range of $E_{\textsc{f}}$ investigated in sample~J. This dispersion relation is strongly non-parabolic for all values of $E_2$.

Equations (\ref{eq:dispersion-E1}) and (\ref{eq:dispersion-E2expanded}) are empirical analytical expressions for the dispersion relation of the HH$-$ and HH$+$ subbands respectively, directly derived from transport experiments, and are another central result of this paper. They are plotted in Fig.~\ref{Fig:dispersion-E(k)}, from $k_{1,2}=0$ to 0.14~nm$^{-1}$. The extrapolations of both dispersion relations merge at $k_1 = k_2 =0$ with $E_1=E_2$, as they should.

\begin{figure}[t]
    \includegraphics[width=1.0\columnwidth]{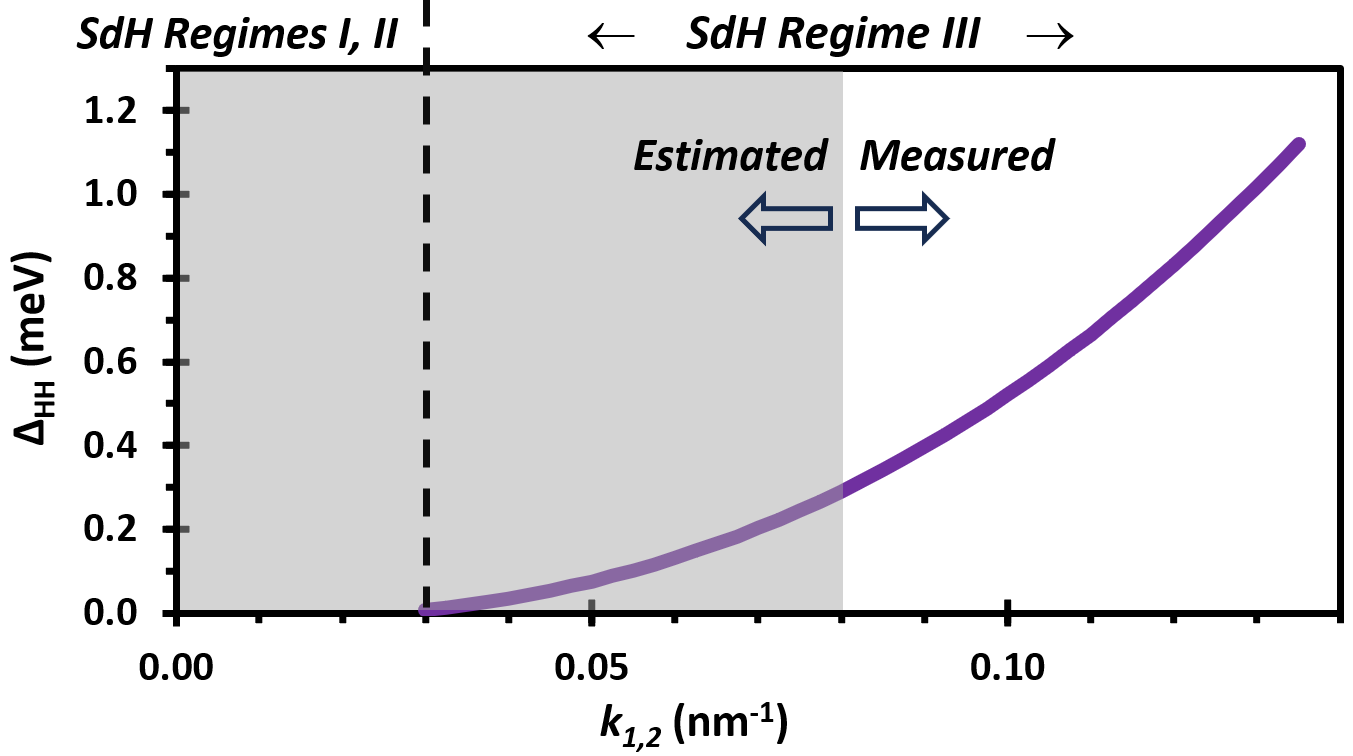}
    \caption{Spin-orbit splitting energy $\Delta E = E_1 - E_2 = \Delta_{\textsc{hh}}$ between the HH$-$ and HH$+$ subbands in sample~J, using Eqs.(\ref{eq:dispersion-E1}) and (\ref{eq:dispersion-E2expanded}).}
    \label{Fig:DeltaHH}
\end{figure}

\textbf{Spin-orbit-splitting energy from dispersions.} Having obtained the empirical dispersions of HH$-$ and HH$+$ from our magnetotransport results, Figure~\ref{Fig:DeltaHH} plots the spin-orbit-splitting energy $\Delta_{\textsc{hh}}$ against crystal momentum. Strictly speaking, the experimental data from both HH subbands in sample~J only overlap with the same crystal momentum between $k_{1,2}=0.080$~nm$^{-1}$ (corresponding to the minimum $k_2$ experimental value) and 0.091 nm$^{-1}$ (corresponding to the maximum $k_1$ experimental value). However, since Eq.~(\ref{eq:dispersion-E1}) is a general result valid for $0.062 < k_1 < 0.158$~nm$^{-1}$, we will consider the experimental window for $\Delta_{\textsc{hh}}$ to range from $k_{1,2}=0.080$ to 0.135~nm$^{-1}$ (the white area in Fig.~\ref{Fig:DeltaHH}), corresponding to the minimum/maximum experimental $k_2$ values in sample J.

For $k_{1,2} < 0.08$~nm$^{-1}$ (they gray area in Fig.~\ref{Fig:DeltaHH}), $\Delta_{\textsc{hh}}$ is estimated from extrapolating the dispersion relations. Near $k_{1,2} \approx 0.03$~nm$^{-1}$ (vertical dashed line in Fig.~\ref{Fig:DeltaHH}), the extrapolated estimate predicts $\Delta_\textsc{hh} \approx 0$, marking the boundary between the SdH regimes II and III (see Table~\ref{tab:SdHregimes}). As a consistency check, recall that Fig.~\ref{Fig:densities}(a) predicted $\Delta_{\textsc{hh}} \approx 0$ when $p_1 = p_2 \approx (0.21\pm0.01) \times 10^{15}$/m$^2$ in sample~J; this $p_1$ subband density would correspond to $k_1 \approx 0.05$~nm$^{-1}$, not very far from $k_1 \approx 0.03$~nm$^{-1}$ predicted by Fig.~\ref{Fig:DeltaHH}.

In principle, one cannot measure $p_2, m_2, \Delta_\textsc{hh}$ down to $k_{1,2}=0$ using SdH oscillations because one must inevitably enter SdH regimes I and II to do so, where $p_2$ and $m_2$ are no longer resolvable.

\textbf{Comparing experiments to Luttinger theory.} Figure~\ref{Fig:Luttinger-vs-experiments} demonstrates that experimental results and Luttinger theory calculations follow the same trends for both $m_1$ and $m_2$. The calculated effective mass of the HH$-$ subband ($m_1$) is near-independent of the Fermi energy [see Fig.~\ref{masses_quantitative} in the Appendix] below the LH/HH anticrossing. Accordingly, the Luttinger-calculated dispersion for HH$-$ is an isotropic parabola (see Fig.~\ref{fermi_surfaces} in the Appendix) below the LH/HH anticrossing. In contrast, the effective mass of the HH$+$ subband ($m_2$) has a strong dependence on the Fermi energy [see Fig.~\ref{masses_quantitative} in the Appendix]; its dispersion is strongly non-parabolic (see Fig.~\ref{masses} in the Appendix) and anisotropic (see Fig.~\ref{fermi_surfaces} in the Appendix). As $E_{\textsc{f}}$ decreases to zero, $m_2$ tends toward the value of $m_1$ (see Fig.~\ref{masses_quantitative} in the Appendix). By far the most important observation, our Luttinger theory calculations underestimate \textit{all} experimental $m_1$ and $m_2$ values, both ours and those from literature, by the \textit{same} common factor ($\approx 2.3$).

Some of the individual trends mentioned above have already been reported in past theoretical studies on 2DHGs hosted in (100) GaAs single heterojunctions, e.g., \cite{Ekenberg1985, Ando1985, winkler2000rashba, winkler2005anomalous, Marcellina17}. However, there is a wide range of reported behaviors, and the only consensus on the properties of $m_1$ and $m_2$ is that $m_1$ is small ($0.15m_e-0.20m_e$), $m_1$ is only weakly dependent on 2DHG density, and the dispersions of all subbands are non-parabolic. Therefore the quantitative disagreement between experiment and theory shown in Fig.~\ref{Fig:Luttinger-vs-experiments} is not limited to our particular implementation of the Luttinger model.

Of course, the larger-than-predicted experimental effective masses mean that the experimental dispersion relations, Fermi energies, and spin-orbit splitting energy gap between HH$-$/HH$+$ will not match theory \cite{Ekenberg1985, Ando1985, winkler2000rashba, winkler2005anomalous, Marcellina17}. For example, Ref.~\onlinecite{Marcellina17} predicts $\Delta_\textsc{hh} \approx 2.2$~meV for a GaAs inversion layer and $\Delta_\textsc{hh} \approx 1.2$~meV for a GaAs accumulation device at $k_{1,2} = 0.1$~nm$^{-1}$ and nominally $p_{2d} = 1 \times 10^{15}$/m$^2$ (see Fig.~3 in \cite{Marcellina17}). Our experiments give $\Delta_\textsc{hh} \approx 0.52$~meV at $k_{1,2} = 0.1$~nm$^{-1}$ and $p_{2d} = 1.1 \times 10^{15}$/m$^2$ (see Figs.~\ref{Fig:dispersion-E(k)} and \ref{Fig:DeltaHH}). Our Luttinger theory calculations predict $\Delta_\textsc{hh} \approx 1.2$~meV at $k_{1,2} = 0.1$~nm$^{-1}$ (see Fig.~\ref{bstruct_V_no_so_2} in the Appendix). We mention both accumulation and inversion layer results from \cite{Marcellina17} because their inversion layer results is the only theory study that reproduces a near-linear dependence of $m_2$ with 2DHG density/Fermi energy, closer to experiment. The predicted $m_2$ from their accumulation device calculation has a very different trend than our experiment: it first increases from $p_{2d} = 0.5 \times 10^{15}$/m$^2$ to a peak value at $1.2 \times 10^{15}$/m$^2$, and then \textit{decreases} with increasing 2DHG density.

The Rashba-split  HH$-$ subband is unique among the subbands in the heavy-hole and light-hole bands: it is the only one with a near-parabolic dispersion at 2DHG densities below the LH$+$/HH$-$ anticrossing (see Figs.~\ref{bstruct_V_no_so_1}$-$\ref{bstruct_V_no_so_3} in the Appendix). All other subbands are non-parabolic (i.e., HH$+$, LH$+$, LH$-$, but also the higher-energy subbands HH2$+$, HH2$-$, LH2$+$, LH2$-$, HH3$+$,~...). The reason for the near-parabolicity of HH$-$ is fundamentally linked to the specific band hybridization associated with SIA, which we describe below.

In a symmetric confining potential (e.g., a QW), the HH$+$ and LH$-$ subbands (HH$-$ and LH$+$ subbands) hybridize, causing band non-parabolicity. However, HH$+$ and HH$-$ (LH$+$ and LH$-$) remain degenerate in energy for all $k_{1,2}$ values. The repulsion between HH$-$ and LH$+$ (or HH$+$ and LH$-$) can be traced to the $R$ terms of the Luttinger Hamiltonian in Eqs.~(\ref{bulk_hamil}) and (\ref{bulk_hamil_4subband_sym}) in the Appendix, and is the cause for subband anticrossings. In an asymmetric confining potential (e.g., a SH), the HH$+$ subband hybridizes with \emph{both} LH$-$ and LH$+$. The HH$-$ subband hybridizes with both LH$-$ and LH$+$ as well. Thus HH$+$ and HH$-$ experience repulsion from each other through their common hybridization with LH$+$/LH$-$. This repulsion can be traced to the $S$ terms of the Luttinger Hamiltonian in Eqs.~(\ref{bulk_hamil}) and (\ref{bulk_hamil_4subband_asym}) in the Appendix; it is the cause for the spin-orbit splitting of HH$+$/HH$-$ and the lifting of their degeneracy at $k_{1,2}\neq0$. All of the above is described in more detail in Appendix~\ref{app:explanation}.

The HH$-$ subband experiences repulsion from the HH$+$ subband below it, and from the LH$-$ subband above it. The near-parabolic dispersion of HH$-$ strongly implies these opposite repulsions almost cancel each other, over a wide range of electric fields and of 2DHG densities below the HH$-$/LH$+$ anticrossing (see Figs.~\ref{bstruct_V_no_so_1}$-$\ref{bstruct_V_no_so_3} in the Appendix). This near-cancellation does not occur in any other valence subband.

\textbf{Possible origins for $m_1$ and $m_2$ disagreement.} There are many possible origins for the quantitative disagreement between theory and experiment on $m_1$ and $m_2$ in (100) GaAs. The most obvious is that Luttinger theory applies to $B=0$~T whereas SdH experiments are performed when $B \neq 0$~T. However, when $B<1$\,T, the magnetic field dependence of $m_1(B)$ and $m_2(B)$ is very weak. This is known from CR studies (see Fig.~2 from \cite{Rachor2009} and Fig.~3 from \cite{stormer1983energy}), from SdH studies (see Fig.~3 from \cite{eisenstein1984effect} and Fig.~2 from \cite{nichele2014spin}), and from theory (see Fig.~8 from \cite{simion2014magnetic}). Extrapolating our $m_1(B)$ and $m_2(B)$ to $B=0$~T still leaves $m_1$ and $m_2$ significantly much larger than predicted. Magnetic field cannot be responsible for the disagreement between Luttinger theory and experiment.

Another possible origin for the disagreement could be an ``incorrect'' choice of input parameters to the Luttinger model, leading to smaller calculated effective masses than in experiments. This would involve choosing different values for the Luttinger parameters ($\gamma_1, \gamma_2, \gamma_3$) in Eq.~(\ref{bulk_hamil}) of the Appendix. For example, enforcing the condition $m_e/(\gamma_1+\gamma_2) = 0.34m_e = m_1$ while otherwise using $\gamma_1, \gamma_2, \gamma_3$ as free parameters to fit the density dependence of the experimental $m_2$ and $\Delta_{\textsc{hh}}$ might deliver apparent agreement with magnetotransport experiments. However, this would be ignoring the fact that in-plane and out-of-plane 2D effective masses for (100) GaAs are intrinsically linked to each other in the Luttinger model via the relations:
\begin{eqnarray}
m^*_{\textsc{hh},\parallel} =  \frac{m_e}{\gamma_1+\gamma_2}
& \quad\text{and}\quad &
m^*_{\textsc{hh},\perp} = \frac{m_e}{\gamma_1 - 2\gamma_2},\quad\\
m^*_{\textsc{lh},\parallel} =  \frac{m_e}{\gamma_1-\gamma_2}
& \quad\text{and}\quad &
m^*_{\textsc{lh},\perp} = \frac{m_e}{\gamma_1+2\gamma_2},
\end{eqnarray}
\noindent where $\perp$ means the [100] crystal direction, $\parallel$ means within the (100) plane, and all masses are at the $\Gamma$ point. The values $\gamma_1=6.85$, $\gamma_2=2.10$, and $\gamma_3=2.90$ used in our calculations are firmly established in the literature \cite{Adachi1985GaAs, Pavesi1994photoluminescence, Winkler1995excitons, Vurgaftman2001, HerbertLi2001}, and were almost entirely obtained from fitting $m^*_{\textsc{hh},\perp}$ and $m^*_{\textsc{lh},\perp}$ in optical experiments. Thus the values for $m^*_{\textsc{hh},\parallel}$ and $m^*_{\textsc{lh},\parallel}$ (and by extension $\gamma_1, \gamma_2$, and $\gamma_3$) are essentially ``locked in,'' lest the agreement between Luttinger theory and optical experiments in GaAs heterostructures is invalidated.

Strain arising from the crystal lattice mismatch between different materials is known to affect the effective mass \cite{Annelise2023}. However, GaAs and AlGaAs have the lowest lattice mismatch of any III\nobreakdash-V semiconductor, and high-quality GaAs hosting the 2DHG in a (non-inverted) single heterojunction should be more relaxed than in any quantum well heterostructure. Strain in our case cannot be responsible for the disagreement because the required amount needed to double the effective mass is unrealistic.

Another possible origin for the disagreement could be that some assumptions underlying the data analysis are not valid. For example, the assumption that Landau levels are equidistant in energy in the SdH Regime III is almost certainly not valid. In another example, the single-particle picture, explicitly assumed in the Luttinger model, could be invalid. In that case, strong many-body interactions could increase the effective mass.

\textbf{Many-body interactions}. The interaction parameter $r_s$ represents the ratio of Coulomb to kinetic energies, and is typically used to characterize the strength of hole-hole interactions \cite{Spivak2010}. In a 2DHG with single subband occupancy, it is given by $r_s = m^*_{\textsc{hh}} e^2 / 4\pi \hbar^2 \epsilon \sqrt{\pi p_{2d}}$ where $\epsilon$ is the dielectric constant. A 2DHG is considered to be in the strongly interacting regime if $r_s>3$. In sample~J, the interaction parameter ranges from $r_s$\,$\approx$\,\,10 (at $p_{2d}$\,=\,1.9\,$\times$\,10$^{15}$/m$^2$) to 16 (at $p_{2d}$\,=\,0.75\,$\times$\,10$^{15}$/m$^2$), using the density-weighted average effective mass $m^*_{\textsc{hh}} = (p_1m_1+p_2m_2)/(p_1+p_2) \approx 0.5m_e$. These $r_s$ values indicate sample~J is deep into the strongly interacting regime. At $p_{2d}$\,=\,5.0\,$\times$\,10$^{15}$/m$^2$, even the sample from Ref.~\onlinecite{stormer1983energy} has $r_s$\,$\approx$\,\,6. In other words, all SH samples listed in Fig.~\ref{Fig:SdH-literature-mass-experiments} are well into the strongly interacting regime, and many-body interactions are certainly plausible.

To simulate the effect of many-body interactions on effective masses $m_1$ and $m_2$, we perturbatively added hole-hole interactions to our Luttinger model (not shown). As a result, both effective masses $m_1,m_2$ increase with increasing interactions. Interestingly, both $m_1$ and $m_2$ became larger by the same common factor, which itself did not depend on $E_{\textsc{f}}$. Intuitively, the HH$-$ and HH$+$ effective masses are intrinsically linked: they belong to the same subband at $k=0$ nm$^{-1}$ in Fig.~\ref{Fig:VB} (or  Figs.~\ref{bstruct_V_no_so_1}$-$\ref{bstruct_V_no_so_3} in the Appendix). It is therefore perhaps not surprising that the same scaling factor applies to both $m_1$ and $m_2$. Our model with many-body interactions was severely limited in its parameter space: only weak hole-hole interactions could be simulated before our model broke down, and only small Fermi energies ($<$\,0.3~meV) could be used as computing resources became too onerous.

In light of all available information, we thus atribute the enhancement of both HH$-$ and HH$+$ effective masses beyond the Luttinger model SOI predictions to be most likely due to many-body interactions, and is a key claim of this paper.

\textbf{2DHG systems other than (100) GaAs}. It remains to be seen whether any of the specific experimental observations reported here could be generalized to other material systems. For example, the uniqueness of the near-parabolic dispersion of HH$-$ does not appear to be only limited to the GaAs/AlGaAs material system. Our Luttinger model calculations of 2DHGs in single heterojunctions from another material system, In$_{0.53}$Ga$_{0.47}$As/In$_{0.53}$Al$_{0.47}$As with stronger SOI than GaAs, also reveal a parabolic dispersion for HH$-$ and non-parabolic dispersions for all other HH and LH subbands (not shown). Luttinger calculations in Fig.~6(b) from Ref.~\onlinecite{Marcellina17} show a near-independence of $m_1$ on the 2DHG density from 0.5 to 3\,$\times$\,10$^{15}$/m$^2$ for inversion layers in GaAs, InAs, InSb, and Ge. This essentially predicts a near-parabolic dispersion for HH$-$ over that range of 2DHG densities for all of these materials. Thus we speculate the near-parabolic dispersion of HH$-$ could be a general feature of \textit{any} asymmetrically-confined 2DHG in zincblende semiconductor materials.

Effects of spin-orbit splitting can be observed in experimental conditions (e.g., temperature, mobilities) well outside the SdH regime. In GaAs single heterojunctions, the HH$-$ and HH$+$ subbands maintain their population imbalance $\Delta p$ at high temperatures in magnetotransport \cite{grbic2004} and cyclotron resonance studies \cite{CRmass-meta-analysis}. We thus believe spin-orbit splitting is relevant to materials other than GaAs, with lower mobilities and operated at much higher temperatures than 100~mK. For example, the effective mass of holes in silicon MOSFETs has recently been suggested to be much heavier than expected \cite{Wendoloski2026PRBaccepted}, to explain the large difference between the mobilities of electrons and holes despite having similar effective masses ($m^*_{\text{elec},\parallel} = 0.19m_e$ versus $m^*_{\textsc{hh},\parallel} = 0.22m_e$). Ref.~\onlinecite{Wendoloski2026PRBaccepted} attributes the implied larger $m^*_{\textsc{hh},\parallel}$ to spin-orbit interactions. This would be consistent with the presence of $m_2$ from HH$+$ in Si \cite{Marcellina17}, even though the absence of SdH oscillations cannot confirm its existence.

\section{Conclusions}
\label{sec:conclusion}

We have studied undoped, accumulation-mode GaAs/AlGaAs (100) single-heterojunction two-dimensional hole gases with exceptionally low disorder, achieving mobilities up to 84 cm$^2$/Vs while sustaining strong structural inversion asymmetry corresponding to out-of-plane electric fields up to 26 kV/cm. In this regime we realize very large Rashba-driven spin–orbit polarization, reaching a subband population imbalance as high as 36\% at $B<1$~T over low total densities $p_{2d}=0.76–1.9\times 10^{15}$/m$^{-2}$.

Using low-field magnetotransport, we extracted the branch-resolved properties of the spin–orbit–split heavy-hole subbands by Fourier analysis resolving the SdH interference pattern and fitting the Lifshitz–Kosevich temperature damping of the corresponding Fourier amplitudes. This procedure direct measurements of both $m_1$ and $m_2$ for HH$-$ and HH$+$ over the same magnetic field range, enabling a transport-based reconstruction of the HH$-$ and HH$+$ dispersions and analytic parameterizations of both branches [Eqs.~(\ref{eq:dispersion-E1}) and (\ref{eq:dispersion-E2expanded})].

Two robust physical conclusions emerge. First, the HH$+$ branch is strongly nonparabolic: its SdH effective mass varies substantially with density (from $\sim 0.58 m_e$ to $\sim 0.76 m_e$ over our range), consistent with the pronounced heavy-hole–light-hole mixing expected in asymmetric (100) hole systems. Second, and more remarkably, the HH$-$ branch exhibits an almost density-independent mass ($\approx 0.34 m_e$), demonstrating that HH$-$ is near-parabolic below the first LH$+$/HH$-$ anticrossing. Practically, this identifies a wide and experimentally relevant regime in which the Fermi energy and density of states of (100) GaAs single-heterojunction 2DHGs can be determined reliably from transport, overcoming a long-standing ambiguity associated with nonparabolic valence-band dispersions; combining our results with prior data suggests this near-parabolic HH$-$ regime persists from approximately 0.75 up to 5$\times 10^{15}$/m$^2$ in the single-subband limit below the anticrossing.

Having obtained both dispersions, we determine the spin-orbit splitting energy $\Delta_{\textsc{hh}} = E_1 - E_2$ from experiment, and we delineate the field and density windows in which both branches are simultaneously resolvable—providing a framework that helps reconcile discrepancies among earlier (100) GaAs studies based on different measurement regimes and disorder levels.

Finally, comparison with parameter-free Luttinger-model calculations reproduces the qualitative trends, particularly the near-constant HH$-$ mass and the strong density dependence of HH$+$, but underestimates both $m_1$ and $m_2$ by an approximately common factor $\approx$ 2. Together with the large interaction parameters ($r_s \approx 16$) in our density range, this points to substantial many-body renormalization of the heavy-hole mass in strongly asymmetric 2DHGs, and establishes an experimental benchmark for future interaction-aware theories of hole spin–orbit physics in GaAs and related material systems.

\begin{acknowledgments}
We acknowledge Drs. Lisa Tracy, John Reno, and Terry Hargett at Sandia National Laboratories for kindly providing one of the GaAs/AlGaAs wafers studied in this work. N.C., F.S., and M.K. contributed equally to this work. J.B. and J.B.K. supervised this work equally. S.R.H. acknowledges support from a NSERC Canada Graduate Scholarships-Doctoral (CGS-D). This research was undertaken thanks in part to funding from the Canada First Research Excellence Fund (Transformative Quantum Technologies), Defence Research and Development Canada (DRDC), National Research Council Canada (NRC), and Canada’s Natural Sciences and Engineering Research Council (NSERC). The University of Waterloo’s QNFCF facility was used for this work. This infrastructure would not be possible without the significant contributions of CFREF-TQT, CFI, ISED, the Ontario Ministry of Research and Innovation, and Mike and Ophelia Lazaridis. Their support is gratefully acknowledged.
\end{acknowledgments}

\appendix

\section{The Luttinger model}
\label{secA:Luttinger}

\subsection{The bulk Hamiltonian}

We consider the valence band of GaAs in the basis of six p-type subbands labeled by the total angular momentum and its projection, $|j,j_z\rangle$. We order the basis states as follows:
$\{|\frac{3}{2},+\frac{3}{2}\rangle, |\frac{3}{2},+\frac{1}{2}\rangle, |\frac{3}{2},-\frac{1}{2}\rangle, |\frac{3}{2},-\frac{3}{2}\rangle, |\frac{1}{2},+\frac{1}{2}\rangle, |\frac{1}{2},-\frac{1}{2}\rangle\}$. The subbands labeled by $j=\frac{3}{2}$, $j_z = \pm \frac{3}{2}$ are the heavy holes, the subbands labeled by $j=\frac{3}{2}$, $j_z = \pm \frac{1}{2}$ are the light holes, and the subbands with $j=\frac{1}{2}$, $j_z = \pm \frac{1}{2}$ form the spin-orbit split-off band. The six-subband Hamiltonian describing the bulk valence band takes the following form~\cite{Novik05}:
\begin{eqnarray}
    &&\hat{H}_{B} = \label{bulk_hamil} \\
    &&\left[
    \begin{array}{cccccc}
        U + V & S_- & R & 0 & -\frac{1}{\sqrt{2}}S_- & -\sqrt{2} R\\
        S_-^* & U - V & 0 & R & \sqrt{2} V & \sqrt{\frac{3}{2}}{S}_- \\
        R^* & 0 & U-V & -S_+^* & \sqrt{\frac{3}{2}}{S}_+ & -\sqrt{2} V\\
        0 & R^* & -S_+ & U+V & \sqrt{2} R^* & -\frac{1}{\sqrt{2}}S_+ \\
        -\frac{1}{\sqrt{2}} S_-^* & \sqrt{2} V & \sqrt{\frac{3}{2}}{S}_+^* &
                \sqrt{2} R & U + \Delta & 0 \\
        -\sqrt{2} R^* & \sqrt{\frac{3}{2}}{S}_-^* & -\sqrt{2} V &
                -\frac{1}{\sqrt{2}} S_+^* & 0 & U+\Delta \\
    \end{array}
    \right] \nonumber
\end{eqnarray}
where
\begin{eqnarray}
    k_{||}^2 &=& k_x^2 + k_y^2, \quad
            k_{\pm} = k_x \pm i k_y, \quad
            k_z = -i \frac{\partial}{\partial z},\\
    U &=& E_v(z) + \frac{\hbar^2}{2m_e}\gamma_1(k_{||}^2 + k_z^2 ),\\
    V &=& \frac{\hbar^2}{2m_e}\gamma_2 (k_{||}^2 -2  k_z^2 ),\\
    R &=& \sqrt{3}\frac{\hbar^2}{2m_e}( \mu k_+^2 - \bar{\gamma}k_-^2),\\
    S_{\pm} &=& - 2\sqrt{3} \frac{\hbar^2}{2m_e} \gamma_3 k_{\pm} k_z.
\end{eqnarray}
Further, $E_v(z)$ is the valence band edge in the growth ($z$) direction, $\gamma_1$, $\gamma_2$, and $\gamma_3$ are the Luttinger parameters, $\mu = (\gamma_3-\gamma_2)/2$, $\bar{\gamma}=(\gamma_2+\gamma_3)/2$, and $\Delta$ is the spin-orbit gap. For GaAs we take $\gamma_1=6.85$, $\gamma_2=2.1$, $\gamma_3=2.9$, and $\Delta = 0.35$ eV.

The Hamiltonian (\ref{bulk_hamil}) is consistent with Ref.~\onlinecite{Luttinger56}, except there is an overall factor of $(-1)$ allowing us to consider higher-excited hole states with a more positive energy (traditionally, the valence band extends towards more negative energies). The Hamiltonian is also consistent with Ref.~\onlinecite{Novik05} with the same factor of $(-1)$.
In Refs.~\onlinecite{liu2018strong,Marcellina17} the Hamiltonian is reduced to a four-subband form, accounting for the four hole subbands with total angular momentum $j=\frac{3}{2}$, and the energy is measured as above (without the additional factor). Furthermore, in Refs.~\onlinecite{Szumniak12,Szumniak13} the $4\times 4$ Hamiltonian is further simplified by setting $S_{\pm}=0$. In all cases, however, all available Hamiltonian terms appear to match.

\subsection{Bulk inversion asymmetry}

Next, we account for the bulk inversion asymmetry of GaAs. We accomplish this by considering the so-called Dresselhaus spin-orbit interaction Hamiltonian. We utilize the following Hamiltonian:
\begin{equation}
    \hat{H}_{BIA} = -\beta_0 \vec{k}\cdot\vec{\Omega}_{\vec{J}}
    - \beta \vec{\Omega}_{\vec{k}}\cdot \vec{J}.
    \label{hBIA}
\end{equation}
Here, the operator $\vec{\Omega}_{\vec{O}}$ is a ``vector operator'', i.e., a short-hand notation for third-order groupings of the components of any vector operator $\vec{O}$. In the Hamiltonian (\ref{hBIA}), we have two instances of this operator, applied to the angular momentum $\vec{J}$ (first term) and the momentum $\vec{k}$ (second term). The meaning of this vector operator is:
$\Omega_{\vec{O}}^x = \{ O_x, O_y^2 - O_z^2 \}$, and the other two components are generated by cyclic permutations, with $\{ A, B\} = \frac{1}{2}(AB+BA)$. In the following, we take the GaAs parameters $\beta_0 = -0.00393$ eV$\cdot${\AA} and $\beta = -81.93$ eV$\cdot${\AA}$^3$. The above Hamiltonian and the parameters are taken from Ref.~\onlinecite{winkler2003spin}, with two lowest-order terms retained. The Hamiltonian has an overall negative sign to account for the overall sign of the Hamiltonian (\ref{bulk_hamil}), a choice identical to that in Refs.~\onlinecite{Szumniak12,Szumniak13}. We consider $\hat{H}_{BIA}$ in the reduced basis of four subbands, $\{ |\frac{3}{2},+\frac{3}{2}\rangle, |\frac{3}{2},+\frac{1}{2}\rangle,
|\frac{3}{2},-\frac{1}{2}\rangle, |\frac{3}{2},-\frac{3}{2}\rangle \}$, because the spin-orbit split-off subband is separated from the heavy and light hole subband by a substantial spin-orbit gap $\Delta_{so}$ and we do not expect it to contribute significantly to the Dresselhaus-related
low-energy subband mixing.

Let us now write the Hamiltonian (\ref{hBIA}) in a more extended form. We have:
\begin{eqnarray}
\hat{H}_{BIA} &=& -\beta_0 \left[ k_x \{ J_x, J_y^2 - J_z^2 \} + k_y \{ J_y, J_z^2 - J_x^2 \} \right. \quad\nonumber\\
~&~& \left. + k_z \{ J_z, J_x^2 - J_y^2 \} \right] - \beta \left[ J_x \{ k_x,k_y^2-k_z^2 \} \right. \nonumber\\
~&~& \left. + J_y \{ k_y,k_z^2-k_x^2 \} + J_z \{ k_z,k_x^2-k_y^2 \} \right].
\end{eqnarray}
This Hamiltonian will contain terms which are linear, quadratic, and cubic in $k_z$, and of course terms independent of $k_z$. In consideration of bulk materials all these terms need to be included. However, we will consider a heterostructure, with a relatively strong confinement in the growth ($z$) direction, as defined by the band edge profile $E_v(z)$ in the Hamiltonian (\ref{bulk_hamil}). The terms linear and cubic in $k_z$ will couple the subbands resulting from this confinement. However, we assume that the energy gaps between the subsequent subbands are
large compared to the energy scale of the Hamiltonian $\hat{H}_{BIA}$ close to the Gamma point of the Brillouin zone. This allows us to simplify our Dresselhaus Hamiltonian by neglecting the terms featuring $k_z$ in odd powers. We take then:
\begin{eqnarray}
\hat{H}_{BIA} &=& -\beta_0 \left[k_x \{J_x, J_y^2 - J_z^2 \} + k_y \{J_y, J_z^2 - J_x^2 \}\right] \nonumber\\ && - \beta k_z^2 (J_y k_y - J_x k_x).
\label{dresselhaus}
\end{eqnarray}
This approximation is recommended in Ref.~\onlinecite{winkler2003spin} and utilized in Refs.~\onlinecite{Szumniak12,Szumniak13}.

Let us now write the Dresselhaus Hamiltonian explicitly in a matrix form in the basis of heavy and light hole subbands. We will utilize the following forms for Pauli matrices for the $j=\frac{3}{2}$ algebra:
\begin{eqnarray}
    J_x &=& \frac{1}{2}\left[
        \begin{array}{cccc}
            0 & \sqrt{3} & 0 & 0 \\
            \sqrt{3} & 0 & 2 & 0 \\
            0 & 2 & 0 & \sqrt{3} \\
            0 & 0 & \sqrt{3} & 0
        \end{array}
    \right], \label{jx} \\
    J_y &=& \frac{1}{2}\left[
        \begin{array}{cccc}
            0 & -i\sqrt{3} & 0 & 0 \\
            i\sqrt{3} & 0 & -2i & 0 \\
            0 & 2i & 0 & -i \sqrt{3} \\
            0 & 0 & i\sqrt{3} & 0
        \end{array}
    \right], \label{jy} \\
    J_z &=& \frac{1}{2}\left[
        \begin{array}{cccc}
            3 & 0 & 0 & 0 \\
            0 & 1 & 0 & 0 \\
            0 & 0 & -1 & 0 \\
            0 & 0 & 0 & -3
        \end{array}
    \right]. \label{jz}
\end{eqnarray}
Utilizing these matrices, we have
\begin{eqnarray}
    \hat{H}_{BIA} &=&
    \frac{\beta_0}{4} \left[
        \begin{array}{cccc}
            0 & \sqrt{3}k_+ & 0 & 3k_- \\
            \sqrt{3}k_- & 0 & -3k_+ & 0 \\
            0 & -3k_- & 0 & \sqrt{3}k_+ \\
            3k_+ & 0 & \sqrt{3}k_- & 0
        \end{array}
    \right] \nonumber\\
    &&+ \frac{\beta}{2} k_z^2 \left[
        \begin{array}{cccc}
            0 & \sqrt{3}k_+ & 0 & 0 \\
            \sqrt{3}k_- & 0 & 2k_+ & 0 \\
            0 & 2k_- & 0 & \sqrt{3}k_+ \\
            0 & 0 & \sqrt{3}k_- & 0
        \end{array}
    \right].
\end{eqnarray}

\subsection{Structure inversion asymmetry}

The spin-orbit interaction due to the structure inversion asymmetry is typically modeled with an assumption that an overall external electric field $\vec{E}$ is applied to the system. Retaining the two lowest-order terms, the SIA (Rashba) Hamiltonian is expressed as:
\begin{equation}
    \hat{H}_{SIA} = -\alpha_1 \vec{k}\times\vec{E} \cdot \vec{J}
    - \alpha_2 \vec{k}\times\vec{E} \cdot \vec{\cal J}.
\end{equation}
Here, $\vec{\cal J} = [J_x^3,J_y^3,J_z^3]$. For GaAs, $\alpha_1=-14.62 e$ \AA$^2$ and $\alpha_2 =-0.106 e$ \AA$^2$ with $e$ being the electron charge ($e>0$) (Ref.~\onlinecite{winkler2000rashba}).
The formula is taken from Ref.~\onlinecite{winkler2000rashba} with an overall minus sign accounting for the overall sign of the Hamiltonian (\ref{bulk_hamil}). The formula is consistent with that in Ref.~\onlinecite{winkler2000rashba} retaining tle largest two terms up to the first order in $\vec{k}$.
Here, as for $\hat{H}_{BIA}$, we will consider the spin-orbit interaction only in the subspace of heavy and light hole subbands, excluding the spin-orbit split-off subband.

For our system, we will assume that the electric field is directed along the growth direction, $\vec{E}=[0,0,E_z]$. In such case, our Rashba Hamiltonian can be written as:
\begin{equation}
    \hat{H}_{SIA} = -\alpha_1 E_z (k_y J_x - k_x J_y)
        - \alpha_2 E_z (k_y J_x^3 - k_x J_y^3).
\end{equation}
We now utilize the Pauli matrices (\ref{jx}), (\ref{jy}) to write the Hamiltonian in the matrix form:
\begin{equation}
    \hat{H}_{SIA} = -i\frac{\alpha_1E_z}{2} \left[
        \begin{array}{cccc}
            0 & -\sqrt{3} k_+ & 0 & 0 \\
            \sqrt{3}k_- & 0 & -2k_+ & 0 \\
            0 & 2k_- & 0 & -\sqrt{3}k_+ \\
            0 & 0 & \sqrt{3}k_- & 0 \\
        \end{array}
    \right] \nonumber
\end{equation}
\begin{equation}
    -i\frac{\alpha_2E_z}{8} \left[
        \begin{array}{cccc}
            0 & -7\sqrt{3} k_+ & 0 & 6k_- \\
            7\sqrt{3}k_- & 0 & -20k_+ & 0 \\
            0 & 20k_- & 0 & -7\sqrt{3}k_+ \\
            -6k_+ & 0 & 7\sqrt{3}k_- & 0 \\
        \end{array}
    \right].
    \label{rashba}
\end{equation}
When we compare the above form to that from Ref.~\onlinecite{winkler2000rashba}, we find that there is a difference in overall complex phase and that our form appears to be a conjugate of the form previously published. We do not understand why this should be so, and suspect that it is an issue
with the definition of $k_{\pm}$: when we use $k_{\pm} = k_y \pm k_x$, everything appears to fit.
We will remain consistent with the notation introduced with the Hamiltonian (\ref{bulk_hamil}) and keep the above final form of $\hat{H}_{SIA}$.

\section{Computational procedure}
\label{secA:computational}

We need to find the eigenenergies and eigenstates of the Hamiltonian
\begin{equation}
    \hat{H} = \hat{H}_B + \hat{H}_{BIA} + \hat{H}_{SIA}.
\end{equation}
Our sample is a quantum well or a heterostructure, which is translationally invariant in the $x$ and $y$ direction, but has a strong confinement along the $z$ (growth) direction. Therefore, $k_x$ and $k_y$ are good quantum numbers, and we only have to contend with the operator form of $k_z = -i \frac{\partial}{\partial z}$. To resolve the states of the system in the $z$ direction, we choose a computational box in the form of an infinite quantum well, such that
\begin{equation}
    V_{BOX}(z) = \left\{
    \begin{array}{cc}
        \infty & z < 0 \\
        0 & 0 \leq z \leq W, \\
        \infty & z > W,    \\
    \end{array}
    \right.
\end{equation}
and $W$ is the width which wholly encompasses any potential defined in the $z$ direction by the valence band edge $E_v(z)$ in the operator $U$ from the Hamiltonian (\ref{bulk_hamil}). We choose our computational basis as the set of eigenstates for our computational box, which are simply
\begin{equation}
    \langle z | l \rangle = \sqrt{\frac{2}{W}}
    \sin \left( \frac{l\pi}{W} z\right),
\end{equation}
where $l=1,2,\dots$ is an integer. These basis function form an orthonormal basis set.

We only need three matrix elements involving these basis functions. The first element is that of the actual confinement potential:
\begin{equation}
    \langle l_1 | E_v (z) | l_2\rangle =
    \frac{2}{W} \int_0^W dz \sin\left( \frac{l_1\pi}{W} z \right)
    E_v (z) \sin\left( \frac{l_2\pi}{W} z \right).
\end{equation}
We will have to calculate these integrals numerically for any pair of numbers $(l_1,l_2)$.
This is because we will deal with a numerical form of the band edge potential $E_v(z)$.

The second matrix element is that of the operator $k_z=-i\frac{d}{dz}$. We have:
\begin{equation}
    \langle l_1 | k_z | l_2 \rangle
    = -i\frac{2}{W}\int_{0}^{W} dz
    \sin \left( \frac{l_1 \pi}{W}z\right)
    \frac{d}{dz}
    \sin \left( \frac{l_2 \pi}{W}z\right).
\end{equation}
Elementary calculus gives
\begin{equation}
    \langle l_1 | k_z | l_2 \rangle = \frac{il_2}{W} \left(
    \frac{\cos(l_1+l_2)\pi - 1 }{l_1+l_2} +
    \frac{\cos(l_1-l_2)\pi - 1 }{l_1-l_2}
    \right).
\end{equation}
The element is zero for $l_1=l_2$ and whenever the two numbers are of the same parity.

The third element needed is that of the operator $k_z^2 = - \frac{d^2}{dz^2}$.
We have:
\begin{equation}
    \langle l_1 | k^2_z | l_2 \rangle = + \frac{l_1^2\pi^2}{W^2} \delta(l_1,l_2).
\end{equation}

We will look for eigenstates of our total Hamiltonian in the form of the Luttinger spinors:
\begin{equation}
   |n,k_x,k_y\rangle = \left[
    \begin{array}{c}
        \sum_{l=1}^{N} A^n_l(k_x,k_y) |l\rangle\\
        \sum_{l=1}^{N} B^n_l(k_x,k_y) |l\rangle\\
        \sum_{l=1}^{N} C^n_l(k_x,k_y) |l\rangle\\
        \sum_{l=1}^{N} D^n_l(k_x,k_y) |l\rangle\\
        \sum_{l=1}^{N} E^n_l(k_x,k_y) |l\rangle\\
        \sum_{l=1}^{N} F^n_l(k_x,k_y) |l\rangle\\
    \end{array}
   \right],
\end{equation}
where $N$ is the number of basis states $|l\rangle$ which we will choose to ensure convergence.
The number $n$ enumerates the eigenstates (and eigenenergies) of the full Hamiltonian.
For each eigenstate $n$, the coefficients $A^n_l(k_x,k_y)$ through $F^n_l(k_x,k_y)$ build the wave function components originating from each subband. The coefficients are not expected to be equal (or indeed even similar). We expect that the low-lying hole states will have predominantly heavy hole character owing to the subtleties of effective masses, and so we expect only
the coefficients $A^n$ and $D^n$ to be large. The large magnitude of coefficients $B^n$ and $C^n$ heralds the light hole character of the $n$-th hole state. As we are interested in the low-energy segment of the hole states only, we expect the coefficients $E^n$ and $F^n$ to be small throughout.

The computational procedure will therefore involve setting up the total Hamiltonian in a block form. We will deal with blocks of the size $N\times N$ in which we will formulate each of the operators from the Hamiltonian (\ref{bulk_hamil}) (e.g. $U+V$, $R$, etc.), as well as the operators $k_+$ and $k_-$ relevant for the BIA and SIA Hamiltonians. Further, these matrix operators are assembled to form the full $6N\times 6N$ Hamiltonian which will then be diagonalized numerically. As a result, we obtain $6N$ eigenenergies and eigenstates, of which we
will only be interested in several states with the lowest energy. This diagonalization will be carried out as a function of the wave numbers $k_x$ and $k_y$ not too far from zero (i.e., within the validity of the six-subband $\vec{k}\cdot\vec{p}$ model).

\begin{figure}[t]
    \includegraphics[width=1.0\columnwidth]{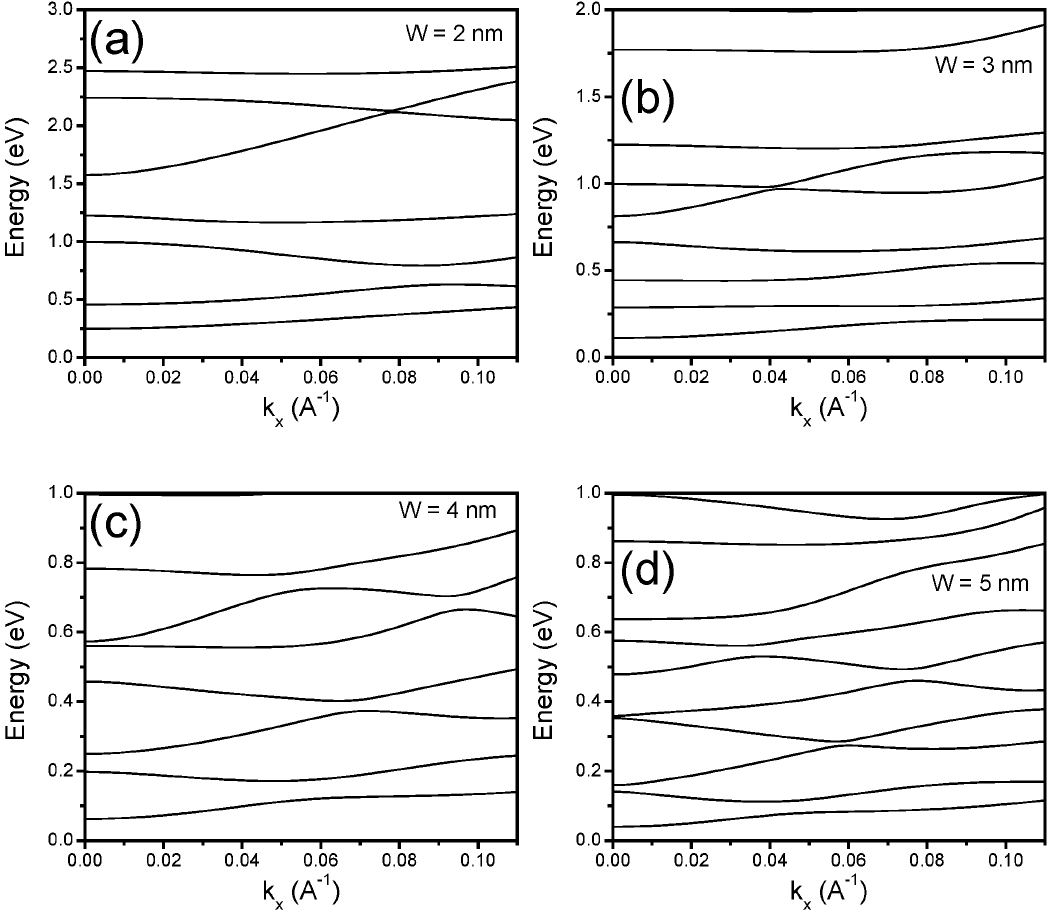}
    \caption{Band diagrams for the infinite quantum well with width $2$ nm (a), $3$ nm (b), $4$ nm (c) and $5$ nm (d). The Rashba term is absent as the quantum well is symmetric. The Dresselhaus term is neglected. }
    \label{subbands_well_noD}
\end{figure}

\section{Band structure of a rectangular quantum well}
\label{secA:subbandstructureQW}

In this Section we discuss the band structure of an infinite, symmetric, rectangular GaAs quantum well. Since this structure has inversion symmetry (i.e., there is no electric field applied to it), the Rashba Hamiltonian $\hat{H}_{SIA}$ will be identically zero and we only need to consider the bulk Hamiltonian $\hat{H}_B$ and the Dresselhaus Hamiltonian $\hat{H}_{BIA}$.We calculate the band diagram for several quantum well widths, $W=2$ nm, $3$ nm, $4$ nm, and $5$ nm.

In Fig.~\ref{subbands_well_noD} we show the band structure calculated by diagonalizing the Hamiltonian (\ref{bulk_hamil}) in the basis of $50$ quantum well functions in the $z$ direction (the quantum number $l=1,\dots,50$). Note that here we have not included $\hat{H}_{BIA}$. The subband diagram is calculated as a function of $\vec{k}=[k_x,0]$ within about $20$\% of the size of the Brillouin zone around the $\Gamma$ point. We show the band diagram for quantum well widths $2$nm (a), $3$ nm (b), $4$ nm (c), and $5$ nm (d). In each panel, the energy scales are different; they are chosen to show several lowest-lying subbands. All levels are doubly degenerate, accounting for the expected Kramers degeneracy in our inversion-symmetric system exhibiting time reversal symmetry. For the narrow well, $W=2$ nm, we find that the subbands are widely spaced. As the quantum well becomes wider, the subbands draw nearer in energy and the characteristic pattern of anticrossings becomes more evident. In particular, the anticrossing of the two lowest subbands occurs closer and closer to the $\Gamma$ point. Because of these anticrossings, the subbands are strongly nonparabolic, and the effective mass corresponding to each of them is strongly $k$-dependent.

\begin{figure}[t]
    \includegraphics[width=0.7\columnwidth]{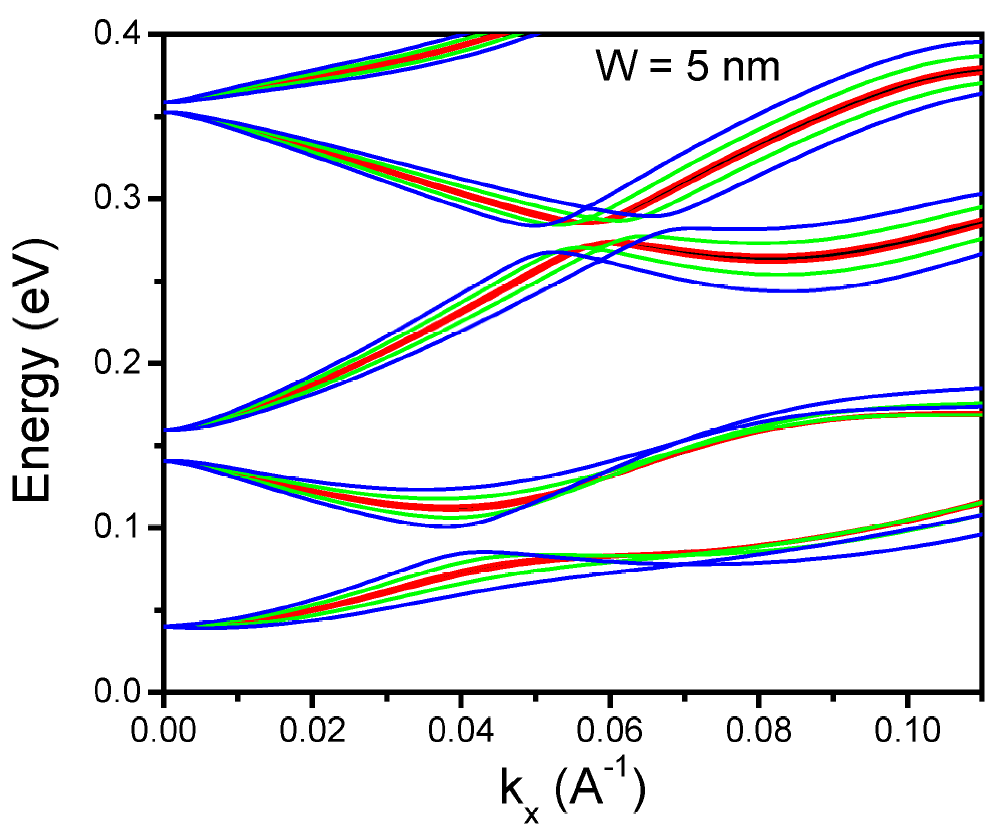}
    \caption{Evolution of the band structure for the infinite quantum well of width $5$ nm as a function of the magnitude of the Dresselhaus coefficient $\beta_0$. Black, red, green, and blue lines correspond respectively to the GaAs value of $\beta_0$, the value increased by a factor of $10$, $50$, and $100$. Note that the black lines are essentially underneath the red lines.}
    \label{subbands_well_BIA}
\end{figure}

Let us now include the Dresselhaus Hamiltonian $\hat{H}_{BIA}$, Eq. (\ref{dresselhaus}). Of the two terms, we will fully account for the first one, scaled by the parameter $\beta_0$. As for the second term, we find that our approach does not handle it correctly when introduced as shown in Eq. (\ref{dresselhaus}). The average $\langle k_z^2 \rangle$ of the $\hat{k_z}$ operator cannot be handled in our basis of quantum-well eigenstates. Its proper handling involves diagonalizing the bulk hamiltonian $\hat{H}_B$ first, extracting the lowest-energy eigenstates, and introducing this cubic Dresselhaus term perturbatively. While, of course, it is possible to do within our presentation, we find that qualitative points can be made by tuning the parameter $\beta_0$ to a higher value. Indeed, the two matrix Hamiltonians composing $\hat{H}_{BIA}$ have a similar
pattern of off diagonal matrix elements, and differ only by the nature and magnitude of coefficients. Therefore, we can examine the influence of the bulk inversion asymmetry on our quantum well band structure as a function of the parameter $\beta_0$. This evolution is shown in Fig.~\ref{subbands_well_BIA}. Black, red, green, and blue lines correspond respectively to the
natural GaAs value of $\beta_0$, the value increased by a factor of $10$, $50$, and $100$. The black lines are obscured by the red lines. We find, in general, that the double degeneracy of the subband levels is removed by the bulk inversion asymmetry. If we tentatively identify the two subband edges with a ``spin'' degree of freedom, we find that the band edges are spin-split without any magnetic field. The degree of this splitting is, however, very small for the unaltered GaAs Dresselhaus parameter $\beta_0$. From the artificially enhanced cases we see that the BIA Hamiltonian splits the edges apart, i.e., as its result, the lower-energy level is shifted down in energy, while the higher-energy level is shifted further up. As the subbands approach the energy anticrossings resulting from the subband mixing, the gaps are not the same for the higher (``spin-up'') and lower-energy (``spin-down'') states. As a result, the curvature of the nondegenerate subbands is different, and we can anticipate that the two ``spin'' components of the same band edge will exhibit a different effective mass. We emphasize that for the accurate GaAs Dresselhaus parameter these effects are subtle and invisible on the energy scale of Fig.~\ref{subbands_well_BIA}.

\section{Band structure of a single heterojunction}
\label{secA:subbandstructureSH}

\subsection{Single heterojunction layout and parameters}

\begin{figure}[t]
    \includegraphics[width=0.8\columnwidth]{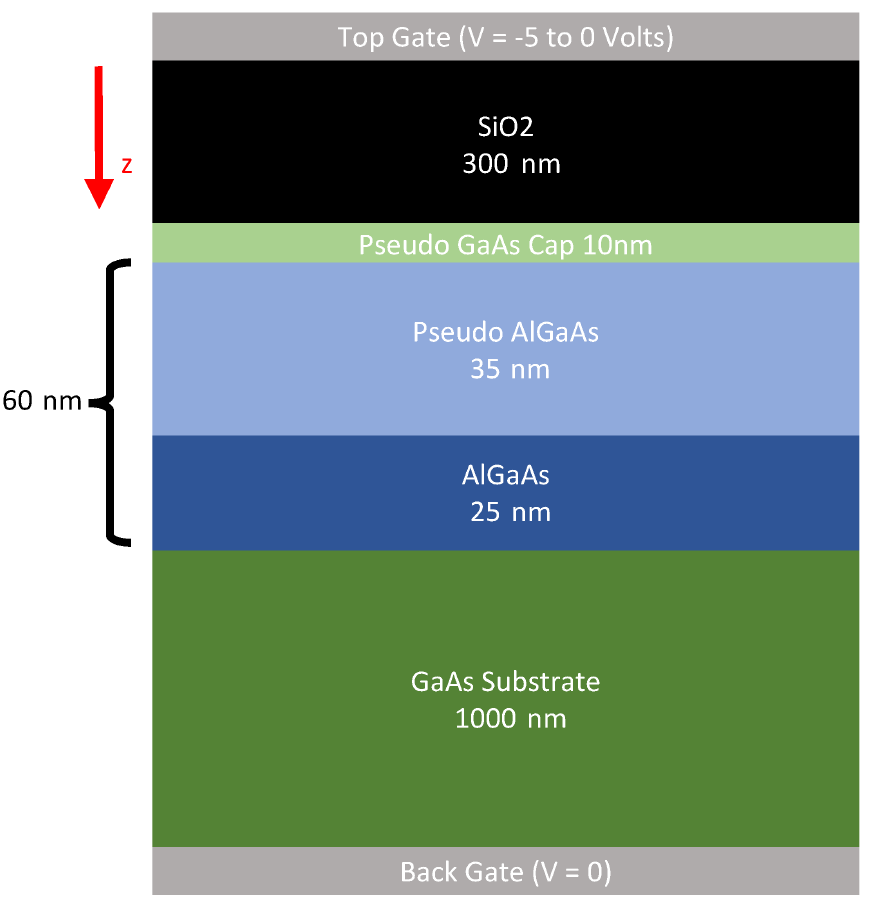}
    \caption{Schematic diagram of the device examined in this work. The quantum well with the two-dimensional hole gas is created at the AlGaAs/GaAs interface. }
    \label{device}
\end{figure}

Figure~\ref{device} shows the schematic cross-section of the device in which the two-dimensional quantum well is created. The coordinate $z$ is measured vertically downwards and the origin is placed at the top of the SiO$_2$ oxide layer. We focus on the interface between the GaAs substrate (the green region) and the Al$_{0.5}$Ga$_{0.5}$As layer (dark blue). For the purpose of modeling, the light blue regions denote the ``pseudo Al$_{0.5}$Ga$_{0.5}$As'' material whose material parameters are the same as Al$_{0.5}$Ga$_{0.5}$As, but the bandgap is artificially enhanced. Application of a negative voltage to the top-gate creates a confinement for the holes close to the GaAs/AlGaAs interface.

\begin{figure}[t]
    \includegraphics[width=0.8\columnwidth]{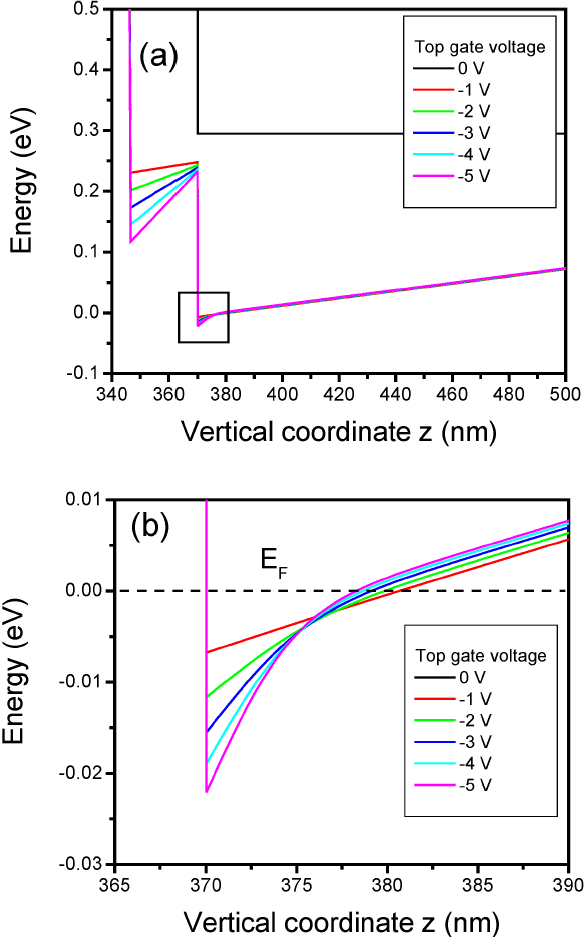}
    \caption{ (a) Band edge $E_v(z)$ of the heavy and light hole subbands as a function of the coordinate along the growth direction. The energies are calculated self-consistently for the top-gate voltage from $0$ V to $-5$ V. (b) The region marked with the black rectangle in (a), zoomed in to show the formation of the triangular quantum well confining the two-dimensional hole gas. The Fermi energy is fixed at $0$ eV.
    }
    \label{subband_edges}
\end{figure}

In Fig.~\ref{subband_edges}(a) we show the valence band edge as a function of the coordinate $z$ extracted from self-consistent calculations using the package \texttt{nextnano++}\texttrademark. At the extreme left of the graph the material is ``pseudo Al$_{0.5}$Ga$_{0.5}$As''; it becomes the actual Al$_{0.5}$Ga$_{0.5}$As at $z=345$ nm. The region of GaAs substrate begins at $z=370$ nm. When the top-gate voltage is set to zero (black line), the band edges are flat and the holes are not confined at the interface. As the top-gate voltage is made more negative, {\em two} triangular quantum wells are created at the two interfaces shown. We see that if the area of true Al$_{0.5}$Ga$_{0.5}$As were thicker, both quantum wells would be deep enough to be populated by  hole carriers. In experiments, we do not observe carriers in the GaAs cap layer and top AlGaAs barrier layer. There is no ohmic reservoir connected to them, and hole carriers therefore cannot populate these layers. To prevent these layers from being automatically populated in the \texttt{nextnano++}\texttrademark simulations, we have to introduce ``pseudo Al$_{0.5}$Ga$_{0.5}$As'' and ``pseudo GaAs'' materials with a larger bandgap than their nominal value, but keeping the same dielectric constant to preserve electric field information. With this approach, we only deal with one triangular quantum well which confined holes in the intended region of the device.

In Fig.~\ref{subband_edges}(b) we show the region of interest. The self-consistent calculations were carried out for all values of the top-gate voltage assuming that the Fermi energy is $E_F = 0$ eV. We find that as the gate voltage is made more negative, the triangular quantum well becomes deeper and is capable of confining the two-dimensional hole gas of increasing density.

\begin{figure}[t]
    \includegraphics[width=0.8\columnwidth]{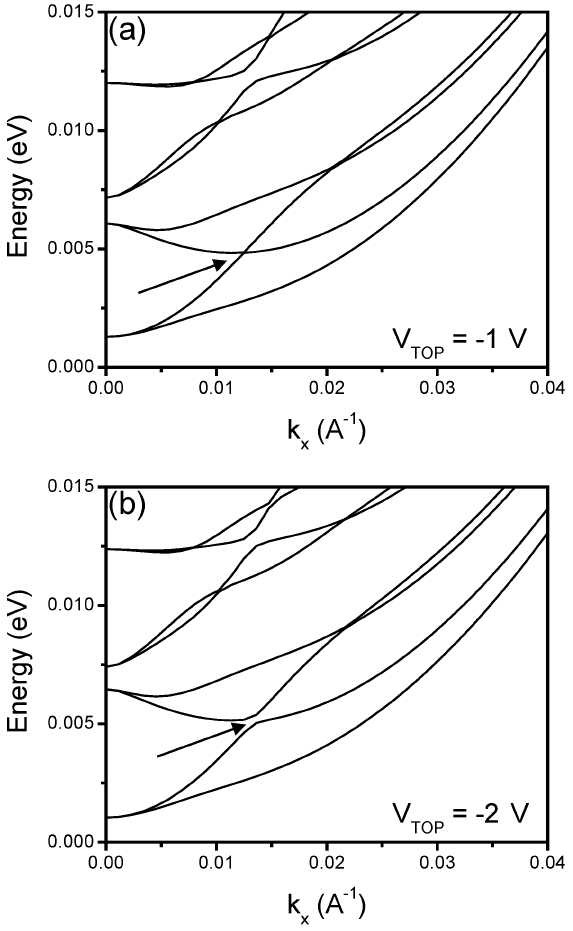}
    \caption{Band diagrams for the two-dimensional hole system confined in the heterostructure with top voltage $V_g=-1$ V (a) and $-2$ V (b). The Dresselhaus and Rashba terms are omitted. }
    \label{bstruct_V_no_so_1}
\end{figure}

\subsection{Computational procedure}

We account for the triangular confinement of holes in the growth direction by including the valence band edge profile $E_v(z)$ in the Hamiltonian $\hat{H}_B$ (Eq.~\ref{bulk_hamil}). From the self-consistent calculations we have the band edge profile for the heavy and light hole subbands as well as the profile for the spin-orbit split-off subband. The Hamiltonian $\hat{H}_B$ assumes that the latter is simply shifted from the former by the spin-orbit gap $\Delta$. In our calculations we will not make that assumption and instead utilize the calculated band edge profiles owing to the self-consistent nature of the \texttt{nextnano++}\texttrademark simulations. The required matrix elements $\langle l_1 |E_v(z) | l_2\rangle$ are computed numerically for a sufficiently large subband basis $l=1,2,\dots,N$. We choose $N=250$. Also, we set the confinement region to be $W=40$ nm wide and stretching from $z=370$ nm to $z=410$ nm along the growth direction in the device. By numerical studies (not shown), we have checked that these choices give a band diagram with well converged energies of several lowest subbands.

Furthermore, we account for the Rashba spin-orbit interaction $\hat{H}_{SIA}$ (Eq.~\ref{rashba}). To this end, we require the magnitude $E_z$ of the ``electric field'', that is, the degree of asymmetry. We will extract this parameter from the band edges shown in Fig.~\ref{subband_edges}(b) simply by taking the approximate slope of the triangular quantum well (see below for a detailed procedure). As this choice is approximate, a degree of fine-tuning of that value may be necessary to fit to any experimental data.

\begin{figure}[t]
    \includegraphics[width=0.8\columnwidth]{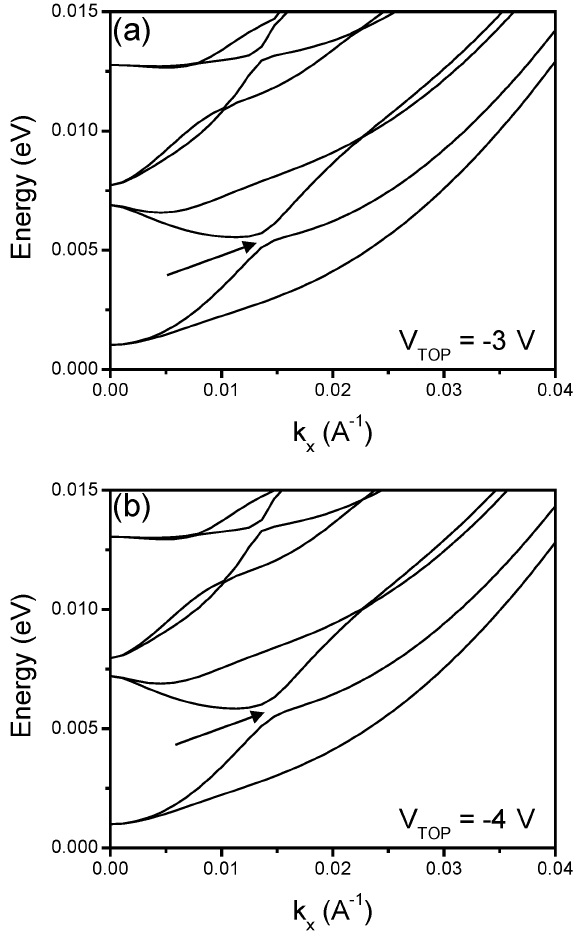}
    \caption{Band diagrams for the two-dimensional hole system confined in the heterostructure with top voltage $V_g=-3$ V (a) and $-4$ V (b). The Dresselhaus and Rashba terms are omitted. }
    \label{bstruct_V_no_so_2}
\end{figure}

\subsection{Band structure in the absence of BIA or SIA}

\begin{figure}[t]
    \includegraphics[width=0.9\columnwidth]{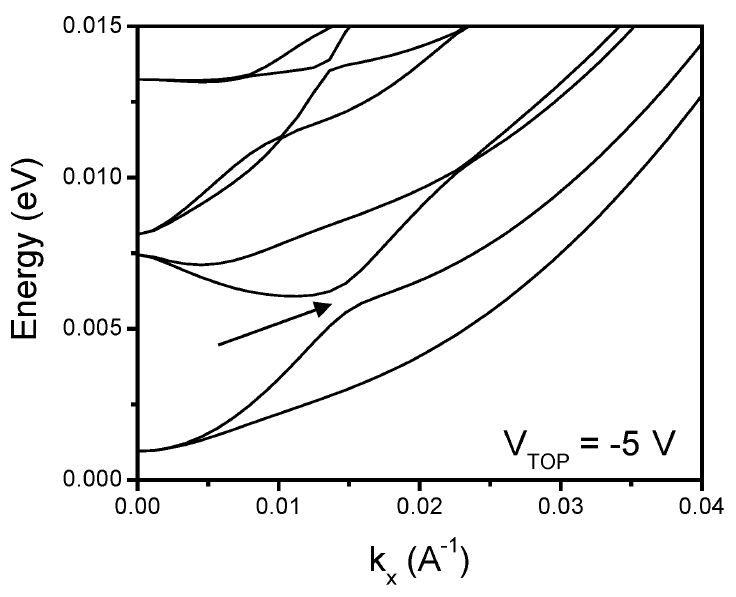}
    \caption{Band diagram for the two-dimensional hole system confined in the heterostructure with top voltage $V_g=-5$ V. The Dresselhaus and Rashba terms are omitted. }
    \label{bstruct_V_no_so_3}
\end{figure}

We begin by calculating the band structure as a function of the top-gate voltage assuming the absence of inversion symmetry, i.e., setting $\beta_0=0$, $\beta_1=0$, $\alpha_1 = \alpha_2=0$. Figures~\ref{bstruct_V_no_so_1}(a), \ref{bstruct_V_no_so_1}(b), \ref{bstruct_V_no_so_2}(a), \ref{bstruct_V_no_so_2}(b), and \ref{bstruct_V_no_so_3} show the results for the top-gate voltage $V_g = -1$ V, $-2$ V, $-3$ V, $-4$ V, and $-5$ V, respectively. Compared to the band structure calculated for the rectangular quantum well [Fig.~\ref{subbands_well_noD}], we find that in all five cases the band edges at $k_x>0$ are not degenerate. Further, in each case, the lowest band edge evolves as a function of $k_x$ in a quasi-parabolic manner. However, the second band edge depends on the wave number nonmonotonically, and exhibits an anticrossing with the third band edge. In each graph, the anticrossing is shown with the black arrow. We see that, from one top-gate voltage to another, the band diagrams are essentially similar with one clear difference: the magnitude of the anticrossing gap increases as the voltage is more negative. The lack of subband degeneracy is a direct result of the breaking of the inversion symmetry of the system, and occurs even without the inclusion of the BIA and SIA terms.

\subsection{Inclusion of BIA and SIA terms}

\begin{figure}[t]
    \includegraphics[width=0.8\columnwidth]{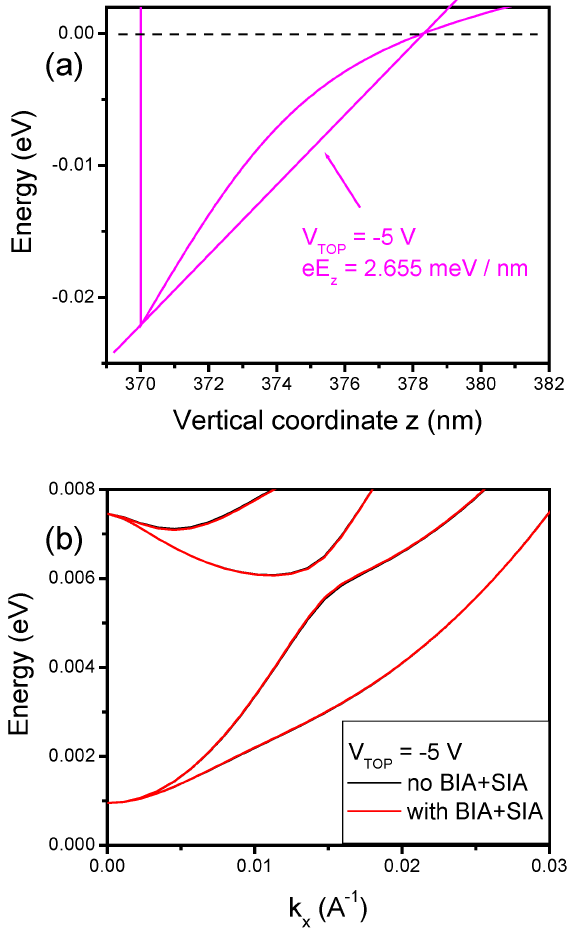}
    \caption{(a) The procedure of extracting the electric field $E_z$ needed for the Rashba SIA Hamiltonian (see text for explanation). (b) Band diagram for the two-dimensional hole system confined in the heterostructure with top voltage $V_g=-5$ V in the absence of either BIA and SIA terms (black lines) and with their full inclusion (red lines). }
    \label{bstruct_V_with_so}
\end{figure}

Let us now account for both Rashba and Dresselhaus Hamiltonians in our calculation. We take the respective parameters with their GaAs values, except for $\beta$ which we set to zero for reasons explained before. As already mentioned, the Rashba Hamiltonian requires the value $E_z$
of the electric field appropriate for our systems. We extract this parameter from the band edge profile in the manner summarized in Fig.~\ref{bstruct_V_with_so}(a). We choose two points $-$ one corresponds to the edge of the heterojunction and sets the point in the band edge profile
with the most negative value. The second point is chosen at the position when the band edge
energy is equal to zero. We choose this value for the second point, because we find from the
and diagrams in Figs.~\ref{bstruct_V_no_so_1}-~\ref{bstruct_V_no_so_3} that the energy corresponding to the lowest band edge is consistently close to zero, meaning that this is the energy at which the ladder of the hole states begins. A line is drawn across the two points and its slope is taken as the value of the parameter $E_z$ In the case of $V_g=-5$ V, shown in Fig.~\ref{bstruct_V_with_so}(a), we find $eE_z = 2.655$ meV$/$nm. We note that this is a ``reasonable approximation'' to what is clearly a nonlinear band edge profile. At some (short) distances within the well, the edge slope is actually larger, whilst towards the right-hand side the potential profile slope is more horizontal. Our choice seems to be an average, which is additionally easy to compute. We note that our procedure works much better for less negative $V_g$, since the resulting quantum well profiles become more linear, as is evident from
Fig.~\ref{subband_edges}(b).

Figure~\ref{bstruct_V_with_so}(b) shows the band structure for our system with $V_g=-5$ V without (black lines) and with (red lines) the BIA and SIA terms. We see that the correction introduced by the inversion asymmetry Hamiltonians is negligibly small. Since for the less negative top-gate voltages the parameter $E_z$ is even smaller, we expect virtually no new elements in the band diagrams introduced by the BIA and SIA Hamiltonians. However, in the remainder of this work we consistently account for these terms in our calculations.

We speculate that the corrections introduced by the SIA Hamiltonian are so small because the effective field $E_z$ is small compared to other characteristic energies and dimensions of the system. In fact, the influence of the top-gate voltage experienced at the GaAs/AlGaAs interface is rather small, as is evident from Fig.~\ref{subband_edges}(a). This can be due to the overall voltages applied to different elements of the device, but also to the large dielectric constants of the SiO$_2$ and AlGaAs layers lying between the top-gate and the interface. Indeed, the change in slope is much more dramatic in the quantum well appearing on the left-hand side of Fig.~\ref{subband_edges}(a), i.e., at the interface between the ``pseudo AlGaAs'' and AlGaAs layers. On a larger energy scale, the band edge sufficiently far below the interface is quite flat and hardly changes at all with the changing top voltage. It appears, therefore, that the removal of the inversion symmetry produces a large removal od subband degeneracy already on the level
of the bulk Hamiltonian $\hat{H}_B$.

\subsection{Discussion of band diagrams}

\begin{figure}[t]
    \includegraphics[width=0.8\columnwidth]{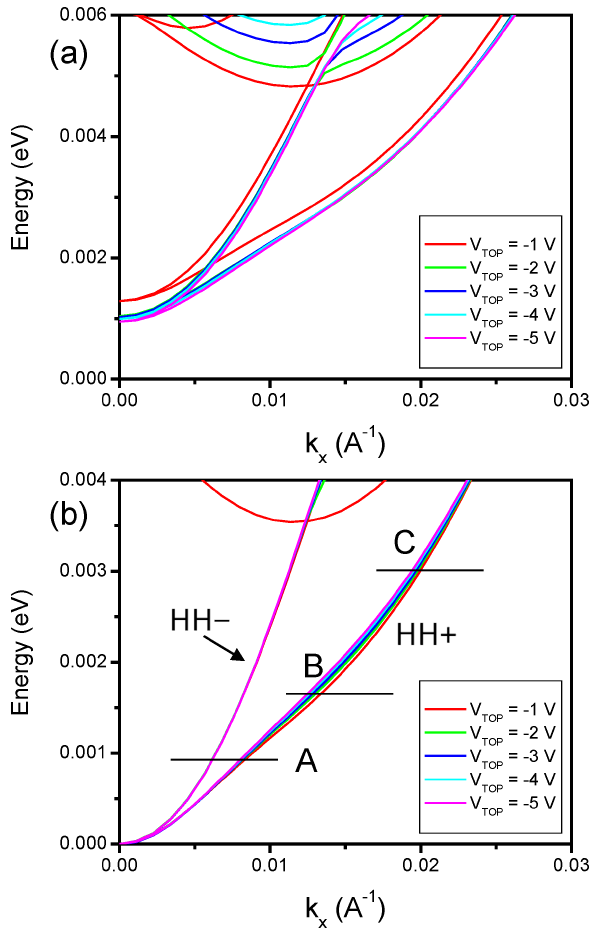}
    \caption{The lowest subband energies as a function of the top-gate voltage $V_g$. Color coding corresponds to that from Fig.~\ref{subband_edges}. Panel (a) shows the energies as they are obtained from the diagonalization of the full Hamiltonian (including SIA and BIA). Panel (b) shows the same energies shifted so that the lowest subband at the Gamma point has zero energy. Regions A, B, and C correspond to a different predicted dependence of the effective mass on the top-gate voltage. }
    \label{masses}
\end{figure}

In Fig.~\ref{masses} we collect the band diagrams calculated using the full Hamiltonian for all five top-gate voltages. The color coding corresponds to that used in Fig.~\ref{subband_edges}.
The panel (a) shows the energies as they emerge from the Hamiltonian diagonalization without any additional shifts. We see that the two lowest subbands shift only slightly as $V_g$
is made more negative, and exhibit a converging tendency. More substantial shifts from one voltage to another are seen in the higher-lying subbands (top of the panel): as the voltage
is made more negative, these energies markedly shift upwards.

In Fig.~\ref{masses}(b) we present the same band edges, but now shifted in such a way that for each $V_g$ the lowest subband energy is zero at $k_x=0$. This graph shows clearly the changes of the slope and curvature of the two lowest subbands for the purpose of the qualitative analysis of the effective masses $m^*$. We will discuss qualitatively the dependence of $m^*$ on the top-gate voltage assuming that the masses are extracted from the subband curvature in the vicinity of the Fermi energy (by, e.g., taking the second derivative locally). Of the two lowest subbands, the higher-energy one (the ``lighter heavy hole'', HH$-$) appears to have a nearly identical dependence on $k_x$ irrespective of $V_g$. Since the higher-energy subband is steeper (has a greater curvature) as a function of $k_x$, it will be characterized by a smaller effective mass
$m^*_{HH-}$. We find that $m^*_{HH-}$ is essentially voltage-independent.

The lowest, ``heavier heavy hole'' (HH$+$) subband, on the other hand, is generally much less steep in its dependence on $k_x$, and therefore the corresponding effective mass $m^*_{HH+}$ However, we find that the HH$+$ band edge curvature depends somewhat on the top-gate voltage. In discussing the resulting effective masses, we distinguish three regions, marked as A, B, and C in Fig.~\ref{masses}(b). In the region A, the band edges very nearly coincide. In the regions B and C, the band edges are different for different $V_g$, with the edge at the least negative voltage being at the lowest energy.

Suppose now that the hole density is such that the Fermi energy is somewhere in the region A.
If the density does not increase too much when we make $V_g$ more negative, then the extracted $m^*_{HH+}$ is expected to be very nearly voltage-independent, since the band edges coincide in that region. However, if the Fermi energy increases such that there is a transition from the region A into the region B across the inflection point, then $m^*_{HH+}$ is expected to increase with $V_g$ becoming more negative, as the band edges become flatter with that transition.

If the Fermi energy happens to fall in the region B for all $V_g$, and change only slightly when $V_g$ is made more negative, then $m^*_{HH+}$ would  decrease, since the band edges are becoming more steep.

Finally, if the Fermi energy happens to fall in the region C, and remain there for all values of $V_g$, then it would appear that $m^*_{HH+}$ should decrease somewhat as $V_g$ is made more negative, because the band edges become less steep (they shift closer together).

\subsection{Quantitative extraction of the effective mass}

Let us now extract the effective masses characterizing the subbands HH$+$ and HH$-$ quantitatively. In Fig.~\ref{masses}(b) we have shown that the dispersion calculated for different top-gate potentials $V_g$ are very similar, and therefore we will present our results for just one value of $V_g = -5$ V. We will now choose several values of the Fermi energy $E_F$ of the hole gas in order to discuss the dependence of the effective masses of the HH$+$ and HH$-$ subbands as a function of the hole density. As we are in a two-dimensional system, the Fermi energy is proportional to the hole density.

\begin{figure}[t]
    \includegraphics[width=1.0\columnwidth]{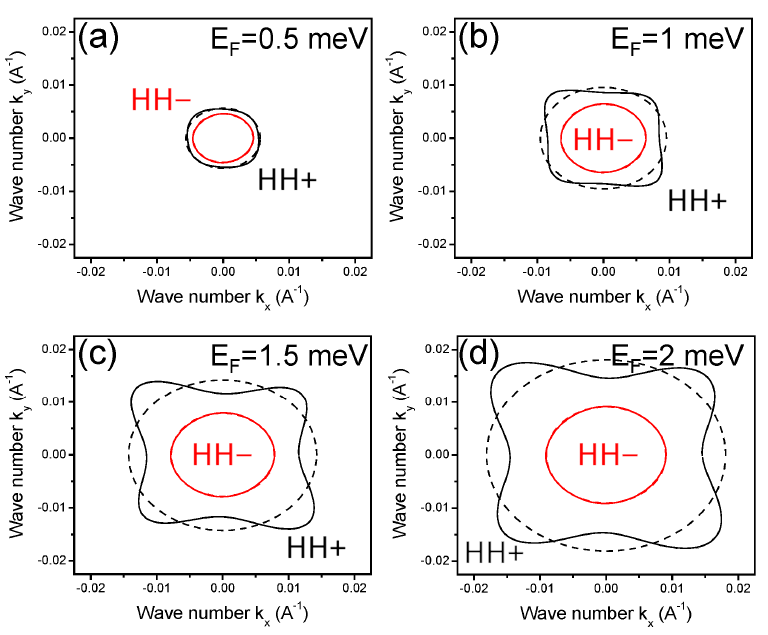}
    \caption{Cross-sections of the dispersion diagram from Fig.~\ref{masses}(a) showing iso-energy lines at the Fermi energy $E_F=0.5$ meV (a), $1$ meV (b), $1.5$ meV (c), and $2$ meV (d) above the band edge for the dispersion calculated at the top-gate potential $V_g=-5$ V. Black (red) color corresponds to the HH$+$ (HH$-$) subband, respectively. Solid lines show the actual subband profile, while dashed lines show the average iso-energy line obtained in the parabolic approximation (see text for details). }
    \label{fermi_surfaces}
\end{figure}

In Fig.~\ref{fermi_surfaces}, we present the cross-sections of the dispersion obtained at $E_F=0.5$ meV (a), $1$ meV (b), $1.5$ meV (c), and $2$ meV (d) counted from the band edge at the $\Gamma$ point. The solid black lines show the iso-energy contours for the HH$+$ subband, whilst the solid red lines show the contours for the HH$-$ subband. We see, in general, that the HH$+$ subband shows a substantial anisotropy (a non-circular shape of the contour), visible particularly clearly for larger values of $E_F$. On the other hand, the HH$-$ subband appears to be isotropic, with nearly circular iso-energy lines. Owing to the anisotropy, strictly speaking the HH$+$ subband cannot be characterized by just a single value of the effective mass. Indeed, depending on the direction of transport in $k$-space, the values of the effective mass can be very different. However, in our experiment we assume that the transport is not completely ballistic, which is clear from the broadening of the Landau levels and not excessively long  alues of scattering times (described in the main text). Therefore, we will now set up an averaging procedure which will enable us to produce a single number for the effective mass of both HH$+$ and HH$-$ subbands.

Our point of departure is the fact that, in the two-dimensional geometry, the area $S$ encompassed by the iso-energy lines from Fig.~\ref{fermi_surfaces} is proportional to the hole density. We wish to approximate the complex cross-sections by simple, circular ones, but encompassing the same area (i.e., corresponding to the same hole density). To this end, we postulate the value of the Fermi momentum $k_F$ such that $S=\pi k_F^2$. For each $E_F$, we find the value of $S$ by extracting the area within each iso-energy line numerically, and simply compute the corresponding $k_F$ in the parabolic approximation. The resulting isotropic iso-energy lines for both HH$+$ and HH$-$ subbands are shown in Fig.~\ref{fermi_surfaces} with dashed lines. We find that for the HH$-$ subband there is nearly no difference between the actual and approximated cross-section, while the differences for HH$+$ are appreciable, particularly at larger values of $E_F$.

\begin{figure}[t]
    \includegraphics[width=0.95\columnwidth]{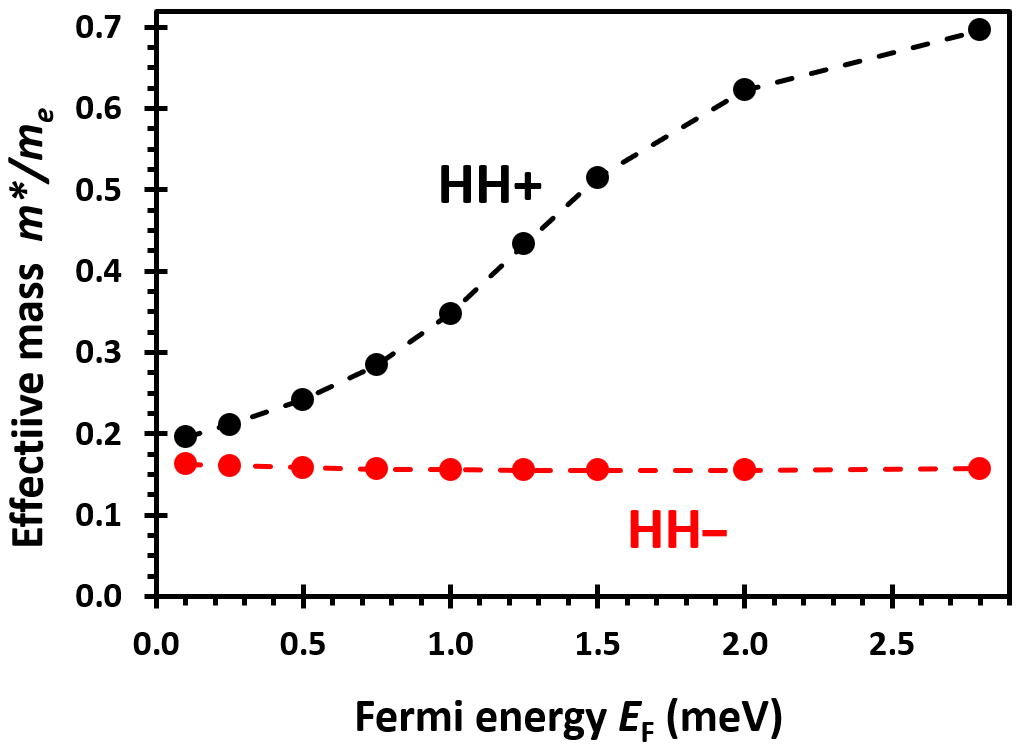}
    \caption{Effective masses for the HH$+$ (black) and HH$-$ (red) subbands extracted with
    the averaging procedure described in the text.}
    \label{masses_quantitative}
\end{figure}

Now, assuming parabolic dispersion, we now utilize the relationship between the Fermi energy
and the Fermi momentum, $E_F = {\hbar ^2 k_F^2} / 2 m^*$, where $\hbar$ is the Dirac's constant,
and $m^*$ is the sought effective mass. Since both $E_F$ and $k_F$ are known, we can simply derive the average effective mass by inverting this relationship. In Fig.~\ref{masses_quantitative} we show the effective masses extracted using this procedure
as a function of the Fermi energy. The HH$-$ effective mass (shown in red) appears to be constant, which reflects the isotropy and parabolicity of this subband seen in Fig.~\ref{masses}. On the other hand, the HH$+$ effective mass (shown in black) increases nearly linearly as the Fermi energy increases. This is consistent with the sub-parabolic shape of the dispersion for that subband seen in Fig.~\ref{masses}.

\section{Simplified analysis of the dispersion of hole subbands - discussion of inversion symmetry}
\label{secA:dispersion}

In this Section we will demonstrate, in the lowest-order $\vec{k}\cdot\vec{p}$ treatment, that the bulk Hamiltonian (\ref{bulk_hamil}) is sensitive to the inversion symmetry of the quantum well even without the spin-orbit interaction terms. In particular, we will demonstrate that if the quantum well is rectangular, i.e., if the system has inversion symmetry, then  the energies for any $\vec{k}$ are doubly degenerate. On the other hand, if the quantum well confinement has a triangular component, the subband degeneracy is seen only at $\vec{k}=\vec{0}$, i.e., at the $\Gamma$ point, whilst at any finite $\vec{k}$ the subbands are not degenerate. Thus, the Hamiltonian (\ref{bulk_hamil}) naturally includes the effect called ``the Rashba spin-orbit interaction'', and we do not need to add any further terms $\hat{H}_{SIA}$ of the form given in Eq. (\ref{rashba}).

In fact, this is the view expressed by Winkler in his theory-only papers. Indeed, in Ref.~\onlinecite{winkler2000rashba}, the Author states explicitly that the Rashba Hamiltonians are ``implicitly contained in the full multi-subband Hamiltonian''. Moreover, as detailed in Ref.~\onlinecite{winkler2003spin}, the characteristic Rashba coefficients $\alpha_1$ and $\alpha_2$ scaling the components of $\hat{H}_{SIA}$ in Eq. (\ref{rashba}) are obtained from
the full bulk Hamiltonian, Eq. (\ref{bulk_hamil}), in a perturbative procedure. As a result, it seems that it is not necessary to add $\hat{H}_{SIA}$ to the full bulk Hamiltonian provided that we solve the bulk multi-subband problem non-perturbatively, as extensively discussed earlier in this Report. The need for the extra $\hat{H}_{SIA}$ in Refs.~\onlinecite{liu2018strong,Marcellina17} appears to be the consequence of the perturbative treatment (including only the lowest subband) of the bulk Hamiltonian. In the following sections it will become apparent why this is so.

The plan of this discussion is as follows. Utilizing the bulk Hamiltonian (\ref{bulk_hamil}) only, we will perform a perturbative analysis (to within the lowest subband) of the quantum well in two cases: (i) inversion symmetry, i.e., a rectangular quantum well, and (ii) lack of inversion symmetry, i.e., a triangular quantum well. We begin with the considerations at the $\Gamma$ point, i.e., $\vec{k}=\vec{0}$, which switches off any subband mixing. In this case we will calculate the lowest-subband wave functions and discuss their symmetry. Then we will extend the discussion to finite $\vec{k}$ and demonstrate that the lack of inversion symmetry activates extra subband mixing, absent in the inversion-symmetric case, and that this mixing is responsible for the removal of the subband degeneracy, i.e., the Rashba effect.

\subsection{The four-subband Hamiltonian}

To begin with, we will not use the full six-subband bulk Hamiltonian as defined in Eq. (\ref{bulk_hamil}), but we will restrict ourselves to four subbands only. In other words, we will take the HH and the LH subbands, and we will neglect the SO-split-off subbands altogether. As a result, in the basis $\{ |3/2,+3/2\rangle, |3/2,-1/2\rangle, |3/2,+1/2\rangle, |3/2,-3/2\rangle \}$, our Hamiltonian will take the form
\begin{equation}
    \hat{H}_{4} = \left[
    \begin{array}{cccc}
        U + V & R & S_- & 0 \\
        R^+ & U - V & 0 & -S_+^+  \\
        S_-^+ & 0 & U-V & R \\
        0 & -S_+ & R^+ & U+V \\
    \end{array}
    \right].
    \label{bulk_hamil_4subband}
\end{equation}
All operators take the same form as those in the first section of this report. In particular, the quantum well confinement $E_v(z)$ is included in the operator $U$. This is the starting point for our analysis. Note that, for reasons that will become apparent soon enough, we have changed the order of the second and third basis state (the light hole subbands) relative to the order employed in Eq. (\ref{bulk_hamil}).

\subsection{Lowest hole subbands at point $\Gamma$}

We now set $k_x = k_y = 0$. This sets all off-diagonal operators $R$ and $S$ to zero, resulting in unmixed subbands. The Hamiltonian (\ref{bulk_hamil_4subband}) retains only the diagonal operators, which take the following form:
\begin{eqnarray}
    U(\Gamma)+V(\Gamma) &=& -\frac{\hbar^2}{2m_e} (\gamma_1 - 2\gamma_2) \frac{\partial^2}{\partial z^2} + E_v(z),\quad\\
    U(\Gamma)-V(\Gamma) &=& -\frac{\hbar^2}{2m_e} (\gamma_1 + 2\gamma_2) \frac{\partial^2}{\partial z^2} + E_v(z).
\end{eqnarray}
We have explicitely accounted for the fact that the system is confined along the $z$ direction,
and therefore the $k_z$ operator present in Eq. (\ref{bulk_hamil}) has to be replaced by the derivative $-i\partial/\partial z$. The first block, i.e., $U+V$, allows us to construct the heavy hole subband edge, while the second one, i.e., $U-V$ will give us the light hole subband edge.
Indeed, the ``inverse effective mass'' $\gamma_1 - 2\gamma_2$ is smaller than $\gamma_1 + 2\gamma_2$, meaning that the actual effective mass for the $U+V$ block is large, hence the heavy hole, and the effective mass for the $U-V$ block is small, hence the light hole.
With the GaAs parameters we have $m_{eff}^{HH}=0.377$ $m_e$ and $m_{eff}^{LH}=0.090$ $m_e$.

Further, let us specify the form of the confinement $E_v(z)$. To make the considerations as simple as possible, let us choose the infinite rectangular quantum well for the inversion-symmetric case:
\begin{equation}
    E_z^{SYM}(z) = \left\{
    \begin{array}{cc}
        0, & \text{for}~~0 < z < W \\
        \infty, & \text{otherwise}
    \end{array}
    \right.
\end{equation}
with $W$ being the quantum well width. For the non-inversion-symmetric case, let us choose an infinite quantum well, but with a triangular bottom:
\begin{equation}
    E_z^{ASYM}(z) = \left\{
    \begin{array}{cc}
        \frac{V_0}{W} z, & \text{for}~~0 < z < W \\
        \infty, & \text{otherwise}
    \end{array}
    \right.
\end{equation}
with $V_0$ setting the linear rise of the quantum well (related, of course, to the external electric field). In what follows, we take the following model parameters: $W=5$ nm and $V_0=100$ meV.

\begin{figure}[t]
    \includegraphics[width=0.8\columnwidth]{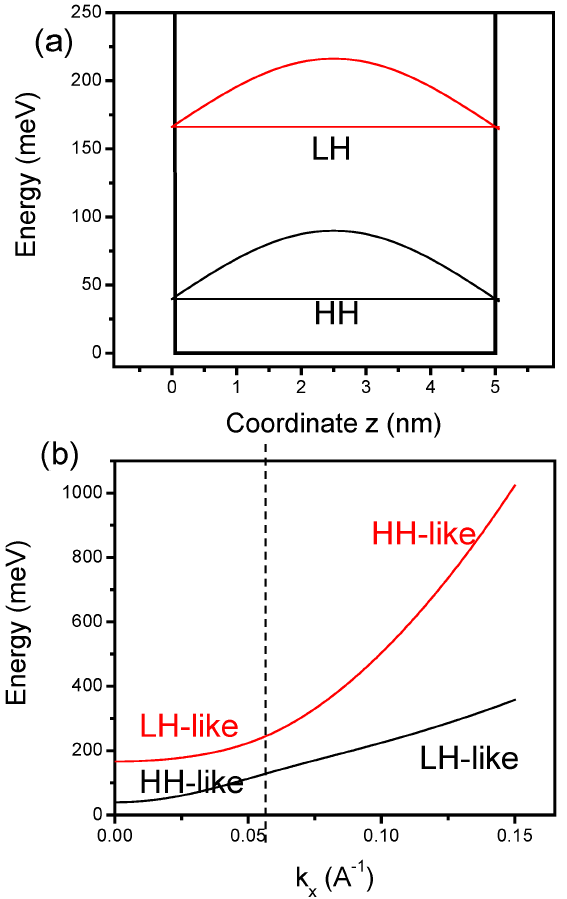}
    \caption{Simplified analysis of two lowest subbands for the rectangular quantum well
    (with inversion symmetry). (a) The shape of the potential with the lowest-energy HH (black) and LH (red) subbands. The horizontal lines denote the energy, while the sinusoids denote the wave functions. (b) Dispersion calculated by taking the two lowest subbands of HH and LH as a function of the wave number $k_x$, with $k_y=0$ throughout. }
    \label{perturb_square}
\end{figure}

The lowest-state energies for each of the subbands are trivially obtained for the rectangular quantum well. Indeed, we have the analytical expressions
\begin{eqnarray}
    E_{HH}^{SYM} = \frac{\hbar^2}{2m_{eff}^{HH}} \frac{\pi^2}{W^2} = 39.859 ~\text{meV},\\
    E_{LH}^{SYM} = \frac{\hbar^2}{2m_{eff}^{LH}} \frac{\pi^2}{W^2} = 166.205 ~\text{meV},
\end{eqnarray}
with our model numerical values. As usual for the infinite rectangular quantum well, the HH and LH lowest-energy states are identical and are expressed as
\begin{equation}
    \psi_{HH}^{SYM}(z) = \psi_{LH}^{SYM}(z) = \sqrt{\frac{2}{W}} \sin \left( \frac{\pi}{W} z  \right)
\end{equation}
for $0 < z < W$, and zero outside this region. We note that in a {\em finite} quantum well these wave functions would not be the same. Indeed, both of them would penetrate into the barriers, but the light hole wave function would penetrate further, as the energy of the LH state is higher.
This, however, introduces only a quantitative correction. In our qualitative discussion we note that the wave functions are {\em symmetric} (even) with respect to the middle of the quantum well, i.e., they obey the inversion symmetry of the Hamiltonian. This is evident in Fig.~\ref{perturb_square}(a) where we show the two wave functions for the HH (black) and the LH (red) subbands, respectively, positioned on top of the corresponding energy
values (horizontal lines).

\begin{figure}[t]
        \includegraphics[width=0.8\columnwidth]{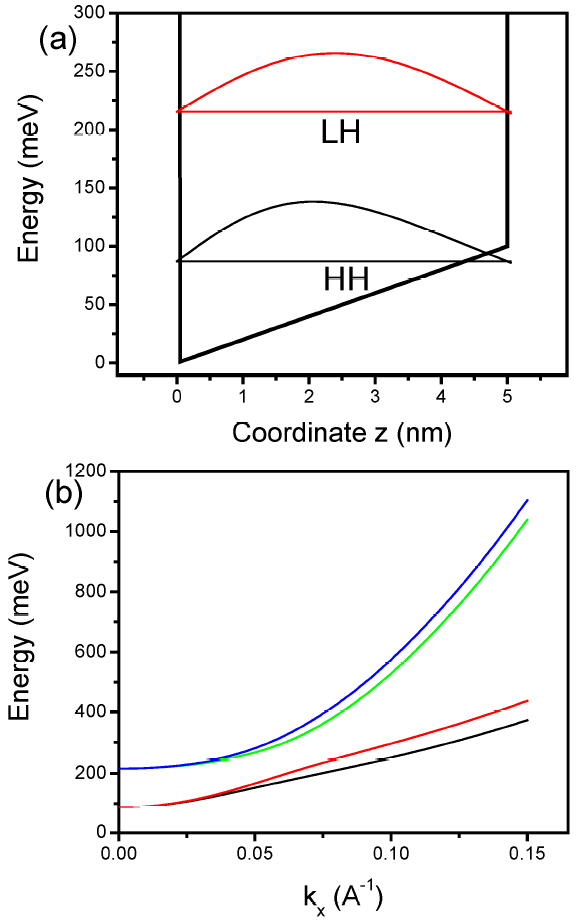}
        \caption{ Simplified analysis of two lowest subbands for the triangular quantum well
        (without inversion symmetry). (a) The shape of the potential with the lowest-energy HH (black) and LH (red) subbands. The horizontal lines denote the energy, while the curves show the wave functions. (b) Dispersion calculated by taking the lowest subbands of HH and LH as a function of the wave number $k_x$, with $k_y=0$ throughout.
        }
        \label{perturb_triang}
\end{figure}

For the case without the inversion symmetry, in our chosen potential, the expressions for energy and wave functions are not analytical. It is true that the triangular quantum well has semi-analytical solutions in terms of the Airy's function, but this presupposes that the linear rise, in our case $\frac{V_0}{W}z$, extends to arbitrary positive coordinates $z$. Our choice is to restrict this rise only up to $z=W$, and set the potential to infinity afterwards. This enables us to make meaningful comparisons with the rectangular quantum well of the same width,
but prevents us from using the Airy's treatment. However, the one-dimensional problem is of course trivial to solve numerically. With our parameters we obtain
\begin{eqnarray}
    E_{HH}^{ASYM} &=& 87.175~\text{meV},\\
    E_{LH}^{ASYM} &=& 215.554~\text{meV}.
\end{eqnarray}
In Fig.~\ref{perturb_triang}(a) we plot these energies as horizontal solid lines, black for the
HH subband and red for the LH subband, within the triangular quantum well considered in this calculation (thick black line). We see that the HH subband energy lies still within the ``rise'' of the triangular quantum well, while the LH subband energy is already well in the region where the walls of the well are vertical. This enables us to suspect that the wave functions corresponding to these subband edges will not be identical, unlike in the case of the rectangular quantum well.
Indeed, the black and red curves show the wave functions corresponding to the HH and LH subband edge, respectively. These wave functions are calculated numerically together with the eigenenergies. We see that the HH wave function $\psi_{HH}^{ASYM}(z)$ does not peak in the middle of the quantum well. Its maximum falls somewhat closer to the left-hand side of the well, because the well is deeper there. On the other hand, the LH wave function $\psi_{LH}^{ASYM}(z)$ is somewhat more symmetric, and peaks closer to the middle point of the quantum well. A closer inspection, however, reveals that even this LH wave function peaks somewhat left of the middle,
reflecting the change in depth of the triangular well. We see that in both cases the wave functions do not have a definite parity with respect to the middle of the quantum well, unlike in the case of the rectangular quantum well.

\subsection{Calculation of dispersion}
\label{app:explanation}

Let us now move on to the discussion of subband mixing in the two cases. To this end, we will work in the basis of not just the Bloch functions, but also the envelope functions for the subbands calculated in the previous Section. That is, our basis will be now
$\{ \psi_{HH}|3/2,+3/2\rangle, \psi_{LH}|3/2,-1/2\rangle,~
\psi_{LH}|3/2,+1/2\rangle,$ $\psi_{HH}|3/2,-3/2\rangle \}$,
with the envelope functions $\psi_{HH}$, $\psi_{LH}$. As a result, we need to compute the matrix elements of the operators $R$, $S_+$, and $S_-$ appearing in the Hamiltonian (\ref{bulk_hamil_4subband}) against the envelope functions. We will have now
\begin{equation}
    R(k_x, k_y) = \sqrt{3} \frac{\hbar^2}{2m_e} (\mu k_+^2 - \bar{\gamma}k_-^2) \langle \psi_{HH}|\psi_{LH} \rangle,
\end{equation}
that is, the operator $R$ will now become a number, whose value depends on the material parameters, the values $k_x$ and $k_y$, and the overlap between the HH and LH wave function in the growth direction. This is the case because the operator $R$ depends only on the in-plane momentum, and does not depend on the momentum $k_z$ in the growth direction. Furthermore,
\begin{equation}
    S_{\pm}(k_x, k_y) = -2\sqrt{3} \frac{\hbar^2}{2m_e} \gamma_3 k_{\pm}
     \langle \psi_{HH}|k_z|\psi_{LH} \rangle.
\end{equation}
We see that the elements $S_{\pm}$ now depend on the material parameters, the in-plane momenta,
but also on a matrix element of the gradient $k_z = -i \partial/\partial z$ calculated against the HH and LH subband wave functions. Evidently, the matrix elements $R$ and $S$ will crucially depend on the symmetries of the subband wave functions, and their correct calculation is the key to understanding the resulting dispersion.

\begin{figure*}[t]
    \includegraphics[width=1.9\columnwidth]{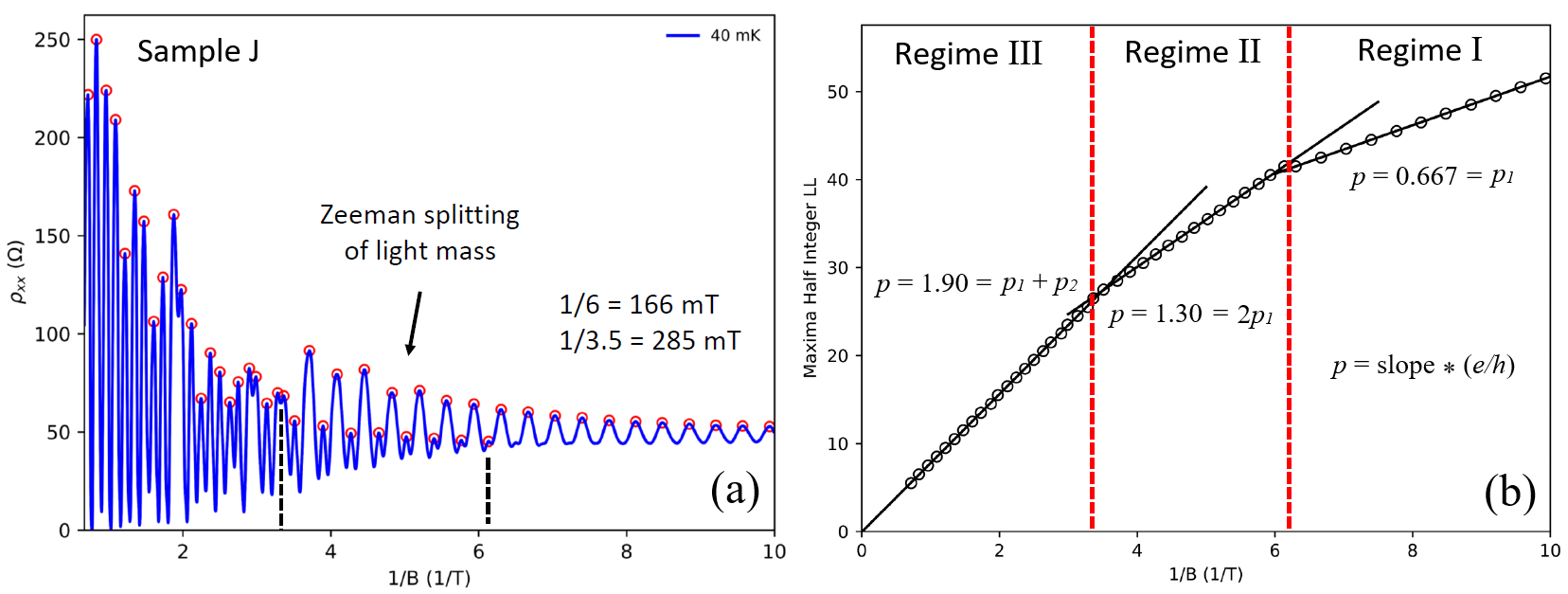}
    \caption{(a) SdH oscillations in sample J plotted versus $1/B$, at $p_{2d} = 1.90 \times 10^{15}$/m$^2$. (b) Index of SdH maxima (red circles) plotted against $1/B$, from data in panel (a). }
    \label{Fig:ZeemanSplit}
\end{figure*}

We begin with the rectangular quantum well, i.e., the case with inversion symmetry. For our infinite well, the HH and LH wave functions in the growth direction are identical and even (symmetric with respect to the middle of the well). Therefore, the overlap needed for the element $R$ is $\langle \psi_{HH}^{SYM}|\psi_{LH}^{SYM} \rangle=1$. On the other hand, the integral needed for the element $S$ is $-i \langle \psi_{HH}^{SYM}|\partial/\partial z|\psi_{LH}^{SYM} \rangle = 0$. Indeed, the gradient of an even function becomes odd, and the integral of an odd function vanishes exactly. This result survives the generalization to the rectangular well with a finite depth. In such case the subband wave functions $\psi_{HH}^{SYM}$ and $\psi_{LH}^{SYM}$ are not identical, but are both of even parity, and therefore the matrix element $S$ must vanish exactly. With this understanding, let us now write explicitely the four-subband Hamiltonian for the case with inversion symmetry for an arbitrary point in the reciprocal space:
\begin{widetext}
\begin{equation}
    \hat{H}_{4}^{SYM} = \left[
    \begin{array}{cccc}
        E_{HH}^{SYM} + E_{HH,||}(k_x,k_y) &
          R(k_x,k_y)   & 0 & 0 \\
        R(k_x,k_y) & E_{LH}^{SYM} + E_{LH,||}(k_x,k_y) & 0 & 0  \\
        0 & 0 & E_{LH}^{SYM} + E_{LH,||}(k_x,k_y) & R(k_x,k_y) \\
        0 & 0 & R(k_x,k_y) & E_{HH}^{SYM} + E_{HH,||}(k_x,k_y) \\
    \end{array}
    \right].
    \label{bulk_hamil_4subband_sym}
\end{equation}
\end{widetext}
Here, the in-plane diagonal dispersion components are
\begin{eqnarray}
    E_{HH,||}(k_x,k_y) &=& \frac{\hbar^2}{2m_e}(\gamma_1+\gamma_2)(k_x^2+k_y^2),\\
    E_{LH,||}(k_x,k_y) &=& \frac{\hbar^2}{2m_e}(\gamma_1-\gamma_2)(k_x^2+k_y^2).
\end{eqnarray}
We see that the even symmetry of both subband energies has lead to a {\em block diagonal} form
of the Hamiltonian, with the two blocks being identical. This leads inevitably to the {\em double degeneracy} of the dispersion relations for any in-plane momenta $(k_x,k_y)$. Within each block, the resulting states will be superpositions of the HH and LH functions, because they are mixed by the element $R$. Of course, $R$ depends quadratically on the in-plane momentum, and therefore the further we are from the $\Gamma$ point, the stronger the intersubband mixing. This is why we can only discuss the resulting states as ``HH-like'' or ``LH-like'', meaning that they are both superpositions of HH and LH components, with HH or LH dominating, respectively. Figure~\ref{perturb_square}(b) shows the dispersion curves obtained by diagonalizing the Hamiltonian (\ref{bulk_hamil_4subband_sym}) as a function of $k_x$, with $k_y=0$ throughout. The two dispersion curves are exactly doubly-degenerate. At low values of $k_x$, the lower-energy edge (black) is HH-like, while the higher-energy edge (red) is LH-like. We see that as $k_x$ increases, the two curves are shifting closer in energy. This is expected: the heavy holes are actually light in-plane, while the light holes are heavy in-plane. Therefore, at some value of $k_x$ they are expected to cross. However, owing to the finite element $R$, describing the subband repulsion, they actually anticross, which is evident for $k_x$ lightly larger than $0.05$ $A^{-1}$, marked with the vertical dashed line. Close to the anticrossing, it is not possible to ascertain the character of the subband edges, as the HH and LH subbands are strongly (and nearly equally) mixed. For even larger values of $k_x$ the character of the two subbands flips, and now
the low-energy edge (black)  is in fact LH-like, while the high-energy edge (red) is HH like.
We stress, however, that in spite of this complexity of the spectral content, the two subband edges are always doubly degenerate.

Let us now move on to the case of the triangular confinement, i.e., without the inversion symmetry. We need two elements: the overlap
$\langle \psi_{HH}^{ASYM}|\psi_{LH}^{ASYM} \rangle$ and the gradient
$-i \langle \psi_{HH}^{ASYM}|\partial/\partial z|\psi_{LH}^{ASYM} \rangle$.
Let us start with the overlap. As is evident from Fig.~\ref{perturb_triang}(a), the subband edge wave functions $\psi_{HH}^{ASYM}$ and $\psi_{LH}^{ASYM}$ are not identical, and we expect their overlap to be somewhat less than one. This is indeed the case - from our numerical calculations we obtain $\langle \psi_{HH}^{ASYM}|\psi_{LH}^{ASYM} \rangle = 0.993738$. So, again, the element $R(k_x,k_y)$ will not be zero, as was the case for the rectangular well.

The key finding, however, is that the gradient element $-i \langle \psi_{HH}^{ASYM}|\partial/\partial z|\psi_{LH}^{ASYM} \rangle = -i 6.013355\cdot 10^{-2} $
nm$^{-1}$, that is, it is {\em not zero}. This introduces the elements $S_+$ and $S_-$ into the Hamiltonian, which takes now the form
\begin{widetext}
\begin{equation}
    \hat{H}_{4}^{ASYM} = \left[
    \begin{array}{cccc}
        E_{HH}^{ASYM} + E_{HH,||} &
          R   & S_- & 0 \\
        R & E_{LH}^{ASYM} + E_{LH,||} & 0 & -S_+^+  \\
        S_-^+ & 0 & E_{LH}^{ASYM} + E_{LH,||} & R \\
        0 & -S_+ & R & E_{HH}^{ASYM} + E_{HH,||} \\
    \end{array}
    \right].
    \label{bulk_hamil_4subband_asym}
\end{equation}
\end{widetext}

\begin{figure}[t]
    \includegraphics[width=1.0\columnwidth]{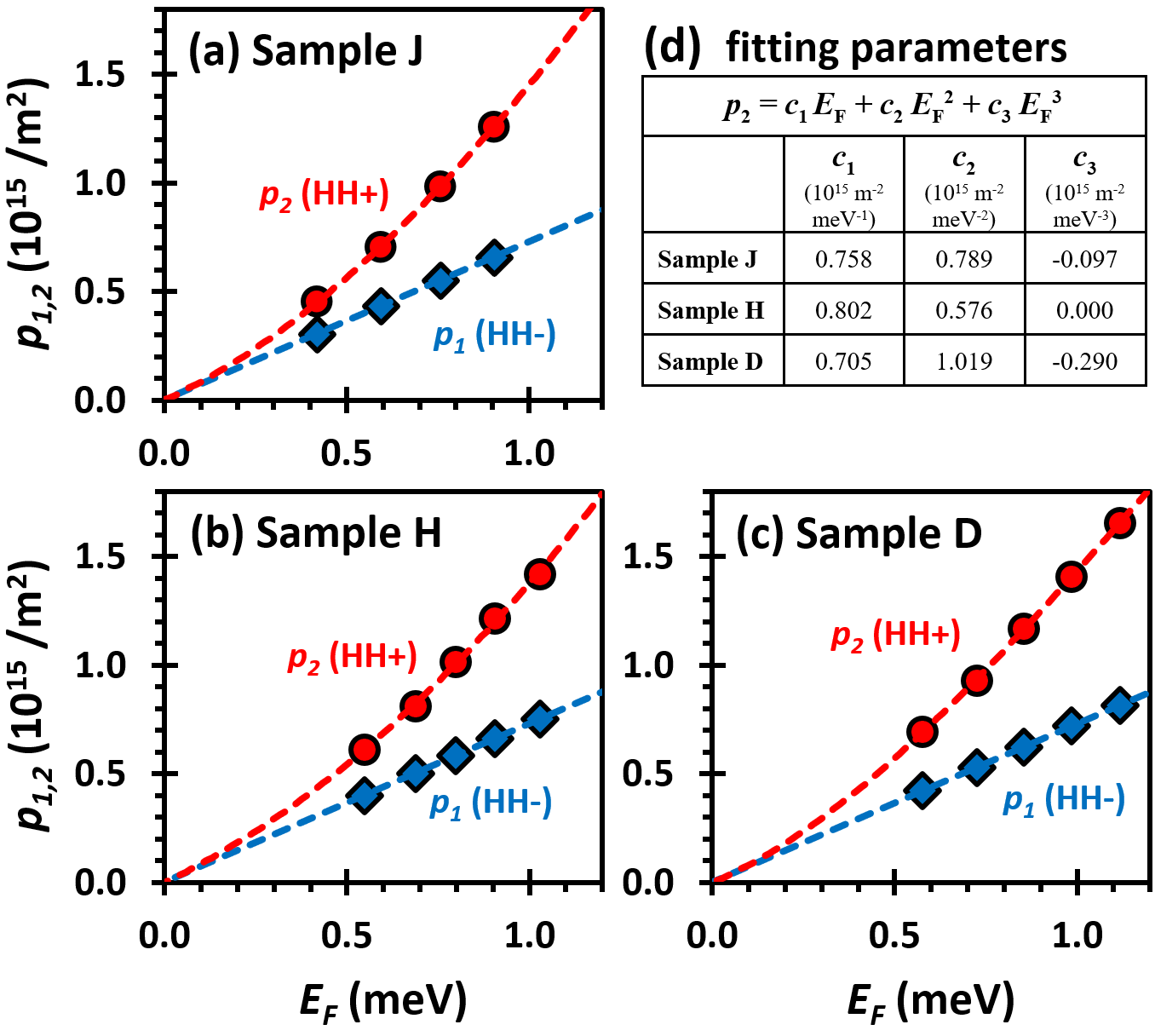}
    \caption{Carrier density versus Fermi energy for HH$-$ subband (blue diamonds) and HH$+$ subband (red circles) in: (a) sample J, (b) sample H, and (c) sample D. Red dashed lines (HH$+$) are polynomial best fits with parameters listed in panel (d). Blue dashed lines (HH$-$) show $p_1 = m_1 E_{\textsc{f}}/2\pi\hbar^2 = c_4 E_{\textsc{f}}$ using $m_1=0.349m_e$ and $c_4 = 0.729 \times 10^{15}$/m$^2$\,meV for all three samples. }
    \label{Fig:dispersion-p(E)}
\end{figure}

If we compare this Hamiltonian to $\hat{H}^{SYM}$ shown in Eq. (\ref{bulk_hamil_4subband_sym}), we see that it is no longer block-diagonal. Moreover, all elements, with the exception of the subband edge energies $E_{HH}^{ASYM}$ and $E_{LH}^{ASYM}$ depend on the in-plane momenta $k_x$ and $k_y$. Let us diagonalize our Hamiltonian $\hat{H}_{4}^{ASYM}$ as a function of $k_x$, keeping $k_y=0$ for the sake of simplicity. The resulting dispersion is shown in Fig.~\ref{perturb_triang}(b). The subbands form degenerate pairs only at $k_x=0$. At arbitrarily small, but finite values of $k_x$ we see the spectrum composed of four nondegenerate states. This is the central result of this analysis, showing that (i) the apparent ``zero-magnetic field spin splitting'' is actually due to the symmetry breaking, and (ii) this splitting (the Rashba effect) can be captured based on the bulk Hamiltonian, without any extra terms. The splitting of the lower-energy pair, as well as the splitting of the higher-energy pair will of course depend on the parameter $V_0$, which describes the slope of the triangular bottom of the confinement. In the limit $V_0\rightarrow 0$ we of course recover the inversion symmetric case shown in Fig~\ref{perturb_square}(b). Moreover, this splitting is also dependent on the width $W$ of the quantum well. As is evident from Fig.~\ref{perturb_triang}(a), taking $W=5$ nm resulted in relatively small differences in wave functions of the HH and LH subbands. In a wider well, these differences can become much larger, resulting in a larger gradient term, and in consequence in a bigger Rashba splitting. Further, on closer inspection of the dispersion relations, we find that their curvature changes at $k_x$ somewhat larger than $0.05$ $A^{-1}$, as was the case for the rectangular quantum well (although the critical value is not precisely the same). This reflects the change of character of the subbands: for smaller $k_x$ the two low-energy subbands are predominantly HH-like and the two high-energy subbands are predominantly LH-like, while for larger values of $k_x$ the spectral content is opposite. However, unlike in the case of the rectangular confinement, now the wave functions of the subbands are superposition of all four Bloch states.\\

\section{About spin splitting in Regime II}
\label{secA:spin-splitting}

Figure \ref{Fig:ZeemanSplit} shows the classic analysis of filling factor index versus $1/B$. In that framework, the sudden doubling of the SdH oscillation frequency between Regime I and II would normally herald the Zeeman spin splitting of carriers (here, the carriers are HH$-$ with light mass $m_1$). The kink appearing between Regime~II and Regime~III would signal the appearance of a second 2D subband. Three clearly distinct linear segments can be identified, one for each of Regimes I, II, and III. Although just about observable in older studies \cite{eisenstein1984effect}, the linear segment of Regime II has never been as clear as in the experiments reported here, for all samples (D, H, and J).

\section{Non-parabolicity of the HH$+$ subband}
\label{secA:parabolic}

Figure~\ref{Fig:dispersion-p(E)} plots $p_1$ and $p_2$ versus Fermi energy for samples D, H, and J, from the densities in Figs.~\ref{Fig:densities}(a), \ref{Fig:Subplot_all}(a), and \ref{Fig:Subplot_all}(b). Unsurprisingly, $p_2(E_{\textsc{f}})$ is supralinear in all three samples: this is another independent experimental confirmation that the HH$+$ subband is non-parabolic.


%

\end{document}